 \documentclass[final,3p,times,twocolumn,authoryear]{elsarticle}

\usepackage{amsmath,amssymb,amsthm,bm}
\usepackage{mathtools}
\usepackage[english]{babel}
\usepackage{graphicx}
\usepackage{smartdiagram}
\usepackage{pdfpages}
\graphicspath{{Figures/}}
\DeclareGraphicsExtensions{.pdf,.png}

\usepackage{tabularx,booktabs,multicol,multirow,makecell}
\usepackage{tikz}
\usetikzlibrary{arrows,shapes,positioning,shadows,trees,decorations.pathmorphing,decorations.pathreplacing,backgrounds,fit,shapes.symbols,chains,shapes.geometric}
\usepackage{caption}
\usepackage{xcolor}

\usepackage{comment}
\DeclareMathOperator*{\spn}{span}

\usepackage{enumitem}
\setlist{noitemsep,topsep = 0pt}
\usepackage{xfrac}

\usepackage[hyphens]{url}
\usepackage[hidelinks]{hyperref}
\hypersetup{breaklinks=true,pdfauthor=author} 

\journal{Astronomy and Computing}

\newcommand{\RR}{{\mathbb{R}}}

\newcommand{\NN}{{\mathbb{N}}}

\renewcommand{\vector}[1]{\bm{#1}}
\newcommand{\tivec}[1]{\tilde{\bm{#1}}}
\renewcommand{\c}[1]{\mathcal{#1}}

\renewcommand{\RR}{{\mathbb{R}}}

\begin{document}

\begin{frontmatter}

\title{1-DREAM: 1D Recovery, Extraction and Analysis of Manifolds in noisy environments.\tnoteref{1DREAM}}
\tnotetext[1DREAM]{This project has received financial support from the European Union's Horizon 2020 research and innovation program under the Marie 
Sklodowska-Curie grant agreement No. 721463 to the SUNDIAL ITN Network.}

\author[Bham]{Marco Canducci\corref{cor1}\fnref{fn1}}
\ead{M.Canducci@bham.ac.uk}%
\author[GroningenA]{Petra Awad}\ead{p.awad@rug.nl}%
\author[GroningenB]{Abolfazl Taghribi}\ead{abolfazl.taghribi@gmail.com}%
\author[GroningenB]{Mohammad Mohammadi}\ead{m.mohammadi@rug.nl}%
\author[Gent]{Michele Mastropietro}\ead{michele.mastropietro@ugent.be}%
\author[Gent]{Sven De Rijcke}\ead{sven.derijcke@ugent.be}%
\author[GroningenA]{Reynier Peletier}\ead{r.f.peletier@astro.rug.nl}%
\author[Korea,Chile]{Rory Smith}\ead{rorysmith274@gmail.com}%
\author[GroningenB]{Kerstin Bunte}\ead{kerstin.bunte@googlemail.com}%
\author[Bham]{Peter Ti{\v{n}}o}\ead{p.tino@cs.bham.ac.uk}%
\address[Bham]{University of Birmingham, School of Computer Science, B15 1TT, Birmingham, United Kingdom}
\address[GroningenA]{University of Groningen, Kapteyn Astronomical Institute, 9747 AD Groningen, The Netherlands}
\address[GroningenB]{University of Groningen, Bernoulli Institute for Mathematics, Computer Science and Artificial Intelligence, 
9700 AK Groningen, The Netherlands}
\address[Gent]{Ghent University, Department of Physics and Astronomy, Krijgslaan 281, S9, B-9000 Gent, Belgium.}
\address[Korea]{Korea Astronomy and Space Science Institute (KASI)
776, Daedeok-daero, Yuseong-gu, Daejeon, 34055,South Korea}
\address[Chile]{Universidad Technica Frederico de Santa Maria, Avenida Vicuña Mackenna 3939, San Joaquín, Santiago}

\cortext[cor1]{Corresponding author}


\begin{abstract}
Filamentary structures (one-dimensional manifolds) are ubiquitous in astronomical data sets. Be it in particle simulations or observations, filaments are always tracers of a perturbation in the equilibrium of the studied system and hold essential information on its history and future evolution. However, the recovery of such structures is often complicated by the presence of a large amount of background and transverse noise in the observation space. While the former is generally considered detrimental to the analysis, the latter can be attributed to measurement errors and it can hold essential information about the structure. To further complicate the scenario, one-dimensional manifolds (filaments) are generally non-linear and their geometry difficult to extract and model. Thus, in order to study hidden manifolds within the dataset, particular care has to be devoted to background noise removal and transverse noise modelling, while still maintaining accuracy in the recovery of their geometrical structure. We propose 1-DREAM: a toolbox composed of five main Machine Learning methodologies whose aim is to facilitate manifold extraction in such cases. 
Each methodology has been designed to address particular issues when dealing with complicated low-dimensional structures convoluted with noise and it has been extensively tested in previously published works. 
However, for the first time, in this work all methodologies are presented in detail, joint within a cohesive framework and demonstrated for 
three particularly interesting astronomical cases: a simulated jellyfish galaxy, a filament extracted from a simulated cosmic web and the stellar stream of Omega-Centauri 
as observed with the GAIA DR2. Two newly developed visualization techniques are also proposed, that take full advantage of the results obtained with 1-DREAM. 
This contribution presents the toolbox in all its details and the code is made publicly available to benefit the community.
The controlled experiments on a purposefully built data set prove the accuracy of the pipeline 
in recovering the real underlying structures. 
\end{abstract}


\end{frontmatter}


\section{Introduction}\label{sec:Intro}

 Physical structures with filament-like shapes are found in a wide variety of astrophysical domains. 
 In this work we focus on examples of filamentary (stream-like) structures, such as jellyfish galaxy tails, filaments of the cosmic web, and the tidal tails of the globular cluster Omega-Centauri ($\omega-$Cen). Since these systems have a high level of complexity, the processes by which they form and evolve are yet to be fully explored and understood, hence 
 the study of filamentary astronomical structures requires powerful computational methods for their detection, modeling and analysis. 

Each of these examples shows clear morphological significance to this work in that they all have a stream-like structure, but our interest in these examples is also inspired from their astrophysical importance. The interest in jellyfish galaxies stems from their importance in studying the evolution of galaxies in dense environments and in determining the details of environmental influences on galaxies \citep{BoselliGavazzi2006, Grossi2018}. In more details, when a dwarf galaxy enters the environment of a galaxy cluster, its interstellar matter (gas and dust) is often blown out of its body producing a long tail of the constituting matter.  Other related studies have been dedicated to examining the effect of ram pressure on the structure and dynamics of these galaxies \citep{MoriBurkert2000, MayerEtal2006, RoedigerBruggen2008,Tonnesen2012, RoedigerEtal2015, SteinhauserEtal2016, YunEtal2019, SteyrleithnerEtal2020}. The second example is that of the cosmic web that is a network of structures that naturally form as a result of gravitational instability within cosmological volumes, and that have provided great insight into gravitational structure formation, cosmological models, as well as the nature of dark matter and dark energy \citep{ParkLee2007, PlatenEtal2008, LeePark2009, LavauxWandelt2010, BosEtal2012, SutterEtal2015, PisaniEtal2015}. In addition, studies of the cosmic web have given important quantitative measures which improved our understanding of the formation and evolution of galaxies \citep{HahnEtal2007b, Hahn2009, CautunEtal2014}. As a final example, we study a stellar stream located in the halo of the Milky Way. Stellar streams are of great importance as they are imprints of past merger events in the Milky Way's formation history \citep{Johnston1996, Majewski1999, McConnachieEtal2009, MartinezDelgadoEtal2010, VeraCasanovaEtal2021}. Since the dynamics of these merger events are largely dictated by the galaxy's gravitational potential, tidal debris including stellar streams belonging to globular clusters have also been used as probes of the galactic potential as they move through it \citep{KupperEtal2015, PearsonEtal2015, ThomasEtal2017, ThomasEtal2018, BonacaHogg2018, MalhanIbata2019}. 

Extracting topological information from point clouds of discrete data forming the above structures is a difficult task, whether these data sets are provided by $N$-body simulations or by observational surveys. For example, algorithms studying the cosmic web should keep in mind its anisotropic, hierarchical nature which forms different morphological structures spanning over six orders of magnitude in density \citep{CautunEtal2014}. Similarly for the other studied objects, structure detection and learning algorithms have to face problems including the very large numbers of high-dimensional data points as well as the handling of noise or outliers that affect the results of manifold learning and dimensionality reduction. That being said, several methods and algorithms have been developed aiming to extract astrophysical information provided by the structures previously mentioned.
Starting with jellyfish galaxies, morphometry, or the quantitative study of morphological properties, has been used to study ram pressure stripping in \citet{McPartland2016} and as a probe of the evolution of these galaxies in \citet{RomanOliveira2021}. A myriad of algorithms has also been developed for the tracing and studying of the various components of the cosmic web (e.g. Multiscale Morphology Filter [MMF], \citet{AragonCalvoEtal2007}; ORIGAMI, \citet{FalckEtal2012}; NEXUS+, \citet{CautunEtal2013}; Minimum Spanning Tree [MST], \citet{AlpaslanEtal2014}; Bisous, \citet{TempelEtal2016}). 
We refer to \citet{Libeskind2018} for a comparison between many of these algorithms. As for stellar streams, several techniques have been used for their detection and analysis, 
exemplified by
: the Matched Filter (MF; \citet{Rockosi2002, Balbinot2011}), detecting co-moving groups of stars \citep{WilliamsEtal2011, ArifyantoFuchs2006, DuffauEtal2006}, the Streamfinder algorithm \citep{MalhanIbata2018}, and several others. 
Despite the large number of techniques used to study these systems, the need to keep up with the growing size of astronomical data sets and with the many complexities of astrophysical systems is ever-present. Therefore, the development of new Machine Learning algorithms with astrophysical applications creates great potential towards handling larger amounts of data as is needed for the exploration of the galactic halo and stellar streams belonging to it, as well as towards handling all the challenging properties of complex structures as required for the analysis of jellyfish tails and the cosmic web. 

When studying noisy low-dimensional manifolds, it is often necessary to distinguish between two kinds of noise. In the case of point-clouds, these can be defined as background and transverse noise, however the actual distinction between the two is potentially difficult. While background noise is usually referred to as a contaminant to the data, corrupting information hidden within it, transverse noise may hold useful information about the sub-structures. Due to the local overlap between these two types of noise in the vicinity of the sub-structures, discerning between the two contributions can be a difficult task. Nevertheless, since most Manifold Learning techniques are not designed to deal with such ``corruption'' of the data, it is generally good practice to address this problem before their application. A number of filtering and denoising techniques have been devised that aim at reducing the noise over a point-cloud (see \cite{HAN2017103} for an extensive review). However, these seem to work efficiently only in mild cases, where the density of background noise is far lower than the one of the structure. On the other hand, the ant system \citep{AntSystem_Dorigo1999} and ant-colony system \citep{AntColony_Gambardella1996} have been applied efficiently to denoising in presence of a moderate amount of background noise \citep{AntColony_ChuShuChuan2004}. 
Nevertheless, due to the adopted distance metric (Euclidean) and the computational complexity, these methods can be ineffective on large data sets presenting noisy, non-linear manifolds. 

When manifolds are expected within a noisy point-cloud, more direct methodologies are usually applied in order to let their mean curve/surface emerge from the noise. 
In this class of methodologies, there is generally no distinction between background and transverse noise. Local smoothing has been successfully applied to noisy manifolds with a mild level of noise, using different approaches. In \cite{JinHyeong_2004} a weighted version of Principal Component Analysis (PCA) is applied to local neighborhoods, estimated in terms of a point's $k-$nearest neighbors. Then, points in the neighborhood are projected onto the hyper-plane defined by the weighted PCA. A different approach is presented in \cite{Haifeng_2006}, where a linear error-in-variables (EIV) is estimated for each local patch and the smoothed coordinates of noise-less points are derived. Global coordinates are then recomputed for each point, merging partially overlapping neighborhoods. 

A different class of methods aims at projecting neighboring points to the locally estimated tangent space to noisy manifolds. The tangent space estimation can be either performed on the high-dimensional sample \citep{ZhonghuaHao_2017} or the low-dimensional projection (e.g. \cite{yao2019manifold}). Manifold Blurring Mean Shift (MBMS, \cite{WeiranWang_2010}) adopts this formalism, by gradually moving neighboring points along the orthogonal direction to the manifold. Again, the local tangent space estimation is performed via PCA, in small neighborhoods centered on each point.  
A slightly different approach is presented in \cite{lyu2019manifold}, where Non-linear Robust PCA is introduced. Here, local patches are decomposed into a low-rank and a sparse component that account for the tangent space and noise information of the patch, respectively. 

Another strand of work uses a diffusion formulation over the noisy cloud to enhance the manifold's spine. In particular, \cite{hein2007manifold} first construct an asymmetric $k-$nearest neighborhood graph of the point cloud and derive its graph Laplacian. This serves as a generator for the diffusion process, which is solved in terms of a differential equation. A similar approach but with a physically inspired formulation is presented in \cite{8046026}. In contrast to the previous methodology, this method does not use a graph to represent the data and gradually moves points towards high density regions in the data. 

As previously mentioned, these methodologies are used as a pre-processing step when the data is corrupted by noise. Further steps are necessary to recover explicit formulations of the hidden manifolds and their low-dimensional representations. This branch of work falls within the scope of Manifold Learning and has been addressed in a variety of different methodologies. The stepping stones to this field are \emph{Locally Linear Embedding} (LLE, \cite{Roweis00nonlineardimensionality}) and \emph{Complete isometric feature mapping} (Isomap, \cite{Tenenbaum2319}). The methodologies have been further refined and new algorithms defined such as Laplacian- and Hessian-eigenmaps \citep{Belkin01laplacianeigenmaps,Donoho5591}, Local Tangent Space Alignment (LTSA, \cite{doi:10.1142/9789812779861_0015}) and Riemannian Manifold Learning \citep{10.1109/TPAMI.2007.70735}. However, their performance is often hampered by the presence of noise. Aided by a precise formulation of transverse noise, Generative Topographic Mapping (GTM, \cite{Bishop1998GTMTG}) solves this issue by modelling the manifold as a Gaussian mixture, having centers constrained to lie on the image of a low-dimensional unit interval
(of the same dimension as the manifold) smoothly embedded in the ambient space. 

To further complicate the scenario, information may be lying on multiple noisy manifolds in real-world data sets. Generalizations of existing methods to this case were proposed accordingly, giving rise to algorithms such as \emph{Multi Manifold Isomap} (M-Isomap \cite{Fan:2016:EIM:2903049.2903101}), \emph{Multi-Manifold LLE} (MM-LLE \cite{Hettiarachchi:2015:MLL:2791619.2792198} and \emph{Hierarchical-GTM} \citep{Tino_PAMI_2002}. However, carrying the same assumptions and design of their predecessors, they suffer from the same problems (e.g. pre-defined manifold's intrinsic dimension, transverse noise, topologically difficult manifolds). Other techniques gave new perspectives on the problem; examples are \emph{Sparse Manifold Clustering and Embedding} (SMCE \cite{NIPS2011_4246}) and \emph{Manifold Deflation} \citep{Ting2020ManifoldLV}, where manifolds (and their low-dimensional representations) are recovered by means of a graph representation of data. More recently, new mathematical tools have been developed and widely used in multiple fields, although mainly constrained to work in three dimensions (3D point clouds for surface recovery). \emph{Computational Geometry} \citep{boissonnat:hal-01615863} represents manifolds as Simplicial Complexes and refines these representations via triangulations and filtrations. Again though, transverse noise may heavily affect the results and hampers their accuracy.

The general assumption for these methodologies is that the intrinsic dimensionality of the manifolds is known a priori. This is not often the case, so much so that particular effort has been devoted into developing methods able to estimate it (semi-)automatically. In \cite{Haro00translatedpoisson}, the \emph{Translated Poisson Mixture Model} (TPMM) is used to estimate the dimensionality of data in local neighborhoods and partition it accordingly, while \emph{Hidalgo} \citep{Hidalgo2020} is a Bayesian extension of \emph{TWO-NN} \citep{Facco2017EstimatingTI}, where the dimensionality information is recovered for each point based on the distance to its closest neighbor. Other methods consist in evaluating the local covariance matrix within a pre-specified small volume centered on each data point and analyzing its eigen-decomposition (e.g. \cite{MordohaiMedioni2005} and \cite{10.5555/1756006.1756018}). 
Despite all 
effort spent in recovering low-dimensional noisy manifolds in a noisy environment, to the best of our knowledge, a complete, coherent formulation that is also flexible and straight-forward to use is still missing in the current scenario. 
In order to address all the issues presented so far in the context of multiple noisy manifolds learning in a noisy environment, we propose a cohesive toolbox for denoising and 1D manifold (filament) extraction. The toolbox consists of five methodologies that have been exhaustively tested in separate works, however this is the first time that they all come to fruition in a single environment.

The aim of this work is to present all methodologies in detail, highlighting their functionalities and main objectives and demonstrating (with the user in mind) how the various tools can be combined in different ways for a variety of astronomical applications, using both observed and simulated data. We show how their application on astronomical data sets may drive scientific inference on the underlying physical processes and main properties of the studied objects. 
The complete implementation of all methods 
can be found at the following online repository: \url{https://git.lwp.rug.nl/cs.projects/1DREAM}.
\subsection{Organization of the paper}
\begin{table}[h]
\centering
\def\myCWidth{0.12}
\caption{Information about adopted data sets for experimental sections.}
\label{tab:Datasets}
\begin{tabularx}{\columnwidth}{X | X | X }
\toprule
\textbf{Data set} & \textbf{Size} & \textbf{Attributes}\\
\toprule
Synthetic & $4\times 10^4$  & $\vector{t},h,T_1,T_2$ \\
Jellyfish & $9\times 10^4$  & $\vector{t},\rho,T,\mathrm{[Fe/H]},m$ \\
Cosmic Web & $\sim 1\times 10^6$ & $\vector{t}, \vector{\nu}$\\
$\omega_{Cen}$ & $\sim 2\times 10^4$ & $\ell, b$\\
\bottomrule
\end{tabularx}
\end{table}
In Section \ref{sec:Synt_Data} we introduce the synthetic data set used for the controlled experiments in the following section. In order to estimate the efficiency of the methodologies, we carefully construct a mock data set to test the success of our tool-kit in fitting and recovering the known properties of the mock. In particular, we create a point-cloud presenting two elongated non-linear filaments. Two variables exhibit well-defined behaviours along and across the two filaments. In Section \ref{sec:Algorithm}, we present the five different algorithms introduced in this work. We describe each algorithm in detail and apply it to the mock data set. In section \ref{sec:Visualization} we outline two powerful visualization techniques that take full advantage of the methodologies described in sec. \ref{sec:Algorithm}. Finally, the whole methodology, aided by the visualization techniques, is applied to three different astronomically relevant data sets, namely:
\begin{enumerate}
    \item A temporal snapshot of a simulated dwarf galaxy falling into the gravitational potential of a Fornax-Cluster-like galaxy cluster. The simulation is performed using a modification to GADGET-2 \citep{Springel_2005}, an $N$-body/SPH (Smoothed-Particle Hydrodynamics) code, where a moving box follows closely the evolution of the simulated object. Our goal in this case is to study the properties of other simulated quantities, such as temperature and metallicity,     
    to assess if the recovered streams are loci of Star Formation $[\sim 9 \times 10^4~\mathrm{points}]$. 
    \item A temporal snapshot of a large scale formation, Dark Matter only, $N-$body simulation performed with the GADGET-3, where we extract filaments of the cosmic web and study the dynamics of the belonging particles. $[\sim 1.99\times 10^6~\mathrm{points}]$ 
    \item Stellar stream filaments from the GAIA data set. We focus on one particular filament as the remnant of a previous interaction between our Galaxy and an external object $[\sim 2\times 10^4~\mathrm{points}]$.
\end{enumerate}
A summary of properties of the individual data sets can be found in tab. \ref{tab:Datasets}, where the number of particles (Size) and attributes per particle are shown.
We describe these data sets and their analysis in detail in Section \ref{sec:experiments} and draw our conclusions in Section \ref{sec:Conclusions}.

\section{Synthetic data set for controlled experiment}\label{sec:Synt_Data}

While the methods can be applied without prior knowledge on the data set at hand, it is essential that the results coming from a carefully constructed case mirror the true nature of the underlying problem. 
This serves as a synthetic representation for which the ground truth is known to demonstrate our toolbox. 
The data set described here and used throughout section \ref{sec:Algorithm} is designed to morphologically resemble a jellyfish galaxy by defining three major noisy manifolds $\c{M}_k$ with $k=1,2,3$. 
$\c{M}_1$ is a Gaussian distribution having mean $\mu_1 = (0,0,0)$ and covariance matrix
\[
\bm{\Sigma}_1 = 
\begin{pmatrix}
 4 & 0 & 0\\
 0 & 8 & 0\\
 0 & 0 & 26
\end{pmatrix} \enspace.
\]
Thus manifold $\c{M}_1$ is a thick, roughly one-dimensional, elongated structure. 
The thickness of the manifold is large enough to connect manifolds $\c{M}_2$ and $\c{M}_3$. In this example, this manifold represents the ``head'' of the jellyfish galaxy.
Manifolds $\c{M}_2$ and $\c{M}_3$ represent two streams departing from the ``head'' of the jellyfish, simulating the effect of a dynamical process disrupting the main body of the galaxy. They serve as two distinct parts of the ``tail'' and they are inherently one-dimensional: their underlying true structure is the unit interval $[-1, 1]$ embedded in $\RR^3$ through mapping functions $f_2: [-1, 1] \rightarrow \RR^3$ and $f_3: [-1 ,1]\rightarrow \RR^3$. The mapping functions take the form:
\begin{equation}\label{eq:Embeddings}
    f_2(\vartheta) = 
    \begin{bmatrix}
    10\vartheta+8\\ 
    -3-(2\vartheta+2)^2 \\
    5\sin(\pi\vartheta)+5
    \end{bmatrix},~
    f_3(\vartheta) = \begin{bmatrix} 
    -10\vartheta-12\\
    7-4\sin^2\left( \frac{\pi\vartheta}{2} \right)\\ 
    5\cos(\pi\vartheta)-3 
    \end{bmatrix}
\end{equation}
where $\vartheta \in [-1, 1]$.
The underlying one-dimensional structure of the manifolds is then convolved with noise in order to obtain a morphological analogy with the structures generally found in astronomical simulations and observations. Both manifolds have an overlapping double noise structure: the thin and dense, inner region (\emph{core}) and the thick, sparser layer (\emph{sparse}), both defined as Gaussian mixtures.
\begin{figure}[t]
  \centering
  \includegraphics[width=\columnwidth,trim = 1.35cm 0.3cm 1.5cm 1cm,clip]{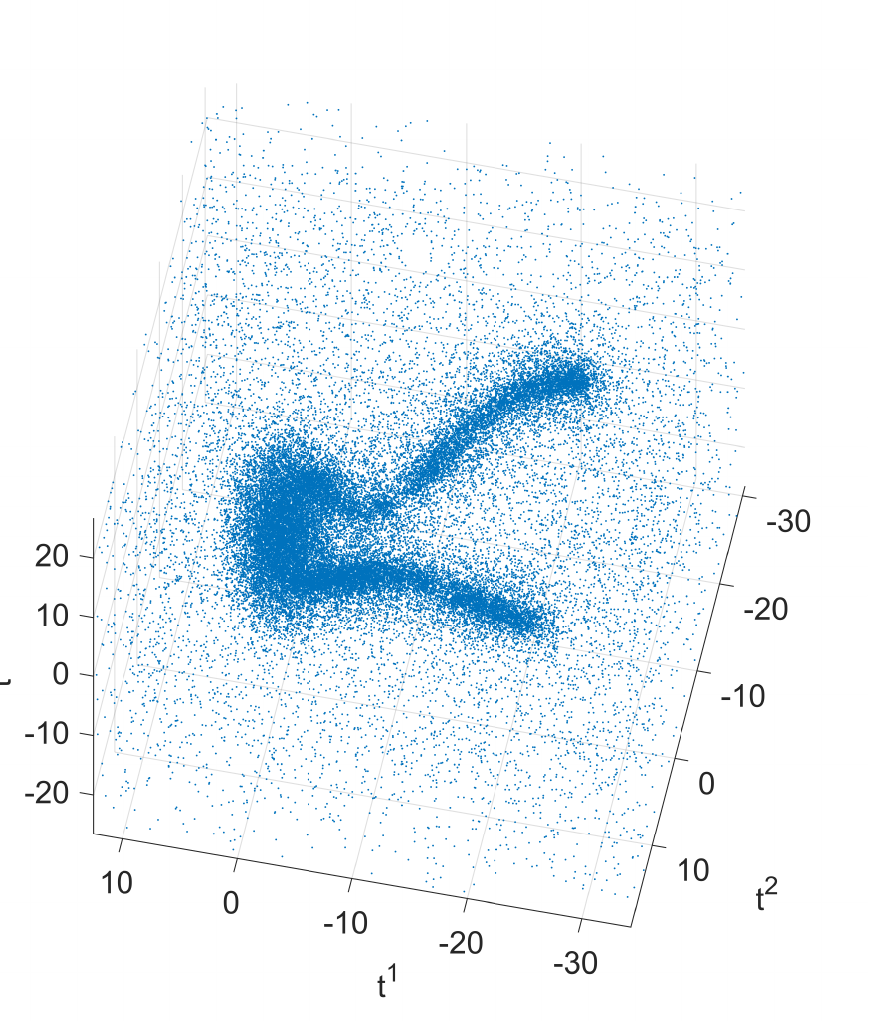}  
  \caption{Final noisy mock data set presenting two noisy elongated filamentary structures, connected to a noisy head and embedded in a noisy environment.}
  \label{fig:JF_Synth_Noisy}
\end{figure}
The dense core of manifold $\mathcal{M}_2$ is defined as a flat Gaussian mixture whose $N^c_2 = 43$ centers lie on $L^c_2 = f_2([-1 ,1])$ (each distancing from the adjacent by $0.5$). The shared covariance matrix for the core Gaussian mixture is the identity matrix $\Sigma^c_2 = \mathbf{I}$ and the component weights are $\pi^c_2 = 1/N^c_2$.
The sparse Gaussian mixture has $N^s_2 = 22$ centers lying on the segment $L^s_2 =L^c_2$ (the distance between adjacent centers is $1$). The shared covariance matrix is $\Sigma^s_2 = 9\mathbf{I}$ and components weights are $\pi^s_2 = 1/N^s_2$.\\
Manifold $\mathcal{M}_3$ has the same overall structure as manifold $\mathcal{M}_2$. The core structure has $N^c_3 = 21$ centers, regularly sampling segment $L^c_3 = f_3([-1, 1])$, while the sparse structure has $N^s_3 = 11$ centers on $L^s_3 = L^c_3$. Consequently, the mixture components for core and sparse structures are $\pi_3^c = 1/N^c_3$ and $\pi_3^s = 1/N^s_3$ respectively. The shared covariance matrices are $\Sigma^c_3 = \Sigma^c_2$ and $\Sigma^s_3 = \Sigma^s_2$ for the core and sparse components, respectively. 
From each manifold $\mathcal{M}_k$, $k = 1,2,3$ we can now generate a point cloud in $\RR^3$, obtaining $\mathcal{P}_1, \mathcal{P}_2$ and $\mathcal{P}_3$. The union $\mathcal{Q} = \bigcup_{k=1,2,3} \mathcal{P}_k$ of all point clouds represents the morphology of the synthetic data set depicted in Figure \ref{fig:JF_Synth_Noisy}. Each manifold and the noisy point distribution are sampled by $10^4$ points, making the size of the synthetic data set $4\times 10^4$ points.

\subsection{Behaviour of simulated physical properties}\label{sec:Quantities}

\begin{figure}[!ht]
\begin{tikzpicture}[node distance = 0cm,nodes = {anchor=north west,inner sep=0cm},font=\footnotesize]
\node[] (a_cap) {\textbf{(a)} Variable $T_1$ for toy data set without background noise.};
\node[anchor=north west] (a) at (a_cap.south west) {\includegraphics[width=\columnwidth,trim = 0.1cm 0.3cm 1.1cm 1cm,clip]{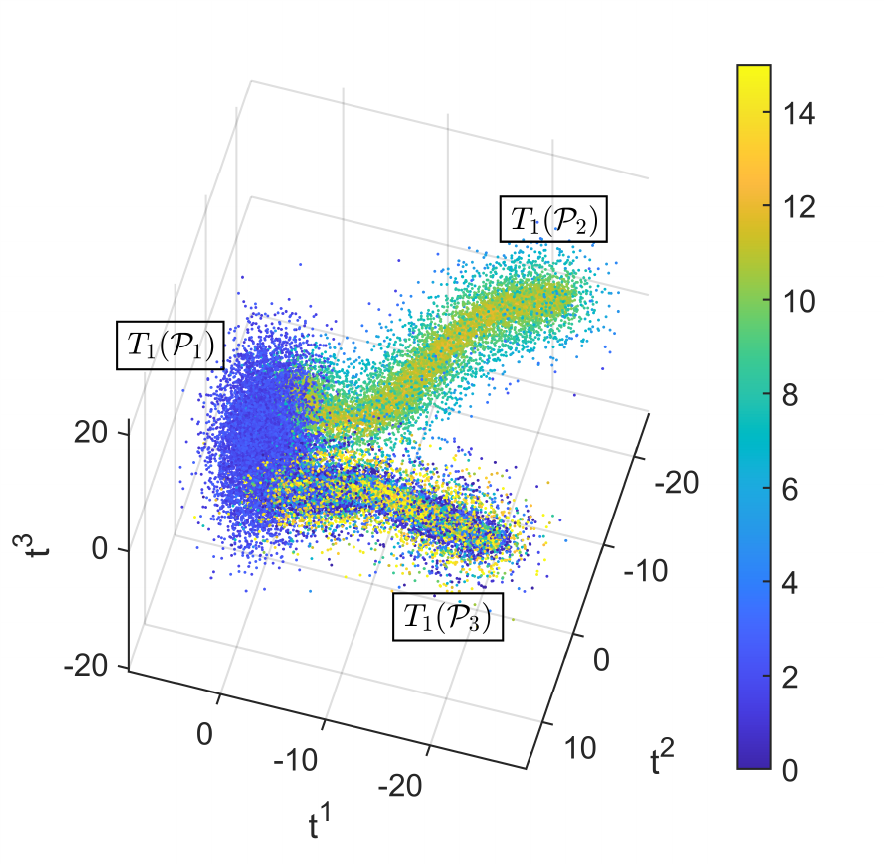}};
\node[] (b_cap) at (a.south west) {\textbf{(b)} Variable $T_2$ for toy data set without background noise.};
\node[] (b) at (b_cap.south west) {\includegraphics[width=\columnwidth,trim = 0.1cm 0.3cm 1.1cm 1cm,clip]{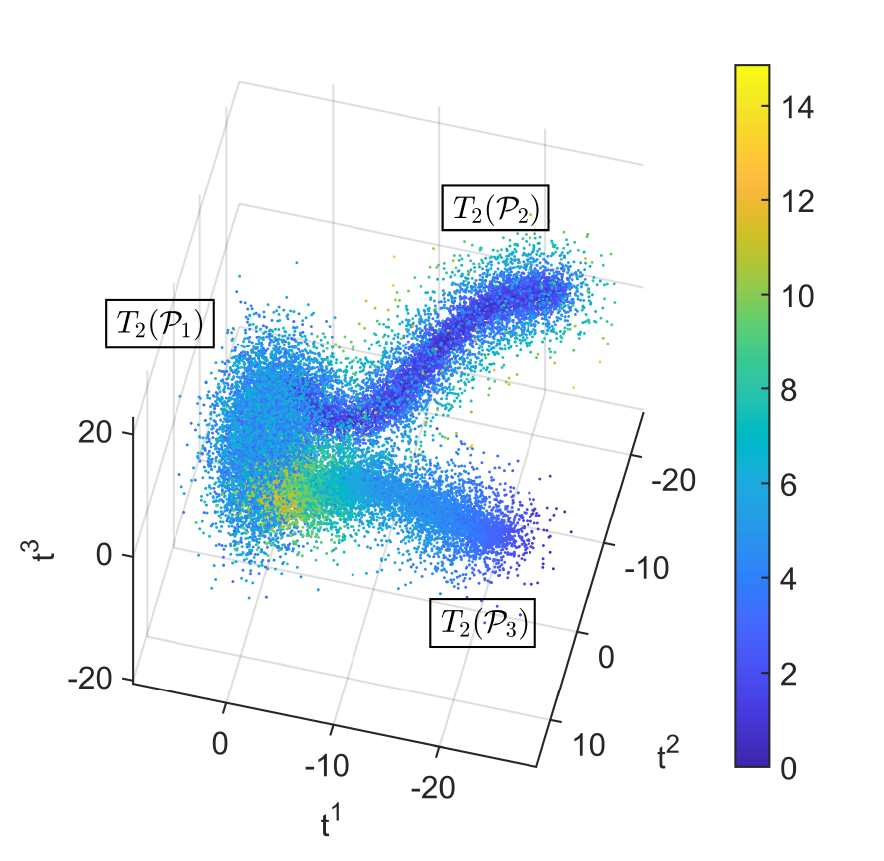}};
\end{tikzpicture}
\caption{Distribution of variables $T_1$ and $T_2$ for manifolds $\mathcal{M}_1$, $\mathcal{M}_2$ and $\mathcal{M}_3$. Panel (a) and (b) mimic a stage of a simulated jellyfish galaxy. Variable $T_1$ shows a sinusoidal behaviour across the radial direction  of manifold $\mathcal{M}_3$ and a decreasing one along the the radial direction of $\mathcal{M}_2$ (panel (a), lower and upper filaments respectively), while being uniformly distributed throughout $\mathcal{M}_1$. Variable $T_2$ is decreasing along the longitudinal direction of manifold $\mathcal{M}_3$ and the radial direction of manifold $\mathcal{M}_2$ (panel (b)). It is again uniformly distributed on $\mathcal{M}_1$. The uniform noise shown in figure \ref{fig:JF_Synth_Noisy} is here omitted for clarity.\label{fig:ToyDataset}} 
\end{figure}

Following the SPH formulation (\cite{PRICE2012759}, \cite{2010PhDT.......301C} and citations therein), we consider each particle $\vector{t}_i$ in data set $\mathcal{Q}$ to sample a spherical volume of radius $r = h_i$. Here $h_i$ is the \emph{smoothing length} equal to the distance of particle $\vector{t}_i$ to its 50-th neighbor in data set $\mathcal{Q}$. Under the assumption of mass preservation \citep{1977MNRAS.181..375G}, we assign a constant value $m_i = 1$ to the mass contained in each volume sampled by particle $\vector{t}_i$. We can now define the density of the sampled volume as:
\begin{equation}\label{eq:densitySPH}
    \rho_i = \frac{m_i}{(4/3)\pi h_i^3}.
\end{equation}
We also define two additional quantities having particular pre-designed behaviours in proximity to the two manifolds in the data set. Quantity $T_1$ is defined to be uniformly distributed in the interval $[T^{\min}_1,T^{\min}_1 +1]$ for particles of manifold $\mathcal{M}_1$, decreasing from the center of manifold $\mathcal{M}_2$ ($f_2([-1, 1])$) and sinusoidally varying depending on the distance to the center (radial) of manifold $\mathcal{M}_3$ ($f_3([-1, 1])$):
\begin{align}
    \label{eq:V1M1}
    T_1(\mathcal{P}_1) &= X_1 \sim \mathcal{U}(T^{\min}_1,T^{\min}_1+1);\\
    \label{eq:V1M2}
    T_1(\mathcal{P}_2) &= \frac{\delta^{~+}_{(2,r)} - d(\vector{t}_i,L^c_2)_{\mathcal{P}_2}}{4} T^{max}_1;\\
    \label{eq:V1M3}
    T_1(\mathcal{P}_3) &= 1.5 T_{\max} \Bigg\{1 + \sin{\Bigg[
    16\pi\frac{d(\vector{t}_i,L^c_3)_{\mathcal{P}_3} - \delta^{~-}_{(3,r)}}
    {\delta^{~+}_{(3,r)} - \delta^{~-}_{(3,r)}}\Bigg]}\Bigg\}.
\end{align}
Where $d(\vector{t}_i,L^c_k)_{\mathcal{P}_k}$ is the distance between $\vector{t}_i \in \mathcal{P}_k$ and the segment $L^c_k = L^s_k = f_k([-1, 1])$, for $j=2,3$; $\delta^{~+}_{(k,r)} = \max_{\vector{t}_i \in \mathcal{P}_k}d(\vector{t}_i,L^c_k)$ and $\delta^{~-}_{(k,r)} = \min_{\vector{t}_i \in \mathcal{P}_k}d(\vector{t}_i,L^c_k)$ are the maximum and minimum radial distances, respectively, within all points $\vector{t}_i \in \mathcal{P}_k$ from the core of manifold $\mathcal{M}_k$. 
Figure \ref{fig:ToyDataset}a 
shows the radial profiles of quantity $T_1$ for the three manifolds.

Quantity $T_2$ is uniformly sampled within the interval $[T^{\min}_2,T^{\min}_2 +1]$ for all particles in $\mathcal{P}_1$, it has an increasing radial profile from the center of manifold $\mathcal{M}_2$ and it decreases along the longitudinal axis of manifold $\mathcal{M}_3$:
\begin{align}
    \label{eq:V2M1}
    T_2(\mathcal{P}_1) &= X_2 \sim \mathcal{U}(T^{\min}_2,T^{\min}_2+1); \\
    \label{eq:V2M2}
    T_2(\mathcal{P}_2) &= \frac{d(\vector{t}_i,L^c_2)_{\mathcal{P}_2} - \delta^{~-}_{2,r}}{4} T^{max}_2;\\
    \label{eq:V2M3}
    T_2(\mathcal{P}_3) &= 3T^{\max}_2\Bigg[ \frac{\delta^{~+}_{3,L} - d(\vector{t}_i,f_3(-1))}{\delta^{~+}_{3,L} - \delta^{~-}_{3,L}} \Bigg] \enspace,
\end{align}
where the quantities $\delta^{~+}_{3,L} = \max_{\vector{t}_i \in \mathcal{P}_3} d(\vector{t}_i,f_3(-1))$ and $\delta^{~-}_{3,L} = \min_{\vector{t}_i \in \mathcal{P}_3} d(\vector{t}_i,f_3(-1))$ are the maximum and minimum (longitudinal) distances, respectively, within all particles $\vector{t}_i \in \mathcal{P}_3$ from point $f_3(-1)$, representing the head of manifold $\mathcal{M}_3$.
In figure \ref{fig:ToyDataset}b 
the radial profiles of quantity $T_2$ for the three manifolds can be found. We omit here the background noise. 
\section{Algorithms}\label{sec:Algorithm}
\begin{figure*}[t]
\begin{tikzpicture}[node distance = 0cm,nodes = {anchor=west,inner sep=0cm},font=\footnotesize]
\node[] (a)             {\includegraphics[width=0.199\linewidth]{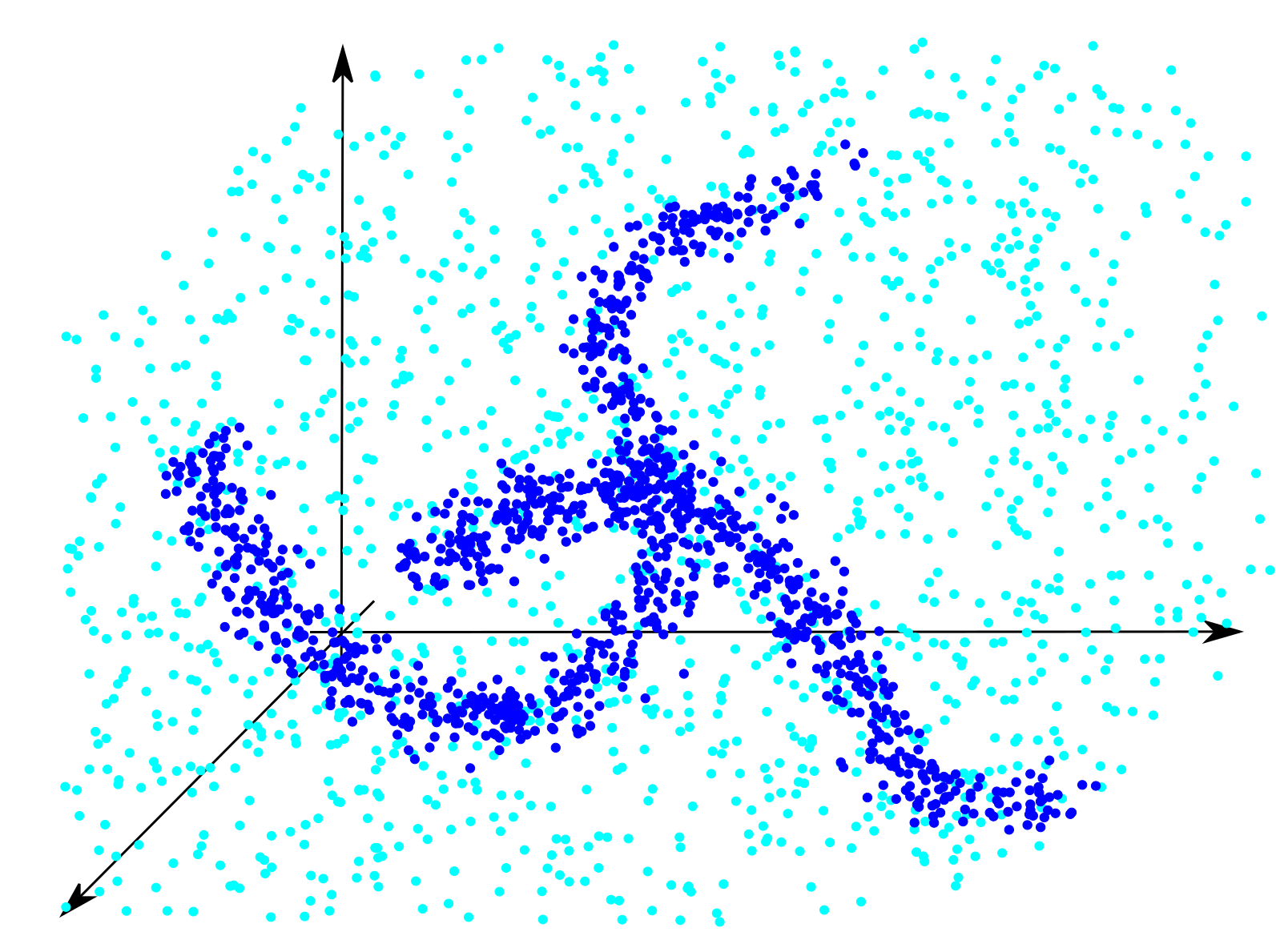}};
\node[] (b) at (a.east) {\includegraphics[width=0.199\linewidth]{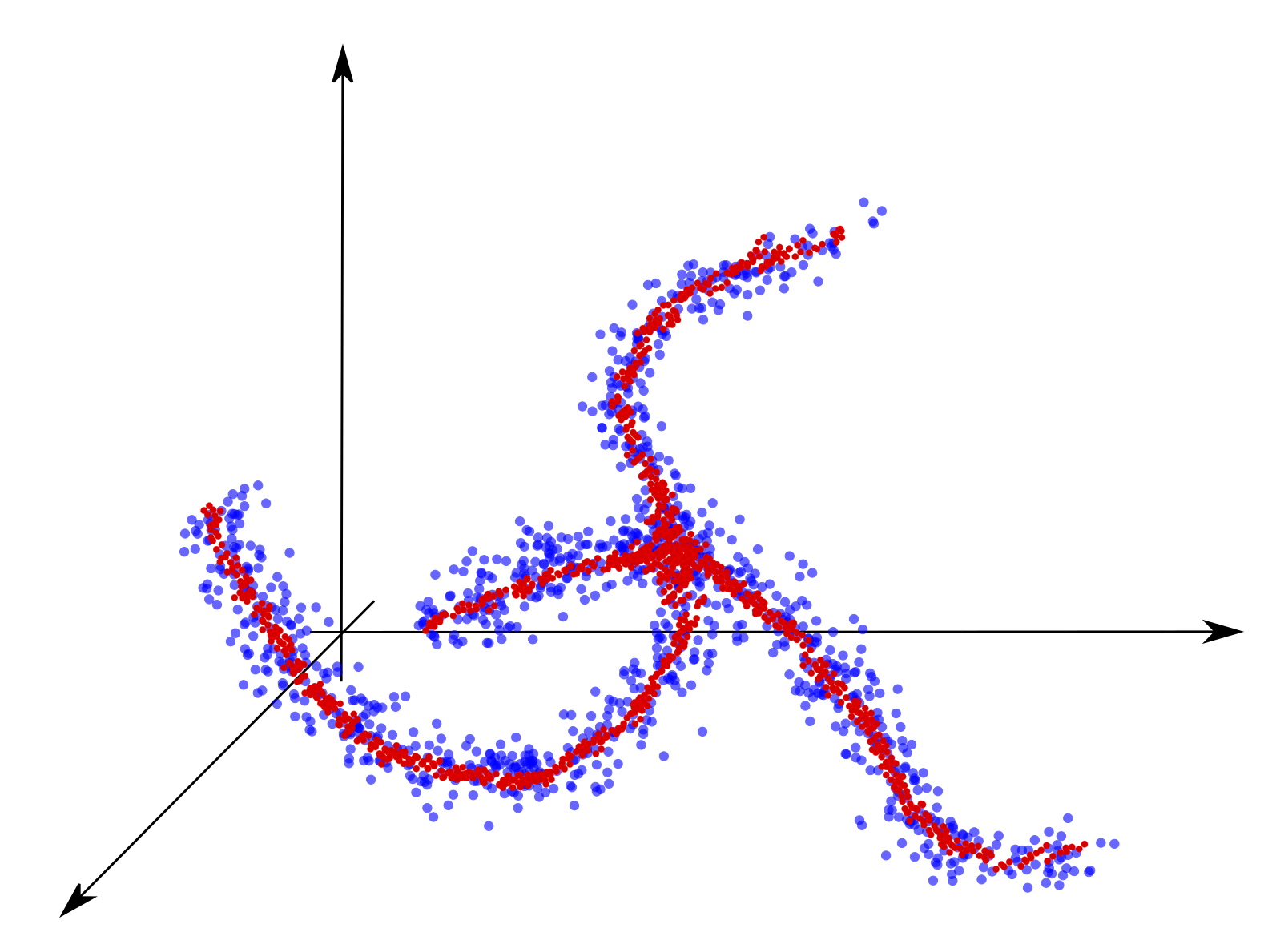}};
\node[] (c) at (b.east) {\includegraphics[width=0.199\linewidth]{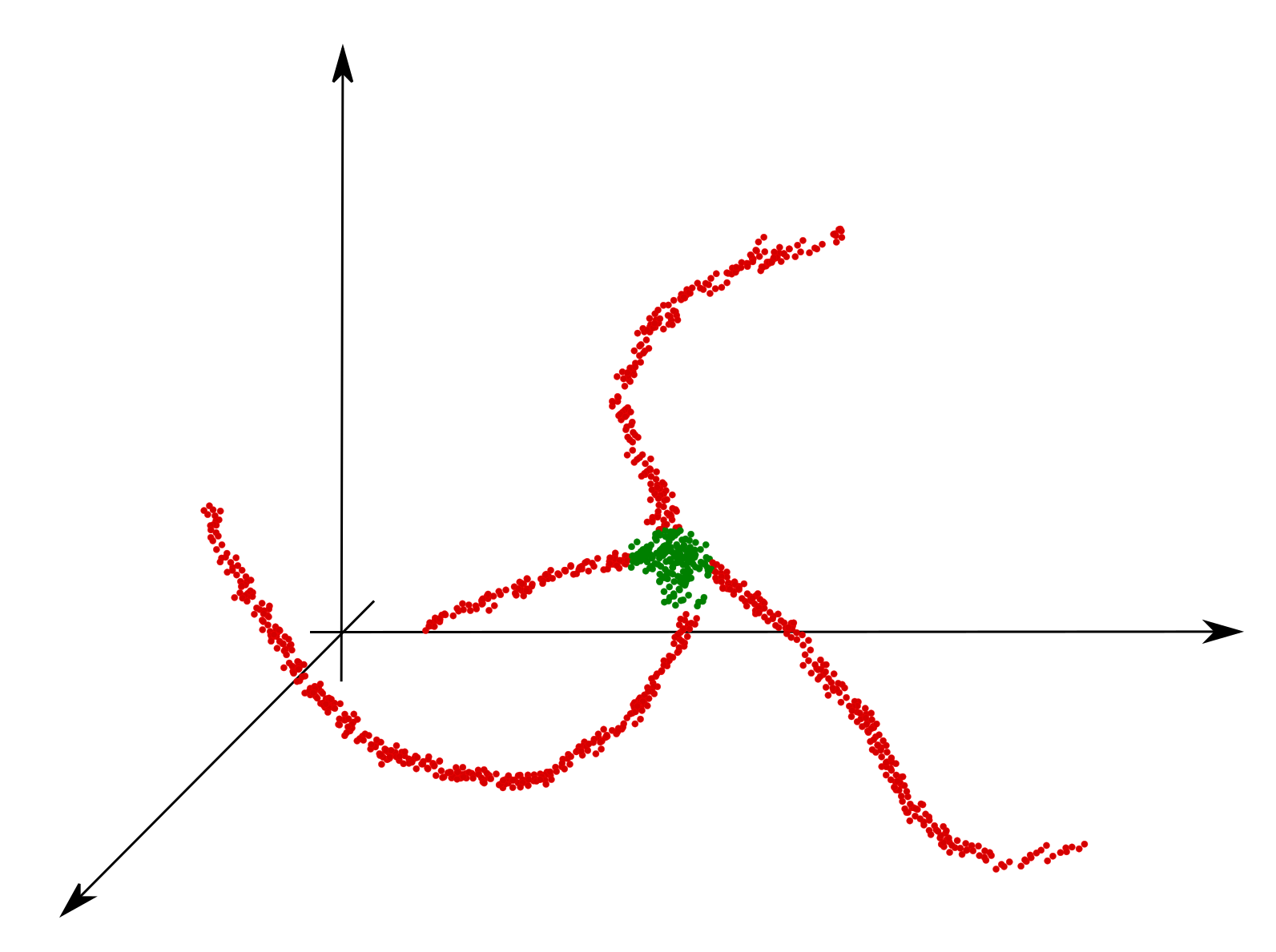}};
\node[] (d) at (c.east) {\includegraphics[width=0.199\linewidth]{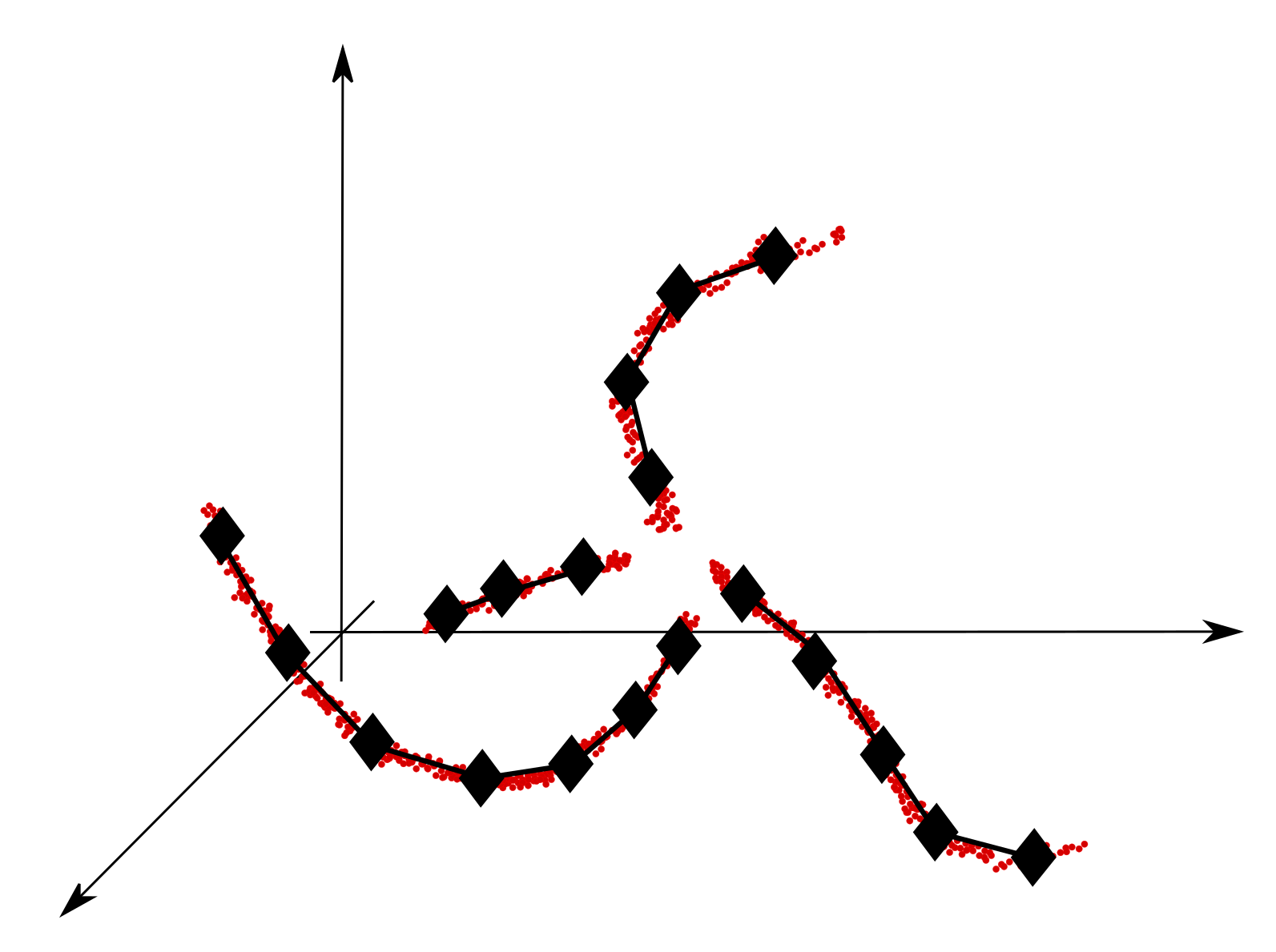}};
\node[] (e) at (d.east) {\includegraphics[width=0.199\linewidth]{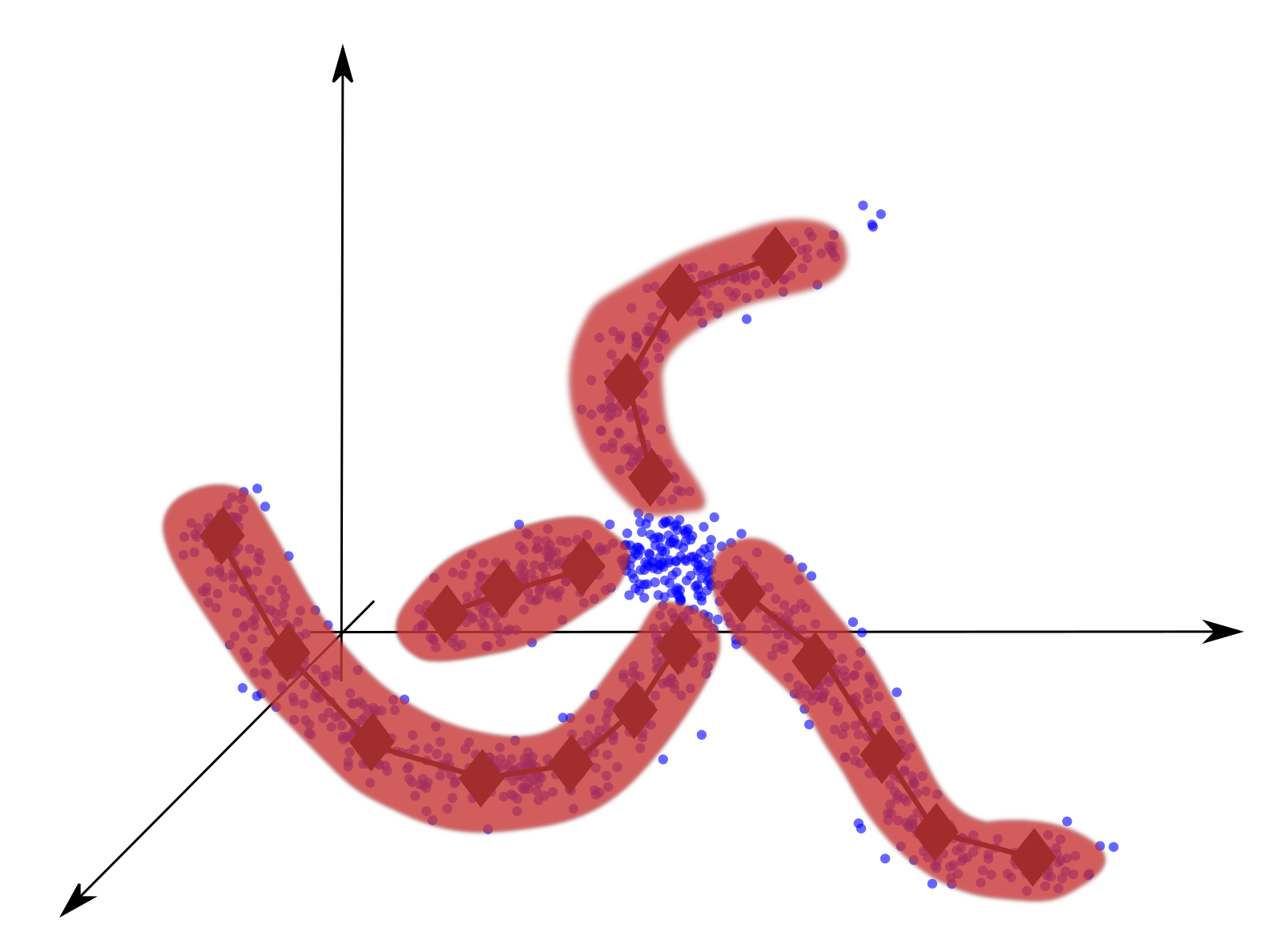}};
\node[anchor=south] (a_cap) at (a.north) {\textbf{(a)} LAAT};
\node[anchor=south] (b_cap) at (b.north) {\textbf{(b)} EM3A};
\node[anchor=south] (c_cap) at (c.north) {\textbf{(c)} Dimensionality Index};
\node[anchor=south] (d_cap) at (d.north) {\textbf{(d)} Crawling};
\node[anchor=south] (e_cap) at (e.north) {\textbf{(e)} SGTM};
\end{tikzpicture}
\caption{Sketches depicting the five methodologies for structure detection (a), denoising (b), dimensionality index (c), crawling (d) and modeling (e) proposed in our toolbox.}\label{fig:Sketches}
\end{figure*}
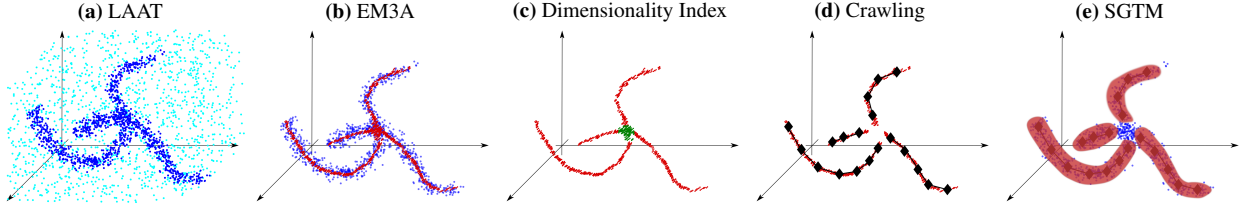
The different methodologies forming the algorithm are here briefly described in order to let the reader have a complete qualitative overview. Each method is further described in detail in the following sections and graphically depicted in figure \ref{fig:Sketches}. 
\begin{description}
\item[LAAT:] (fig. \ref{fig:Sketches}a) Taking advantage of the Ant Colony Optimization methodology, \emph{Locally Aligned Ant Technique} (LAAT) aims at enhancing the contrast between high and low density regions in a point cloud via the use of pheromone. This scalar field is used to distinguish between high and low density regions in the data set. By selecting a threshold in the pheromone value, it is possible to filter out particles while preserving those lying in a dense environment. If sub-structures are hidden within the point-cloud, LAAT helps in uncovering them with an adjustable parameter (threshold);
\item[EM3A:] (fig. \ref{fig:Sketches}b) 
The 
\emph{Evolutionary  Manifold  Alignment  Aware Agents} (EM3A) algorithm aims at enhancing density contrasts in the data set by pushing particles in high density regions towards an empirically estimated mean curve of the hidden sub-structures, using a similar framework as LAAT. The result of this procedure is a ``diffused'' point-cloud where the transverse noise to sub-structures is greatly reduced, enabling a more efficient application of the subsequent methodologies;
\item[Dimensionality Index:] (fig. \ref{fig:Sketches}c) The resulting points in the diffused data set may belong to structures of different, low, intrinsic dimension. By eigen-decomposition of local neighborhoods, the \emph{Dimensionality Index} assigns to each point in the diffused (and respectively, the noisy) data set, an integer label denoting its intrinsic dimension. The labels are used to partition both the noisy and the diffused data sets into their low-dimensional counterparts;
\item[1D Multi-Manifold Crawling:] (fig. \ref{fig:Sketches}d) operates on the one-dimensional partition of the diffused data set.
As an iterative procedure, it discovers filaments in the point-cloud and constructs their corresponding skeletons in the data space, while building their low-dimensional representations. 
The procedure operates a small agent walking (crawling) along the diffused point-cloud following the direction given by the local tangent space estimation. 
The procedure ends by depletion of the data set of the visited regions by crawling. Its end result is an atlas of structures recovered from the data set, each one with its low- and high- dimensional representations. The previously described methodologies are unable to distinguish between different substructures. Their aim is to enhance the possibility of their detection by filtering out (LAAT) or reduce (EM3A) background and transverse noise respectively. It is only with 1D Multi-Manifold Crawling that the low-dimensional structures are detected, separated and pre-modelled (via their low-dimensional representation). While the outcome of the previous methodologies is a global point-cloud, the result of 1D Multi-Manifold Crawling is a set of partitions of the data set, where each partition is a detected structure that carries a low-dimensional representation.
\item[Stream GTM:] (fig. \ref{fig:Sketches}e) Since ultimately all structures are initially noisy and only by pre-processing we are able to recover their skeleton, \emph{Stream Generative Topographic Mapping} (Stream GTM) builds a probabilistic model for each extracted sub-structure, describing the transverse noise distribution along the manifold itself as a constrained Gaussian mixture model. 
This allows for a more natural representation of the filaments and serves as a tool for further analysing the recovered structures.
\end{description}
Via the use of the methodologies composing 1-DREAM it is then possible to identify varying density, low-dimensional regions in any particle data set and to recover filament-like structures hidden within a noisy environment. Furthermore, the structure of individual filaments is regularized via their probabilistic formulation introduced in SGTM. Transverse noise along a detected manifold is here used to achieve smoothness of the manifold structure by modelling it as a mean curve plus noise. While the comparison with other methodologies is not the focus of this work (and will be presented in an upcoming paper\footnote{This additional work can be found at \url{https://git.lwp.rug.nl/cs.projects/1DREAM}.}), 1-DREAM has in this regularization property and advantage with respect to its competitors. The structures recovered via 1-DREAM are indeed more robust to noise, even when taking the stochasticity of the methodologies into account.
\subsection{Noise attenuation}\label{subsec:Noise}

We first describe the methodologies developed for the reduction of background (LAAT) and transverse (EM3A) noise. The pheromone value recovered by LAAT informs us about the background noise level. By thresholding it at specific values (dependent on the data set) the underlying coherent structures emerge from the noisy environment. Filtering out the points with low pheromone value, we obtain the data set to be fed to EM3A. This methodology reduces transverse noise on the manifolds by pushing points towards their spine (mean curve/surface). These steps are often necessary for further analysis of the hidden structures.

\subsubsection{LAAT: Locally Aligned Ant Technique}\label{subsec:Ants}

The Locally Aligned Ant Technique \citep{Taghribi2020LAATLA} aims 
to extract manifolds from noisy data sets in reliance on the idea of Ant Colony Optimization (ACO; \cite{dorigo_new_1999}). 
The latter is a computational method used for revealing the shortest path between two given data points when multiple routes between them are possible. 
In this section, we clarify the methodology of LAAT that links between the natural behaviour of ants as they follow their own pheromone trails in search of the shortest path to food and the capturing of points that are locally aligned with the major directions of a given manifold.  For a complete view of LAAT, we refer the reader to Algorithm 1 in \cite{Taghribi2020LAATLA}.

Consider a data set $\mathcal{Q} = \{ \vector{t}_1, \vector{t}_2, \dots, \vector{t}_n \}$ consisting of $n$ points such that $\vector{t}_i \in \mathbb{R}^D$, then there exists $D$ principle components in a spherical neighborhood $\mathcal{N}^i_r \coloneqq \mathcal{B}(\vector{t}_i,r)$ of radius $r$ around a point $\vector{t}_i$. We denote $\vector{v}_d$ and $\lambda_d$ the local eigenvectors and the corresponding ordered eigenvalues respectively with $d=1,2,...,D$. That being introduced, LAAT then consists of a random walk in which artificial ``ant'' jump from a point belonging to the data set to the next where high preference is given to: jumps along the dominant eigenvectors, and paths where an amount of artificially deposited pheromone is accumulated \citep{dorigo_new_1999}.
\begin{figure}[t]
  \centering
  \includegraphics[width=\columnwidth,trim = 1.7cm 0.2cm 0.4cm 1.8cm,clip]{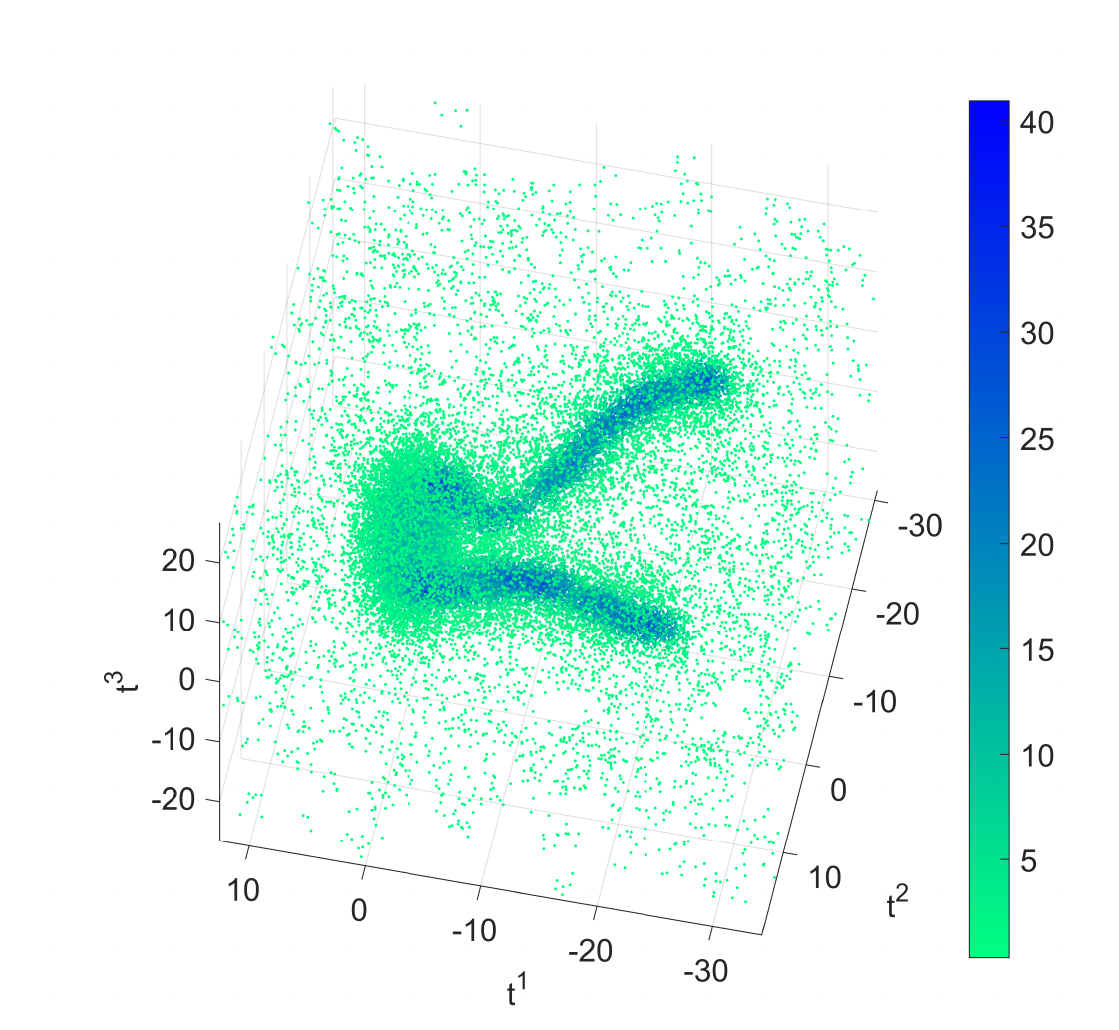}  
  \caption{Pheromone value for every point in the data synthetic data set emulating a Jellyfish galaxy at the end of the procedure.}
  \label{fig:LAAT_SynthJF}
\end{figure}
Given a path $(\vector{t}_j-\vector{t}_i)$ between points $i$ and $j$, the relative normalized weighting of the alignment of this path with a local eigenvector $\vector{v}_d$ can be given as follows:  
\begin{equation}
    w_{d}^{(i,j)} = 
    \frac{|\cos \alpha_d^{(i,j)}|}{\sum\limits_{d^\prime=1}^D |\cos \alpha_{d^\prime}^{(i,j)}|} \enspace .
    \label{cosineweight}
\end{equation}
Where $\alpha_d^{(i,j)}$ is the angle between $(\vector{t}_j-\vector{t}_i)$ and  $\vector{v}_d$. Furthermore, the normalized eigenvalues show the relative importance of the different eigenvectors, and are given by the following:
\begin{equation}
    \tilde{\lambda}_{d}^{(i)} = 
    \frac{\lambda^{(i)}_d}{\sum\limits_{d^\prime=1}^D \lambda^{(i)}_{d^\prime}} 
    \label{normalizeeigenvalues}
\end{equation}  This then allows us to define the preference of the jump from $\vector{t}_i$ to $\vector{t}_j$ that is aligned with the local eigen-directions. The preference is given by: 
\begin{equation}
E^{(i,j)} = \sum\limits_{d=1}^D w_{d}^{(i,j)} \cdot \tilde{\lambda}_{d}^{(i)} 
\label{Eformula}
\end{equation}
Preferences are normalized ($\tilde{E}^{(i,j)}$ ) so that they sum to $1$ within the neighbourhood $\mathcal{N}^i_r$.
Moreover, by defining an amount of pheromone $F^j(t)$ for a point $\vector{t}_j$ at a time $t$, the above preference will allow for the accumulation of pheromone on the points aligning with the manifold. As inspired by nature, an evaporation rate  $0<\zeta<1$ is incorporated in the definition of the pheromone thus serving the purpose of decreasing its amount on less visited points. The pheromone quantity is:

\begin{equation}
F^{j}(\tau+1) = (1-\zeta)\cdot F^{j}(\tau) \enspace ,
\label{pheromonenormalized}
\end{equation}
Again, pheromone quantities are normalized ($\tilde{F}^j(\tau)$) so that they sum to $1$ within $\mathcal{N}^i_r$.
Combining equations (\ref{Eformula}) and (\ref{pheromonenormalized}) allows us to define the total preference of the jump from $\vector{t}_i$ to $\vector{t}_j$: 
\begin{equation}
V^{(i,j)}(\tau) = (1-\kappa)\tilde{F}^j(\tau)+\kappa \tilde{E}^{(i,j)} \enspace. 
\label{vformula}
\end{equation}
Where $\kappa \in [0,1]$ is a parameter which adjusts the relative importance of the two terms. Finally, the jump probabilities can be defined as: 
\begin{equation} 
P(j|i,\tau) = \frac{\exp(\beta V^{(i,j)}(\tau))}{\sum\limits_{j^\prime\in \mathcal{N}^{(i)}_r} \exp(\beta V^{(i,j^\prime)}(\tau))} \enspace ,
\label{pformula}
\end{equation}
Here, $\beta>0$ is a parameter that is analogous to the inverse of temperature in statistical physics \citep{Taghribi2020LAATLA}. Afterwards, choosing a set of hyper-parameters given by the number of ants $N_\mathrm{ants}$, epochs $N_\mathrm{epoch}$, and steps of each ant $N_\mathrm{steps}$ is necessary for the completion of the random walk. A random starting point is chosen such that the random walk begins from the denser neighborhoods. In practice, this means that given the median $\widetilde{\mathcal{N}}$ of the set of neighborhoods $\mathcal{N}=\{|\mathcal{N}_r^{(i)}|\ |  \vector{x_i} \in \mathcal{Q}\}$, the condition for a random starting point $i$ is: 
\begin{equation}
\label{startingcond}
    |\mathcal{N}^{(i)}_r| \geq \widetilde{\mathcal{N}}
\end{equation}
In a given epoch, the ants will then perform the random walk on the points in $\mathcal{Q}$ for $N_\mathrm{steps}$ and with the jump probabilities defined in ($\ref{pformula}$). To update the value of the pheromone quantity, the indices of the points visited by a given ant $\ell$ is stored in a route multi-set $A^\ell$ which allows us to count the multiplicity of visits to the points in the data set. The value of the pheromone quantity on a given point $j$ is updated according to the following formula: 
\begin{equation}
\label{pheromoneupdate}
F^{j}(\tau) = F^{j}(\tau-1)+\nu (j) \gamma \qquad \forall j\in A^\ell \enspace,
\end{equation}
where $\gamma$ is a constant value denoting the amount of pheromone deposited, and $\nu(j)$ is the multiplicity of element $j$ in $A^{\ell}$. Therefore, with the enforced pheromone evaporation rate and the defined jump probabilities, the pheromone will accumulate along the points aligned with given manifolds, and will dissipate in more scattered regions, hence highlighting the structures in a noisy data set. We refer the reader to \citet{Taghribi2020LAATLA} for a demonstration of the high robustness of results to changes in the previously defined parameters, and a comparison with state-of-the-art methods of similar purpose. The application of the LAAT methodology to the synthetic data set described previously is shown in figure \ref{fig:LAAT_SynthJF}. For this analysis we used a radius $r = 2$ and a neighborhood size threshold of $\widetilde{\mathcal{N}} = 5$. The blue inner structure is revealed from within the whole noisy data set. While visually, the same structure can be identified in fig. \ref{fig:JF_Synth_Noisy}, LAAT introduces a scalar field on the data set as a proxy of relevant dense regions. It is then straightforward to provide a pheromone threshold (e.g. in our case $F_\mathrm{Th}^j = 20$) that enables filtering of the noisy data set. The filtered data set will contain only the most relevant regions. A complete list of all parameters in the LAAT methodology is given in tab. \ref{tab:Param_LAAT}. Free-parameters are denoted by a $*$ symbol and their values specified in each experimental study (sections \ref{subsec:JellyFish}, \ref{subsec:CosmicWeb} and \ref{subsec:GAIA}).
\begin{table}[t]
\centering
\def\myCWidth{0.12}
\caption{Full list of parameters for LAAT.}
\label{tab:Param_LAAT}
\begin{tabularx}{\columnwidth}{@{\extracolsep{\fill}}
ll>{\raggedleft}p{\myCWidth\columnwidth}>{\raggedleft\arraybackslash}p{\myCWidth\columnwidth}}
\toprule
    $r~^\ast \in \RR$      &  Neighborhood radius \\
    $\zeta \in \RR ~(\zeta = 0.1)$  &  Evaporation rate\\
    $\kappa \in \RR ~(\kappa = 0.5)$ &  Shape v. Pheromone\\
    $\beta \in \RR ~(\beta = 10)$  &  Inverse temperature\\
    $\gamma \in \RR~(\gamma = 0.05)$ & Deposited pheromone\\
    $ F_\mathrm{Th}^j~^\ast \in \RR$ & Pheromone threshold\\
    $\widetilde{\mathcal{N}} \in \NN~(\widetilde{\mathcal{N}} = 5) $ & Neighborhood threshold\\
    $N_\mathrm{epoch} \in \NN~(N_\mathrm{epoch} = 10)$ & Epochs\\
    $N_\mathrm{steps} \in \NN~(N_\mathrm{steps} = 2500)$ & Steps per epoch\\
    $N_\mathrm{ants} \in \NN~(N_\mathrm{ants} = 500)$ & Ants\\
\bottomrule
\end{tabularx}
\end{table}
The values suggested for the rest of the parameters are the ones used throughout this work.

\subsubsection{EM3A: Evolutionary Manifold Alignment Aware Agents}
\label{subsec:diffusion}
\begin{figure}[t]
  \centering
  \includegraphics[width=\columnwidth,trim = 0.1cm 0.5cm 0.5cm 1cm,clip]{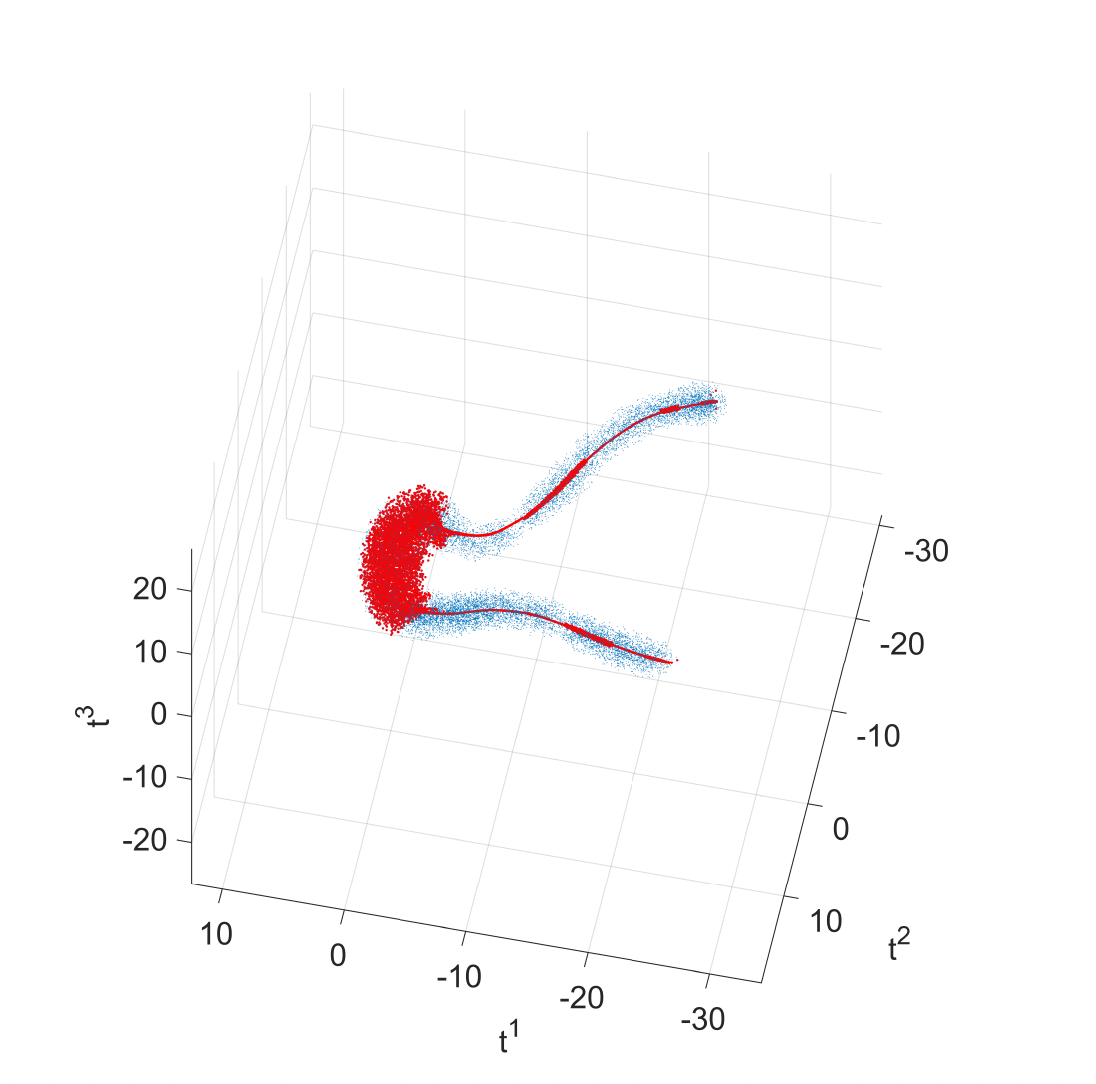}  
  \caption{Data set of points selected s.t. 
  $F^j(t = t_\mathrm{end}) >= F_\mathrm{Th}$ (blue dots) and diffused data set obtained with EM3A (red dots).}
  \label{fig:EM3A_SynthJF}
\end{figure}
While LAAT isolates points in proximity to high-density regions, ideally the true manifold lies within the noisy point-cloud that samples it. In order to find the mean curve of manifolds, by reducing the amount of transverse noise, a secondary step in this analysis is necessary. Intuitively, we would like to be able to push points in proximity to over-dense regions in the data space, towards the unknown mean curve of the manifold. We propose a procedure, devoted to approximate the true nature of manifolds by pulling nearby points towards an empirically estimated, local mean curve. 
This solution is also inspired by ant colony behaviour for picking up and dropping carried objects. 
This section explains the steps followed by EM3A, a method introduced in \citet{Mohammadi2020_EM3A} to employ this behaviour by instructing agents walking through the data space to pick up the data points and place them in a closer proximity to the manifolds. Since the net effect of the methodology is to move points towards high density, low-dimensional regions, the methodology shares similarities with the work presented in \cite{8046026}, where the task is modelled as a diffusion process. For this reason we will often refer to the data set obtained via EM3A as a ``diffused'' one.

The first step in accomplishing the above is to define an approximation strategy to recognizing the manifold structure. In that pursuit, local PCA is employed again to estimate the local eigenvalues and eigenvectors within neighborhoods of radius $r$ centered on the data points. Given data set $\mathcal{Q}$ and $\vector{t}_i \in \mathcal{Q}$, the eigenvectors $\vector{v}^i_d$ and normalized eigenvalues $\tilde{\lambda}^i_d$ for the neighborhood $\mathcal{N}^i_r \coloneqq \mathcal{B}(\vector{t}_i,r) \cap \mathcal{Q}$. Defining the saliencies \citep{10.5555/1756006.1756018} $S_i = i \times (\tilde{\lambda}^{(i)}_d - \tilde{\lambda}^{i+1}_d)$ as the expansion coefficient for the local covariance matrix, the intrinsic dimensionality of the manifold is estimated as given in \citet{WangEtal2008}, \citet{MordohaiMedioni2005} by: 

\begin{equation}
\widehat{d} = \underset{i}{\arg\max 
S_i} \enspace.
\label{intrinsicDim}
\end{equation}

The set $\{\vector{v}^i_1,\dots \ , \vector{v}^i_{\widehat{d}} \}$ can then be used as an approximation to the tangent space of manifold $\mathcal{M}$ at $\vector{t}_i$.

Next, a random walk is initiated in which agents are reinforced to move data points closer to the underlying manifolds in the data set. Let $\bm{U}$ be the matrix whose columns are the first $d$ eigenvectors of $\mathcal{N}^i_r$, and let $\vector{\mu}$ be the kernel average of $\vector{t}_i$'s neighbors, then the distance to a manifold $\mathcal{M}$ is estimated by:

\begin{equation}
        \delta^{\mathcal{M}}(i) = \lVert(I - UU^\top)(\vector{\mu} - \vector{t}_i)\rVert.
    \label{DistToManifold}
\end{equation}

Here $\lVert.\rVert$ is the Euclidean norm. We now define the weights and probabilities associated with the random walk. For each point $\vector{t}_j$ within $\mathcal{N}^i_r$, the following weights are defined: 

\begin{equation}
    w(\vector{t}_i, \vector{t}_j) = 
      \begin{cases} 
      1-\frac{\delta^{\mathcal{M}}(i)}{\iota} & \delta^{\mathcal{M}}(i) \leq \iota \\
      0 & \delta^{\mathcal{M}}(i) > \iota   
      \end{cases}.
    \label{EM3Aweights}
\end{equation}

The parameter $\iota$ is chosen such that $50\%$ of neighbors have none-zero weights. The agent jump probability to the next destination is then given by: 

\begin{equation}
    P(\vector{t}_i, \vector{t}_j) = \frac{w(\vector{t}_i, \vector{t}_j)}{\sum\limits_{m \in \mathcal{N}^i_r}w(\vector{t}_i, \vector{t}_m)}.
    \label{EM3Aprobs}
\end{equation}

Having defined this jump probability, the agents are encouraged to remain close to the manifold. Since the agents are not only walking, but also picking up and dropping down points, what is then required is to define a pick-up probability for the agent of data point $\vector{t}_j$: 

\begin{equation}
    P_\mathrm{pick}(\vector{t}_j) = \frac{1 - w(\vector{t}_i, \vector{t}_j)}{\sum\limits_{m \in \mathcal{N}^i_r}
    \left(1 - w(\vector{t}_i, \vector{t}_m)\right)}.
    \label{pickprobs}
\end{equation}

In other words, the probability to be picked up increases with the distance of the point from the tangent space. Since we aim at enhancing the density contrast between points more likely belonging to the manifold and all others, at each time $t$ we move these points towards the manifold along the complement of the tangent space. The displacement update reads:

\begin{equation}
        \vector{t}^\mathrm{new}_j= \vector{t}^\mathrm{old}_j +  \eta(I - UU^\top)(\vector{\mu} - \vector{t}^\mathrm{old}_i).
    \label{EM3Aupdate}
\end{equation}

Here $\eta>0$ is the learning rate controlling the amount of displacement produced. In addition to the denoising, as mentioned previously, the topological nature of the manifold should also be preserved under the above steps. 
To ensure that, over-smoothing of the manifold is avoided by introducing a threshold such that the agents can only change the neighborhood if the mean distance of neighbors to the tangent space is larger than the threshold.

The performance of the above method is clearly dependent on the chosen radius of the neighborhood \citep{KaslovskyMeyer2014}. Since highly curved manifolds require smaller radii while the suppression effect of the noise requires larger radii, it is difficult to choose a proper value for $r$ without prior knowledge of the manifold properties. Therefore, as a solution to this problem, EM3A combines the above procedure with Evolutionary Game Theory (EGT) concepts to automatically adapt the radius parameter. 

Since $r$ is a continuous variable, we assume that it lies within the range $ [R_{\min}, R_{\max}]$ discretized into $m$ smaller intervals. To link this step to EGT, each interval is therefore viewed as an evolutionary strategy within a population of $m$ strategies. Taking $p_1, \dots , p_m$ as the frequencies of each strategy, we let $\vector{p} = \{ p_1, \dots , p_m \} $ denote the distribution of strategies within the population. In a generation $t$, each agent randomly selects a strategy while following a population share distribution $\vector{p}^{(t)} = [p_\ell^{(t)}]_{m \times 1}$, where $p_\ell^{(t)}$ is the population share (frequency) of the $\ell$-th strategy at the current generation. For this agent with a strategy $\ell$, the neighborhood radius is uniformly selected from the interval $[r_\ell, r_{\ell+1}]$. 

Next, a copy of the data set is provided for all the agents on which they perform the random walk with $N_s$ steps according to the above described rules. The output of each walk is averaged to form the updated data set and the fitness of each strategy is computed. Letting $S_\ell$ denote the number of agents with strategy $\ell$ and $N_\ell$ the number of times they change the data set at a given generation, then the fitness $f_\ell$ of strategy $\ell$ follows: 

\begin{equation}
f_\ell = \frac{N_\ell}{S_\ell \cdot N_s}.    
\label{fitness}
\end{equation}

For a given strategy $\ell$, the rate of change of its frequency $\dot{p_\ell}/p_\ell$, measures its evolutionary success. This measure can be equated to the difference between the fitness of the strategy and the average fitness of the population. In other words for a generation $t$ we can define:

\begin{equation}
    \frac{\dot{p_\ell}}{p_\ell} = f_\ell - \bar{f}.
    \label{replicator}
\end{equation}

The above equation can be rewritten in the following iterative form: 

\begin{equation}
    p^{(t+1)}_\ell = p^t_\ell + p^t_\ell \big(f^t_\ell - \bar{f}^t \big).
    \label{replicator2}
\end{equation}

Therefore, in the following generations, the agents will choose their strategy from the distribution $\vector{p}^{(t+1)} = [p_\ell^{(t+1)}]$. Iterating for a given number of generations will thus lead to the denoising of the manifolds in a noisy data set while maintaining the properties of the embedded structures. Results of EM3A applied to the synthetic data filtered according to the pheromone value are shown in figure \ref{fig:EM3A_SynthJF} (red dots). The processed data is sensibly denser along the mean curve of the two elongated manifolds, while it is roughly unchanged for manifold $\mathcal{M}_1$ (sampled by point-cloud $\mathcal{P}_1$). The  over-densities in the middle of the two manifolds (close to the high-curvature regions) are likely to be caused by slight variations in the pheromone value recovered by LAAT. The crisp selection of particles with high pheromone may have caused varying point-density along the manifolds, and this is converted by EM3A in a stronger push towards the over-dense regions. Indeed, the denser red regions are always delimited by gaps (under-dense regions) in the filtered point-cloud (blue dots). 
All parameters of EM3A are listed in tab. \ref{tab:Param_EM3A}. The free parameters used for the synthetic data set are $R_{min} = 1$ and $R_{max} = 2$
\begin{table}[t]
\centering
\def\myCWidth{0.12}
\caption{Full list of parameters for EM3A.}
\label{tab:Param_EM3A}
\begin{tabularx}{\columnwidth}{@{\extracolsep{\fill}}
ll>{\raggedleft}p{\myCWidth\columnwidth}>{\raggedleft\arraybackslash}p{\myCWidth\columnwidth}}
\toprule
    $R_{\min}~^\ast \in \RR$      &  Minimum radius \\
    $R_{\max}~^\ast > R_{\min}$  &  Maximum radius \\
    $\iota \in \RR ~(\iota > 0)$  & Jump weight (adaptive) \\
    $\eta \in \RR ~(\eta = 1e{-2})$ &  Learning rate \\
    $N_s \in \NN ~(N_s= 500)$  &  Number of steps\\
    $N_g \in \NN ~(N_s= 5)$  &  Number of generations\\    
\bottomrule
\end{tabularx}
\end{table}

\subsection{Modelling}

In the following we describe the methodologies aimed at identifying the local intrinsic dimensionality of structures in the data set (Dimensionality Index) on a point-by-point basis (sec. \ref{subsec:DimIndex}). The one-dimensional points are used for detection of filaments via Crawling (sec. \ref{subsec:Crawling}) and its results as initialization for SGTM (\ref{subsec:SGTM}). Since these methodologies aim at detecting and recovering individual sub-structures in the data set, we refer to this step as ``modelling''.

\subsubsection{Dimensionality Index}\label{subsec:DimIndex}

While in first approximation, the approach used in sec. \ref{subsec:Ants} is enough to roughly estimate the local dimensionality of a neighborhood, we dedicate this section to a more refined version of dimensionality index. In light of the new data set obtained via LAAT and EM3A, the new estimate of local dimensionality incorporates continuity information about the local structure, previously hidden by noise.
\begin{figure}[t]
\centering
\includegraphics[width= \columnwidth,trim = 0.1cm 0.5cm 0.1cm 0.5cm,clip]{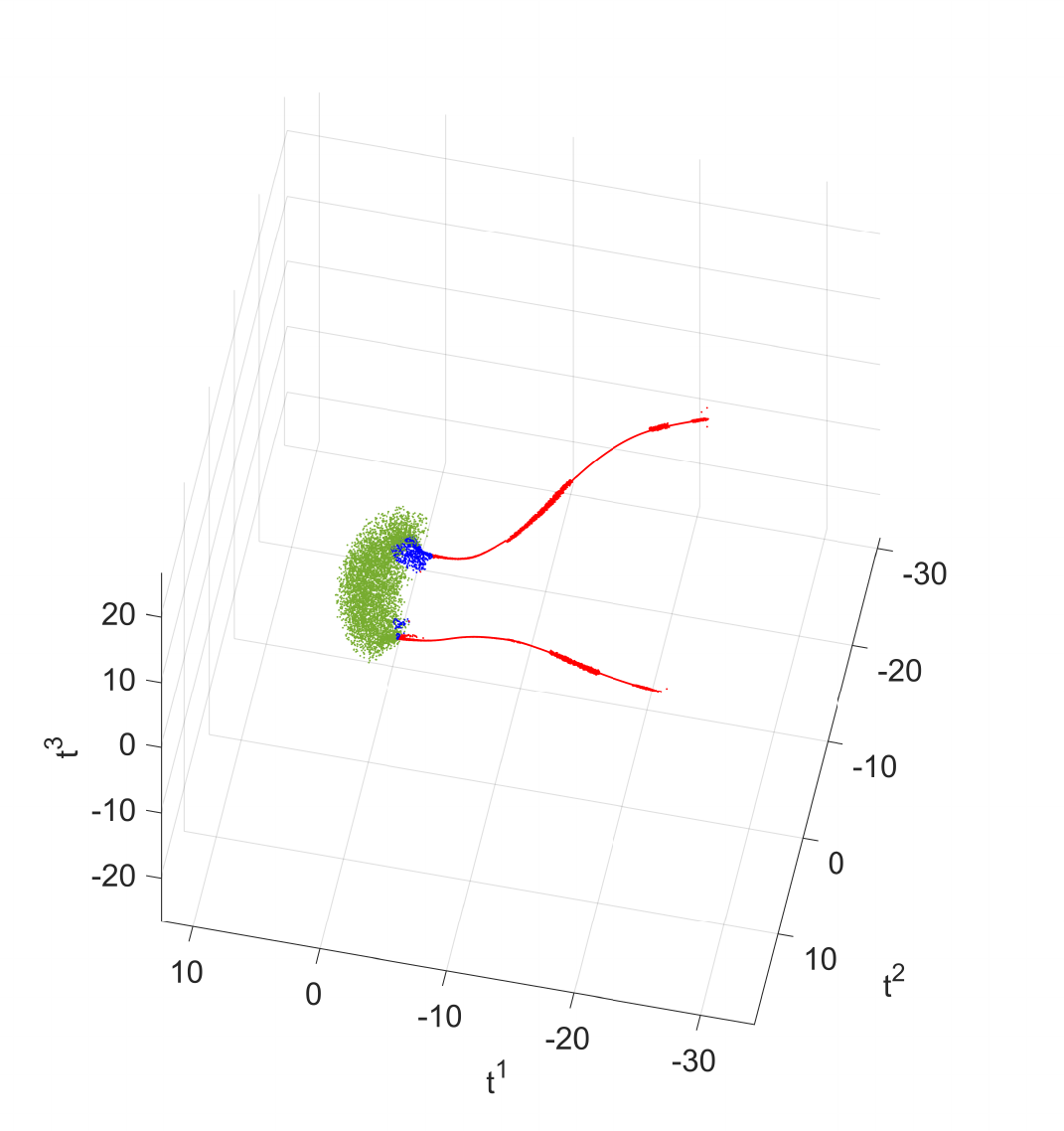}
\caption{Partition of data sets $\tilde{\mathcal{Q}}$ into its corresponding 
one, two and three-dimensional subsets: 
$\tilde{\mathcal{Q}_1}$ (red) 
$\tilde{\mathcal{Q}_2}$ (blue) 
and $\tilde{\mathcal{Q}_3}$ (green).
}
\label{fig:Dim_Index}
\end{figure}
Around each $\tivec{t}_i \in \tilde{\mathcal{Q}}$ we perform local PCA using points from $\mathcal{N}^i_r = \mathcal{B}(\tivec{t}_i;r) \cap \tilde{\mathcal{Q}}$, obtaining eigenspectrum $\lambda_{i,1} \ge \lambda_{i,2} \ge \dots \ge \lambda_{i,d}$.

The dimensionality index $\Delta^O_i$ of $\tivec{t}_i \in \tilde{ \mathcal{Q}}$ used in \cite{WangEtal2008} (limited to 3-dimensional data) is obtained as in eq. \eqref{intrinsicDim}.
However, a more accurate dimensionality index can be found in \cite{CANDUCCI2022103579}. 
We summarize it in the current section. The normalized eigen-spectrum $\tilde{\Lambda}_i$ (see eq. \eqref{normalizeeigenvalues}) of each point $\tivec{t}_i$'s neighborhood is mapped onto the \emph{Simplex} of multinomial distributions. 
The geodesic distance of each $\Lambda_i$ with respect to the vertices $\{\vector{e}_1, \vector{e}_2, \vector{e}_3\}$ is evaluated on the simplex, where $\vector{e}_1 = (1, 0 ,0)$, $\vector{e}_2 = (\sfrac{1}{2}, \sfrac{1}{2} ,0)$, $\vector{e}_3 = (\sfrac{1}{3}, \sfrac{1}{3} ,\sfrac{1}{3})$ represent the eigen-spectra of ideal $1-$,$2-$ and $3-$dimensional neighborhoods respectively. Then, the dimensionality index of point $\tivec{t}_i$ is the index $j$ corresponding to the closest vertex, under the geodesic distance $d_J(\Lambda_i,\vector{e}_j)$:
\begin{align}
    \Delta_i^G &= \arg\min_j d_J(\tilde{\Lambda}_i,\vector{e}_j), \\ d_J(\tilde{\Lambda}_\ell,\tilde{\Lambda}_m) &= 2\arccos{\left(\sum_{k=1}^{D} \sqrt{\tilde{\Lambda}_\ell^k \cdot \tilde{\Lambda}_m^k}\right)} \enspace.
\end{align}
We also propose a ``soft'' version of dimensionality index, by imposing a kernel $K(\tilde{\Lambda}_i;\vector{e}_j)$ on each prototypical vertex of the Simplex. We chose to use a Gaussian smoothing kernel s.t.:
\begin{equation}
    K(\tilde{\Lambda};\vector{e}_j) = \exp\left[- \frac{d_J(\tilde{\Lambda},\vector{e}_j)^2}{2s^2} \right] \enspace,
\end{equation}
where $s$ is the geodesic distance on the simplex between any vertex and the equidistant point on the Simplex with respect to all vertices. This kernelization of the geodesic distances on the Simplex imposes a distribution:
\begin{equation}
    P_i(j) = \frac{K(\tilde{\Lambda}_i;\vector{e}_j)}{\sum_{k=1}^{D}K(\tilde{\Lambda}_i;\vector{e}_k)} \enspace.
\end{equation}
In order to take into account the smoothness of manifolds in data space, we impose a smoothing kernel in that space on each point $\tivec{t}_i \in \tilde{\mathcal{Q}}$ s.t.:
\[
c(i,l) = \exp\left[-\lVert \tivec{t}_i - \tivec{t}_l \rVert^2 / (2 r^2)\right] \enspace.
\]
The smoothed normalized index distribution reads:
\begin{equation}
\label{eq:SmoothIdx_P}
    P^S_i(j) = \frac{1}{\sum_{\tivec{t}_l \in \mathcal{B}(\tivec{t}_i,r)} c(i,l)}    
    \sum_{\tivec{t}_l \in \mathcal{B}(\tivec{t}_i,r)} c(i,l) \cdot P_l(j) \enspace,
\end{equation}
where the sum is taken over diffused points $\tivec{t}_l$ in the spherical neighborhood of $\tivec{t}_i$ of radius $r$. 
The smoothed dimensionality index of $\tivec{t}_i$ is then
\begin{equation}\label{eq:Final_D}
\Delta^S_i = \arg\max_j P^S_i(j) \enspace.
\end{equation}
Every point in $\tilde{\mathcal{Q}}$ (and its noisy counterpart $\mathcal{Q}$) can be assigned to the respective $d-$dimensional subset $\tilde{\mathcal{Q}_d}$ ($\mathcal{Q}_d$), creating a partition of the original set into:
\begin{align}
\tilde{\mathcal{Q}_d} &= \{\tivec{t}_i \in \tilde{\mathcal{Q}} \quad|\quad \Delta^S_i = d\} \label{eq:Dim_IdxG}\\
\mathcal{Q}_d &= \{\vector{t}_i \in \mathcal{Q} \quad|\quad \Delta^S_i = d\} \enspace,
\end{align}
such that $\bigcup_{d=1}^D \mathcal{Q}_d = \mathcal{Q}$. The results of the application of dimensionality index to the point cloud defined in section \ref{sec:Synt_Data} are shown in figure \ref{fig:Dim_Index} and were obtained with the same radius used for LAAT ($r = 2$). For a better visualization of the results we only show here the diffused points as divided into $1-$,$2-$,$3-$dimensional sets, represented in red, blue and green color respectively ($\mathcal{Q}_1$, $\mathcal{Q}_2$ and $\mathcal{Q}_3$ respectively). 
Although the synthetic data set described in section \ref{sec:Synt_Data} does not contain intrinsically two-dimensional points, the dimensionality index proposed in eq. \eqref{eq:Final_D} recovers $\tilde{\mathcal{Q}_2} \neq \emptyset$. This discrepancy can be attributed to the EM3A algorithm diffusing points too strongly towards high-density regions. However, since we are interested in detecting the one-dimensional structures (streams) in the data set, upon visual inspection, the recovered estimation of $\tilde{\mathcal{Q}_1}$ and $\mathcal{Q}_1$ (red points in fig. \ref{fig:Dim_Index}) is acceptable and the contamination of $2-$dimensional (blue) points is minimal.

Note that the only parameter required by the dimensionality index is the neighborhood radius $r$.

\subsubsection{Crawling: Multiple Manifolds \texorpdfstring{$1D$}{1D} Crawling}\label{subsec:Crawling}
\begin{figure}[t]
\centering
\includegraphics[width=\columnwidth,trim = 0.1cm 0.5cm 1cm 1cm,clip]{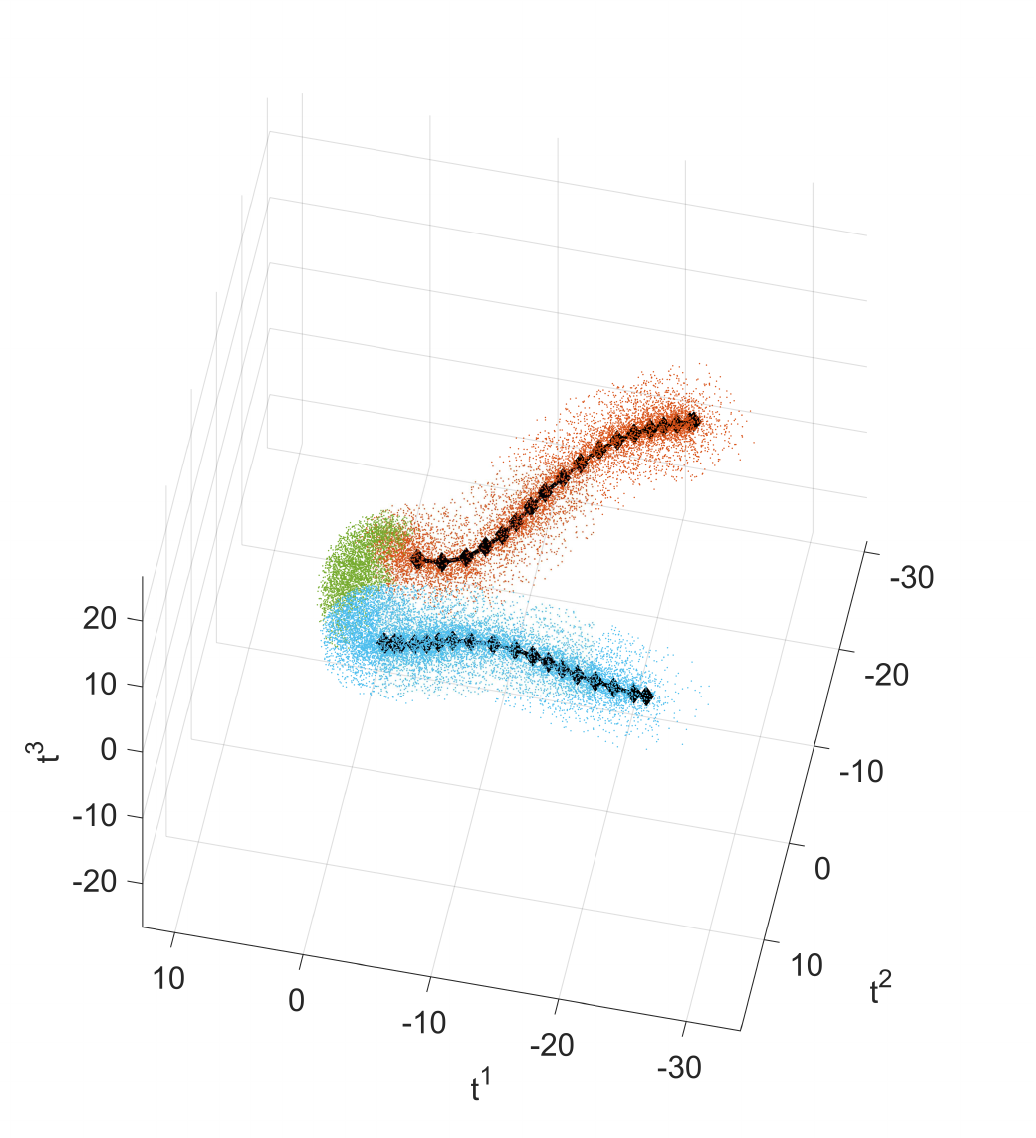}
\caption{Extracted skeletons (black diamonds and line) of the two manifolds after crawling, overlay 
the 1D data set extracted by means of the dimensionality index (red lines
). The point-cloud surrounding each skeleton is the corresponding recovered noisy manifold.}
\label{fig:SynthJF_Crawling}
\end{figure}

We describe here a recursive algorithm that enables us to separate all distinct one-dimensional manifolds contained in data sets $\mathcal{Q}_1$ and $\tilde{\mathcal{Q}}_1$ while sampling them in representative sets of points and building their respective low dimensional representations. 
The end result of this technique (see figure \ref{fig:SynthJF_Crawling}), as opposed to LAAT (sec. \ref{subsec:Ants}), is a discrete skeleton for each one-dimensional manifold in the data set. In fact, LAAT only highlights over-densities within a point-cloud, without partitioning into its low-dimensional components. With Crawling, each skeleton is assumed to lie on a high dimensional embedding of the unit interval (a bent and stretched version of it), sampled by a finite set of points representing the ``steps'' taken by the agent while walking (crawling) on $\tilde{\mathcal{Q}}_1$.
The results from this algorithm are used to initialize Stream GTM through a parametric mapping function $\vector{f}: [-1; 1]~\longrightarrow~\RR^D$ via linear regression applied to the parameters $\vector{W}$.
The only main assumption of the algorithm is that at the selected size (given by the radius parameter, in this case $r = 2$), the tangent space to each manifold is isomorphic to $\RR$.

\subsubsection*{Initialisation}

We first initialize the residuals set $\mathcal{R} = \tilde{\mathcal{Q}}_1$. This data set is used as a reference for regions that have been visited by crawling, leaving $\tilde{\mathcal{Q}}_1$ unscathed. 
Initially, a single ``seed'' $\tivec{t}_0 \in \tilde{\mathcal{R}}$ is randomly selected and PCA applied to its local neighborhood of radius $r$: $\mathcal{B}(\tivec{t}_0,r)\cap \tilde{\mathcal{R}}$. 
The unit eigen-vector $\hat{\vector{v}}_0 = \frac{\vector{v}_0}{\lVert\vector{v}_0\rVert}$, associated to the largest eigen-value $\lambda_1$, spans the tangent space to manifold $\mathcal{M}_k$ at point $\tivec{t}_0$: $\mathrm{T}_{\tivec{t}_0}\mathcal{M}_k$. Eigen-vector $\hat{\vector{v}}_0$, being the direction where most of the neighborhood's variance is preserved, gives us the initial preferential direction for crawling. We can now estimate two new points along direction $\hat{\vector{v}}_0$ at distance $\beta \cdot r$ from $\tivec{t}_0$:
\begin{equation}\label{eq:Cr_Estimates}
\vector{z}_n^{\pm} = \tivec{t}_0 \pm \beta \cdot r \cdot \hat{\vector{v}}_0 \enspace,
\end{equation}
where $\beta = 0.75$ is a regularization parameter aimed at mitigating the effect of outliers on PCA and index $n$ identifies the iteration number (during the initialisation $n = 1$). In order to keep the crawling adherent to manifold $\mathcal{M}_k$, for every new candidate $\vector{z}_n^{\pm}$ we compute its closest neighbor in data set $\tilde{\mathcal{R}}$:
\begin{equation}\label{eq:ClosestN}
\tivec{t}_n^\pm = \arg\min_{\tivec{t}~\in~\tilde{\mathcal{R}}}( \lVert\tivec{t} -  \vector{z}_n^{\pm} \rVert)
\end{equation}
under the condition that $\lVert\tivec{t}_n^\pm - \tivec{t}_0\rVert \leq r$.
This condition enforces the maximum length between two adjacent points on manifold $\mathcal{M}^k$ to never exceed the neighborhood radius $r$. 
After the estimation step, we initialize the set of representative points of manifold $\mathcal{M}^k$ as $\overline{\mathcal{P}}^k = \{\tivec{t}_0, \tivec{t}_1^+, \tivec{t}_1^-\}$ and the low-dimensional counterpart $\mathcal{P}^k = \{0, 1, -1\}$. The direction $\hat{\vector{v}}_0$, being the tangent space to $\mathcal{M}^k$ at point $\tivec{t}_0$, is preserved as a member of the tangent bundle (see 
\cite{tu2010introduction}) to manifold $\mathcal{M}^k$: $TM^k$.
The last step in the initialization phase removes the neighborhood of point $\tivec{t}_0$ from $\mathcal{R}$ overwriting the set.

\subsubsection*{Crawling Update}

After the initialisation phase is completed, at every iteration $n$, crawling is recursively applied to every point identified in the previous iteration $n-1$ following: 
\begin{description}
\item [1. Seed selection:] From $\overline{\mathcal{P}}^k$ select point $\tivec{t}_{n-1}^+$ and compute the neighborhood $\mathcal{N} = \mathcal{B}(\tivec{t}_{n-1}^+,r)\cap \mathcal{R}$. Applying PCA to $\mathcal{N}$, compute the unit principal component $\hat{\vector{u}}_{n-1}$.
\item [2. Parent point recovery:] Recover $\tivec{t}_{n-2}^+$ and $\hat{\vector{v}}_{n-2}^+$ the parent point and corresponding tangent vector of $\tivec{t}_{n-1}^+$. For ``parent'' we mean the point that generated $\tivec{t}_{n-1}^+$ in iteration $n-2$ of crawling.
\item [3. Tangent space projection:] Since PCA is a rotationally invariant method, $\hat{\vector{u}}_{n-1}$ is not a priori identifiable with the current crawling direction (it could be oriented as its inverse vector). We solve this issue by computing the angle $\theta$ between $\hat{\vector{u}}_{n-1}$ and $\hat{\vector{v}}_{n-2}^+$
\begin{equation}
    \theta = \arccos(\hat{\vector{u}}_{n-1} \cdot \hat{\vector{v}}_{n-2}^+) \enspace,
\end{equation}
where $(\cdot)$ denotes the scalar product. The new crawling direction $\hat{\vector{v}}_{n-1}^+$ is given by:
\begin{equation}\label{eq:projection}
    \hat{\vector{v}}_{n-1}^+ =
    \begin{cases}
    +\hat{\vector{u}}_{n-1} &\text{if}~ -\frac{\pi}{2} \leq \theta \leq \frac{\pi}{2} \\
    -\hat{\vector{u}}_{n-1} &\text{if}~  \frac{\pi}{2} < \theta < \frac{3\pi}{2}
    \end{cases} \enspace.
\end{equation}
\item [4. Updates:]  Using equations \eqref{eq:Cr_Estimates} and \eqref{eq:ClosestN}, a new point lying on manifold $\mathcal{M}^k$ in direction  $\hat{\vector{v}}_{n-1}^+$ is found and added to $\overline{\mathcal{P}}^k$. The latent space $\mathcal{P}^k$ and tangent bundle $TM^k$ are also updated coherently. Finally, the neighborhood $\mathcal{N}$ is subtracted from data set $\mathcal{R}$.
\end{description}
When the condition $\lVert\tivec{t}-\vector{z}_n^{\pm}\rVert\leq r$ does not hold for any $\tivec{t} \in \mathcal{R}$ an end of the manifold is encountered 
and crawling is suppressed along the current 
direction. 
The same procedure is applied to point $\tivec{t}_{n-1}^-$, using parent direction $\hat{\vector{v}}_{n-2}^-$, until the second end of the manifold is found.\\ 
Once both ends of a manifold are detected, we repeat the \emph{Initialization} and \emph{Update} phases on data set $\mathcal{R}$, which, at the end of iteration $k$ of the procedure contains all points of data set $\tilde{\mathcal{Q}}_1$ except the ones extracted from manifolds up to $\mathcal{M}^k$. Since data set $\mathcal{R}$ is recursively depleted of points, we expect its size to converge to zero after a certain number of iterations of the whole procedure. This consideration gives us a criterion for halting the crawling algorithm. 
When $|\mathcal{R}| \leq \nu$, $\nu \geq 0$ being a user-specified threshold, assuming the processing of $\tilde{\mathcal{Q}}_1$ took $K$ runs, the procedure results in a collection of extracted one-dimensional manifolds represented by the sets $\left\{\overline{\mathcal{P}}^k\right\}_{k=1}^{K}$,
containing sampled points from data set $\tilde{\mathcal{Q}}_1$ representative of manifolds $\left\{\mathcal{M}^k\right\}_{k=1}^{K}$, $\left\{\mathcal{P}^k\right\}_{k=1}^{K}$, the associated low-dimensional counterparts, and $ \left\{TM^k\right\}_{k=1}^{K}$, the respective tangent bundles.

\subsubsection*{Noisy manifolds recovery}
Assuming that set $\overline{\mathcal{P}}^k$ has sampled manifold $\mathcal{M}^k$ with $L^k$ points, for every point $\tivec{t}^\ell \in \overline{\mathcal{P}}^k$ we compute $\mathcal{N}_\ell^k \coloneqq \mathcal{B}(\tivec{t}_l,r)\cap \mathcal{Q}_1$. We define the noisy sample of manifold $\mathcal{M}^k$ as the union of all neighborhoods of radius $r$ computed on all points $\tivec{t}_\ell,~\ell=1,\dots,L^k$: $\mathcal{A}^k = \bigcup_{\ell=1}^{L^k} \mathcal{N}_\ell^k$. By performing this assignment for every set $\overline{\mathcal{P}}^k$, we obtain the $K$ sets containing samples of the noisy manifolds detected in the different runs of crawling.
We now have $K$ manifolds and for each $\mathcal{M}^k$ in $\mathcal{Q}_1$, with $k=1,\dots,K$, three unique sets:
\begin{itemize}
\item $\overline{\mathcal{P}}^k$: contains 
all points sampled from $\tilde{\mathcal{Q}}_1$ at distance of at most $r$, forming a skeleton for the manifold;
\item $\mathcal{P}^k$: the low-dimensional representation of set $\overline{\mathcal{P}}^k$;
\item $\mathcal{A}^k$: the set of all points describing its noisy structure, sampled from $\mathcal{Q}_1$.
\end{itemize}
We also recover the tangent bundle $TM^k$ associated to manifold $\mathcal{M}^k$, by collecting all tangent directions to the manifold on points in $\overline{\mathcal{P}}^k$. For completeness, the list of parameters used in Crawling are presented in tab. \ref{tab:Param_Crawl}.
\begin{table}[t]
\centering
\def\myCWidth{0.12}
\caption{Full list of parameters for Crawling.}
\label{tab:Param_Crawl}
\begin{tabularx}{\columnwidth}{@{\extracolsep{\fill}}
ll>{\raggedleft}p{\myCWidth\columnwidth}>{\raggedleft\arraybackslash}p{\myCWidth\columnwidth}}
\toprule
    $r~^\ast \in \RR$      &  Neighborhood radius \\
    $\beta \in \RR ~(\beta = 0.75)$ & Jump tolerance \\
\bottomrule
\end{tabularx}
\end{table}

\subsubsection{SGTM: Stream GTM}\label{subsec:SGTM}

The Generative Topographic Mapping (GTM, \cite{Bishop1998GTMTG}) is a generative algorithm used generally for dimensionality reduction and density modelling of high dimensional, noisy data sets. 
It is the probabilistic formulation of Self-Organizing Map (SOM, \citet{Kohonen1982}) and it aims at modelling low dimensional structures in high-dimensional data sets as constrained Gaussian Mixtures. 
In its original formulation, the Gaussian centers are constrained to lie on the principal components of the data set, while the noise model is assumed to be spherical. 
Our formulation differs from the original by imposing a structure on the centers co-linear with the manifold's skeleton and a manifold-aligned noise model (replacing the spherical Gaussian formulation). 
The proposed methodology is a simplification of the one proposed in \cite{CANDUCCI2022103579}, by noting that non-intersecting one-dimensional graphs can always be embedded on the unit segment of the real line via a non-linear, parametric mapping.. 
\begin{figure}[t]
\centering
\includegraphics[width= \columnwidth,trim = 0.3cm 0.5cm 1cm 1cm,clip]{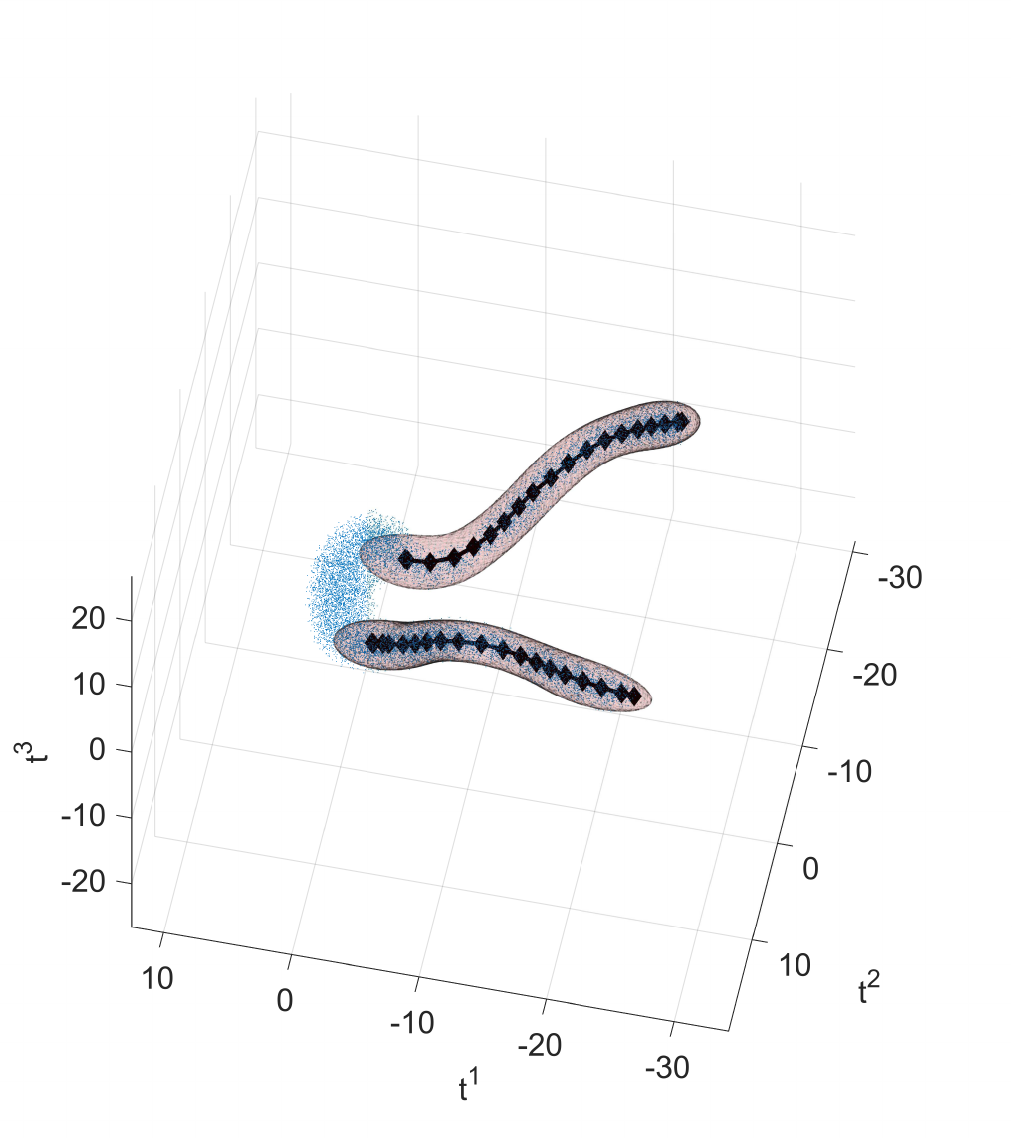}
\caption{Iso-surfaces of the Probability Density Function (PDF) of the individual probabilistic models obtained via SGTM for each manifold in the synthetic data set.}
\label{fig:SynthJF_SGTM}
\end{figure}

Let us consider manifold $\mathcal{M}^k$ found in data set $\mathcal{Q}_1$ by crawling on data set $\tilde{\mathcal{Q}}_1$.
In the following, we describe Stream GTM applied to a single manifold $\mathcal{M}^k$, dropping superscript $k$ (for readability reasons) on every component derived by crawling. However, this methodology is performed for every detected $\mathcal{M}^k$, with $k=1,\dots,K$.\\
We first initialize the latent one-dimensional structure of the manifold by scaling set $\mathcal{P}$, so that it lies on the interval $[-1;1]$:
\begin{equation}
    x_\ell = -1 + \frac{p_\ell - \min(\mathcal{P})}{\max(\mathcal{P}) - \min(\mathcal{P})} \qquad \forall p_\ell \in \mathcal{P} \enspace.
\end{equation}
Calling $\mathcal{X} = \{x_\ell, \ell = 1,\dots,L\}$ the scaled $\mathcal{P}$, we can define a set of $S$ radial basis functions (RBFs) $\phi_1,\dots,\phi_S$, centered on a subset of $\mathcal{X}$:
\begin{equation}\label{eq:RBFs}
    \phi_s(x_\ell) = \exp\left[-\frac{(x_s - x_{\ell})^2}{2\sigma^2} \right] \enspace.
\end{equation}
Here $\sigma$ is computed as the mean distance between adjacent centers: $\sigma = \sum_{s=1}^{S-1} \lVert x_s - x_{s + 1}\rVert/(S-1)$. The centers $x_s$ of the RBFs are sampled regularly from $\mathcal{X}$.

The mapping of points from the latent space $\mathcal{X}$ to embedded points in $\overline{\mathcal{P}}$ is achieved by the function $\vector{y}(x;\bm{W})$, governed by the $S \times D$ matrix of parameters $\bm{W}$. Given the definition of RBFs in equation \eqref{eq:RBFs} we can define the mapping function as:
\begin{equation}
    \vector{y}(\vector{x};\vector{W}) = \bm{\Phi}(\vector{x})\bm{W}
\end{equation}
where $\vector{x}$ is the column vector containing points in $\mathcal{X}$ and $\bm{\Phi}(\vector{x})$ is a $L \times S$ matrix, having $\bm{\Phi}_{\ell s} = \phi_s(x_\ell)$. 
The manifold aligned probabilistic model is a flat mixture model
\begin{equation}\label{eq:sum_int}
 p(\vector{t}|\bm{W},\Sigma_{\ell}) = \frac{1}{L}\sum_{\ell = 1}^{L} 
p(\vector{t}|x_\ell,\bm{\Sigma}_\ell,\bm{W}),
\end{equation}
where the mixture components are locally manifold-aligned multivariate Gaussians centered at the embedded points $\tivec{t}_\ell \in \overline{\mathcal{P}}$:
\begin{equation}\label{eq:NoiseGTM}
 p(\vector{t}|x_\ell, \bm{\Sigma}_\ell,\bm{W})  =  
\frac{1}{[(2\pi )^D |\bm{\Sigma}_\ell|]^\frac{1}{2}} \exp{\left( -\frac{\Delta\vector{t}^\top \bm{\Sigma}_\ell^{-1} \Delta\vector{t}}{2} 
 \right)}
\end{equation}
with $ \Delta\vector{t} = \vector{y}(x_\ell;\bm{W}) - \vector{t}$.
As proposed in \cite{Bishop1998DevelopmentsOT}, we model the local manifold-aligned covariance 
matrix by computing the derivatives of the mapping function with respect to the latent variables:
\begin{equation}
    \bm{\Sigma}_{\ell} = \frac{1}{\upsilon} \mathbf{I} + \omega \frac{\partial \vector{y}^\top}{\partial x}\bigg|_{x_{\ell}} \frac{\partial\vector{y}}{\partial x}\bigg|_{x_{\ell}},
\end{equation}
where the purpose of parameter $\upsilon$ (in this work $\upsilon = 1e3$) is avoiding singularity of $\bm{\Sigma}_{\ell}$ and $\omega$ is a scaling factor equal to the distance between neighboring nodes in the latent space. The derivatives of the mapping function with respect to the latent space coordinates are:
\begin{equation}\label{eq:MappingDer}
    \vector{g}(x_\ell) = \frac{\partial \vector{y}}{\partial x}\bigg|_{x_{\ell}} = \frac{\partial \bm{\Phi}}{\partial x}\bigg|_{x_{\ell}}\bm{W} = \sum_{s=1}^S \frac{(x_{\ell} - x_s)}{\sigma^2}\vector{\phi}(x_\ell)\vector{w}_s \in \mathbb{R}^D,
\end{equation}
where we denote by $\vector{w}_s$ the $s-$th row of matrix $\bm{W}$.
As a latent variable model, Stream GTM can be trained to maximize log-likelihood
\begin{equation}\label{eq:CompLogL}
 \mathcal{L} (\vector{W}) = \sum_{n = 1}^{N} \ln \left\{\frac{1}{L}\sum_{\ell 
= 1}^{L} p(\vector{t}_n|x_\ell,\bm{\Sigma}_\ell, \bm{W}) \right\}
\end{equation}
via the E-M algorithm 
outlined in \cite{Bishop1998DevelopmentsOT}.
All parameters in SGTM are listed in tab. \ref{tab:Param_SGTM}.
\begin{table}[t]
\centering
\def\myCWidth{0.12}
\caption{Full list of parameters for SGTM.}
\label{tab:Param_SGTM}
\begin{tabularx}{\columnwidth}{@{\extracolsep{\fill}}
ll>{\raggedleft}p{\myCWidth\columnwidth}>{\raggedleft\arraybackslash}p{\myCWidth\columnwidth}}
\toprule
    $r~^\ast \in \RR$      &  Neighborhood radius \\
    $S~^\ast \in \NN$      & Number of RBFs \\
    $\upsilon \in \RR~(\upsilon = 1e^3)$    &  Regularization Cov. matrix\\
    $\omega \in \RR$  &  Scaling factor Cov. matrix (adaptive)\\
\bottomrule
\end{tabularx}
\end{table}
The application of SGTM to the individual manifolds in the synthetic data set results in two probabilistic model which can be visualized by their iso-surfaces corresponding to an iso-value of their Probability Density Functions (PDFs), see fig. \ref{fig:SynthJF_SGTM}, pink surfaces.\\

From the discussion presented, the toolbox is mainly dependant on one single hyper-parameter: the neighbourhood radius $r$. It is advisable to choose the parameter after visual inspection of the data set at hand. The choice of this hyper-parameter influences the computational cost of the whole methodology, since LAAT, EM3A, Dimensionality Index, Crawling and SGTM all rely on this for the computation of local PCA. If the radius $r$ is chosen so that a sphere of radius $r$ encloses the estimated thickness of the filaments within the data set, slight variations of this hyper-parameter from the designated value do not influence the results significantly. A more detailed analysis of the stability of the toolbox w.r.t. $r$ will be presented in an additional work, in preparation.
It is not straightforward to estimate the computational cost of the toolbox once the radius $r$ has been chosen, because of the recursive nature of most algorithms. A more detailed analysis of this on specific data sets can be found in the respective papers: \cite{Taghribi2020LAATLA} for LAAT, \cite{Mohammadi2020_EM3A} and for EM3A. A detailed theoretical analysis of EM3A is also presented in \cite{EM3A_MohammadiTheory} where convergence bounds are outlined together with optimal estimation of hyper-parameters. Quantitative analysis of the effect of changes in the parameters, as well as computational costs for rest of the methodologies (Dimensionality Index, Crawling and SGTM) can be found in \cite{CANDUCCI2022103579}. In the respective papers, each methodology is proofed against  state-of-the-art comparable techniques. 

\section{Visualization techniques}\label{sec:Visualization}

\subsection{Bi-dimensional profiles}\label{subsec:BidimProfiles}

After optimization of SGTM through the E-M algorithm, every manifold $\mathcal{M}_k$ is represented as a Gaussian mixture with manifold aligned noise, whose updated centers $\{\tivec{t}_{\ell}; \ell = 1,\dots,L^k\}$ are constrained to lie on a one-dimensional subspace of $\RR^3$. We now describe a methodology that, taking full advantage of the probabilistic nature of SGTM, simultaneously recovers the behaviour of properties along the manifold's elongation and its thickness within the simulated volume. This methodology gives a comprehensive view of the extracted manifold in a single frame, allowing for a better understanding of its main radial and longitudinal features.

\begin{figure}[t]
\centering
 \includegraphics[width = 0.45\textwidth]{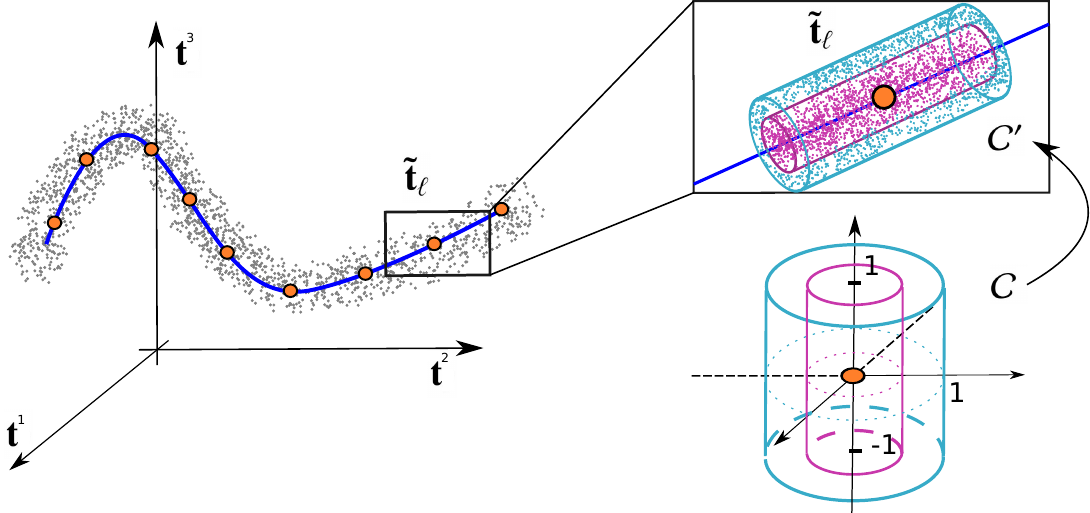}
 \caption{Sketch depicting the formation of the uniformly sampled concentric cylindrical volumes, aligned with the local tangent space of manifold $\mathcal{M}_k$ on the SGTM center $\tivec{t}_\ell$.}
 \label{fig:BDimPlot}
\end{figure}
The centers obtained at the end of optimization have most likely shifted along the manifold, due to local variations of point density within the manifold itself. In order to take this shift into consideration we update the tangent bundle to manifold $\mathcal{M}_k$ by computing the derivative of the (trained) mapping function with respect to latent center $x_\ell$: 
\begin{equation}\label{eq:TB_Update}
 \hat{\vector{\xi}}_{\ell} =  \vector{g}(x_\ell) \enspace,
\end{equation}
where $\vector{g}(x_\ell)$ is derived in equation \eqref{eq:MappingDer}. Doing this for every center $x_\ell \in \mathcal{P}^k$ we obtain the updated subset of the tangent bundle $TM^k = \{\hat{\vector{\xi}}_{\ell}; ~\ell=1,\dots,L^k - 1\}$ of manifold $\mathcal{M}_k$.
Let us now consider a point cloud $\mathcal{C}$ containing points as row vectors, uniformly sampling a cylindrical volume of radius $2$, aligned along the $z-$axis and centered at the origin $\mathcal{O} = (0,~0,~0)$. In order to radially partition the point cloud $\mathcal{C}$ into concentric cylindrical shells (bottom right panel of figure \ref{fig:BDimPlot}, cyan and magenta cylinders), we first create $\overline{\mathcal{C}}$ by projecting $\mathcal{C}$ onto the $x-y$ plane:
\begin{equation}
\overline{\mathcal{C}} = \mathcal{C} 
\begin{pmatrix}
1 & 0\\
0 & 1\\
0 & 0
\end{pmatrix}
\end{equation}
For every point $\overline{\vector{p}} \in \overline{\mathcal{C}}$ we compute its distance to the projected origin $d(\overline{\vector{p}},\overline{\vector{0}} ) = \|\overline{\vector{p}} - [0,0]\|$. We can now group points in $\overline{\mathcal{C}}$ so that
\begin{equation}
    \mathcal{I}_i = \left\{j~|~ d^r_{i-1} \leq d(\overline{\vector{p}}_j,\overline{\vector{0}})< d^r_{i}\right\}\quad \forall \overline{\vector{p}}_j \in \overline{\mathcal{C}},
\end{equation}
where $d^r_i = (i - 1)\times r_M/(c-1)$ is the low extreme of the interval defined by two consecutive concentric rings on the cylinder, $r_M$ the maximum allowed distance from the mean curve of the manifold, $c$ the desired number of bins across the radial direction and $i=1,\dots,c-1$. The sets $\mathcal{I}_1,\mathcal{I}_2,\dots,\mathcal{I}_{N^r}$ contain all indices of points in $\overline{\mathcal{C}}$ (and thus in $\mathcal{C}$) belonging to specific cylindrical shells concentric with respect to the origin $\overline{\vector{0}}$ ($\vector{0}$) and the $z-$axis.
It is always possible to scale, translate and rotate the point cloud so that the cylindrical axis is oriented as vector $\hat{\vector{\xi}}_{\ell}$, the origin over-posed to center $\tivec{t}_{\ell}$ and the axis length equal to $d_{\ell,~\ell+1}$ (as shown in figure \ref{fig:BDimPlot}, top, right panel).

\paragraph{Scaling} In matrix notation, the scaling operator is $\bm{S}$
\[
\bm{S} = 
\begin{pmatrix}
 1 & 0 & 0\\
 0 & 1 & 0\\
 0 & 0 & d_{\ell,~\ell+1}
\end{pmatrix}
\]
\paragraph{Rotation}
  We can compute the quaternion $\vector{q}$ (\cite{hamilton1866elements}) where the first three components are given by $\hat{\vector{\xi}}_{\ell} \times (0, ~0, ~1)$ and the $4-$th component by $\hat{\vector{\xi}}_{\ell} \cdot (0,~0,~1)$. Its matrix representation is given by $\bm{R}$:
\[
\bm{R} = 
\begin{pmatrix}
 1 - 2q_2^2 - 2q_3^2  &    2q_1q_2 - 2q_3q_4  &   2q_1q_3 + 2q_2q_4\\
2q_1q_2 + 2q_3q_4   &    1 - 2q_1^2 - 2q_3^2	&   2q_2q_3 - 2q_1q_4 \\
2q_1q_3 - 2q_2q_4	&   2q_2q_3 + 2q_1q_4   &	1 - 2q_1^2 - 2q_2^2
\end{pmatrix}
\]
\paragraph{Shift} 
We can then shift the scaled and rotated point cloud so that its origin is on center $\tivec{t}_{\ell}$.
\begin{figure*}[!t]
\centering
\begin{tikzpicture}[draw,node distance = 0cm,nodes = {anchor=north west,inner sep=0cm},font=\footnotesize]
\node[] (a_cap) {\textbf{(a)} Estimated manifold $\mathcal{M}_3$, normalized weighted mean, as recovered after the application of the toolbox on the initial noisy data set.};
\node[anchor=north west] (a) at (a_cap.south west) {\includegraphics[width=0.97\textwidth,trim = 1cm 0.5cm 0.5cm 0.5cm,clip]{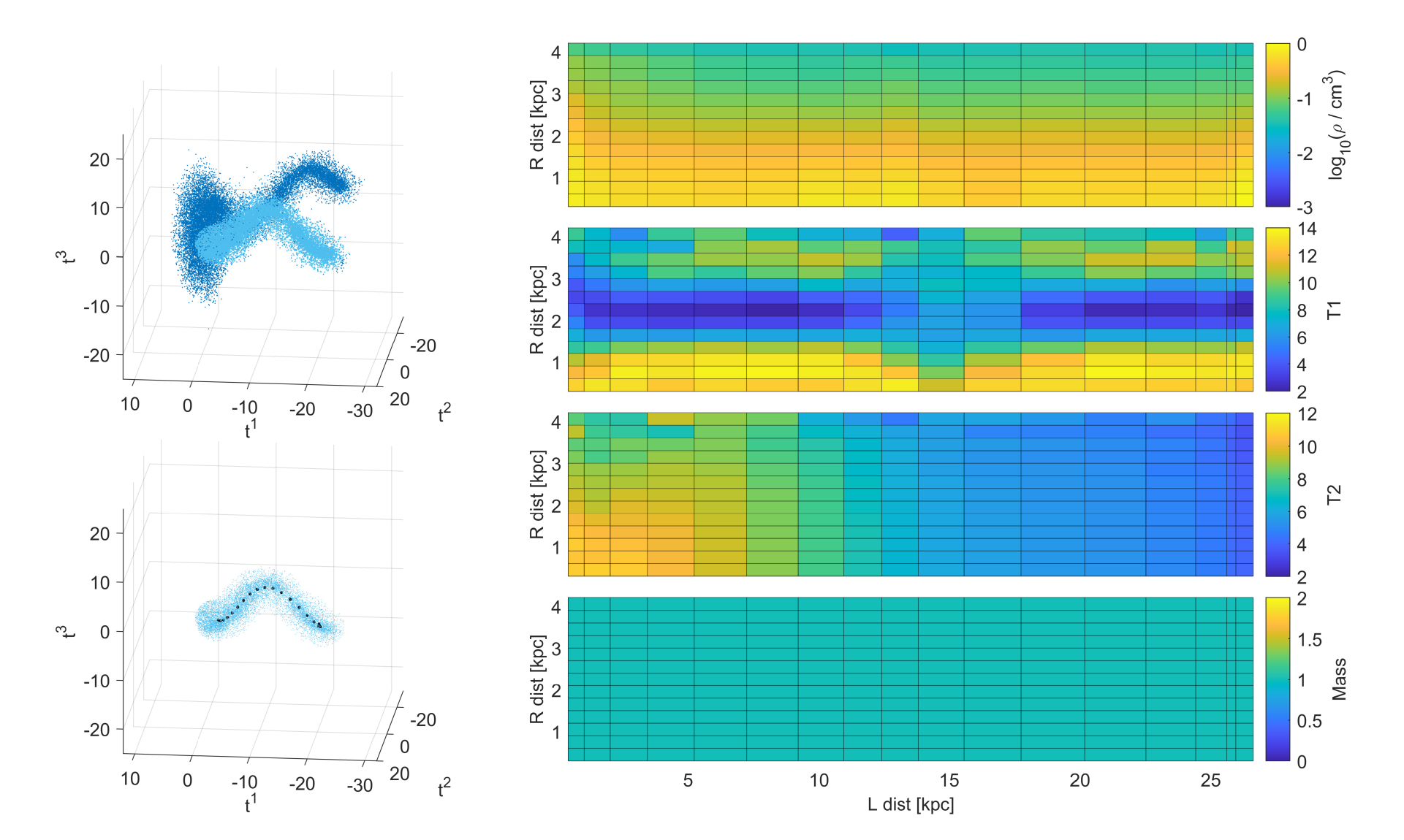}};
\node[] (b_cap) at (a.south west) {\textbf{(b)} Ground Truth manifold $\mathcal{M}_3$, normalized weighted mean. The mean curve of the manifold is the Ground Truth used to construct the data set.};
\node[] (b) at (b_cap.south west) {\includegraphics[width=0.97\textwidth,trim = 1cm 0.5cm 0.5cm 0.5cm,clip]{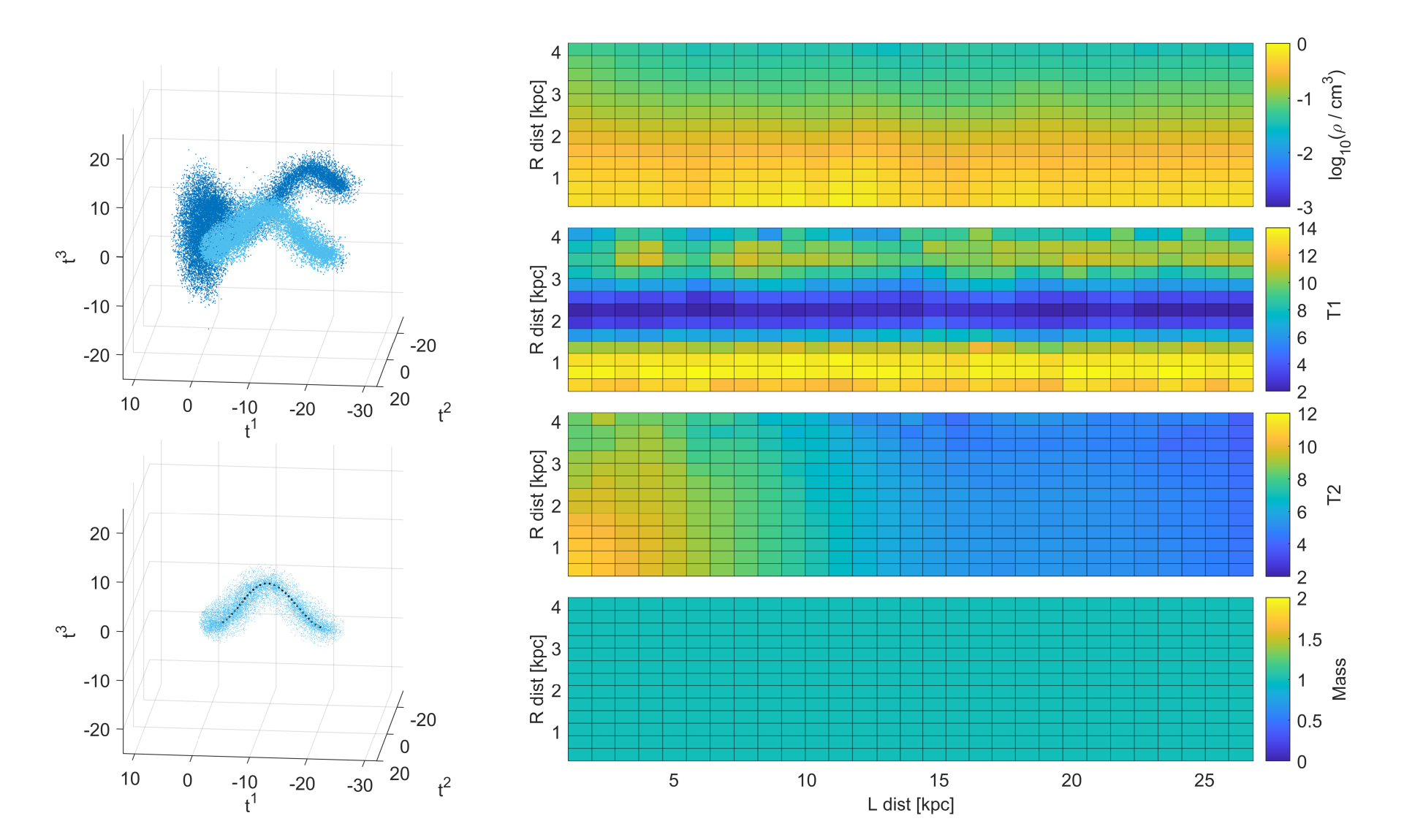}};
\end{tikzpicture}
\caption{Comparison of the estimated manifold $\mathcal{M}_3$ (a) and its Ground Truth (b) on the synthetic Jellyfish data.
Left: top panels show the noisy data set points (blue) with background noise removed and including the noisy manifold identification (cyan) both for SGTM (a) and the Ground Truth (b). Correspondingly, the bottom panels depict the skeleton of $\mathcal{M}_3$ (black) recovered via Crawling and SGTM  in (a) and its Ground Truth in (b). Right: contains the estimated (a) and true (b) bi-dimensional profiles (normalized with equation \eqref{eq:WeightedMean}) for 
variables $\overline{\rho}_i, \overline{T}_{1i}, \overline{T}_{2i},$ and $\overline m_i$ (from top to bottom).}
\label{fig:MeanPlots_M1}
\end{figure*}
\begin{figure*}[t!]
\centering
\begin{tikzpicture}[draw,node distance = 0cm,nodes = {anchor=north west,inner sep=0cm},font=\footnotesize]
\node[] (a_cap) {\textbf{(a)} Estimated manifold $\mathcal{M}_2$, normalized weighted mean, as recovered after the application of the toolbox on the initial noisy data set.};
\node[anchor=north west] (a) at (a_cap.south west) {\includegraphics[width=0.97\textwidth,trim = 1cm 0.5cm 0.5cm 0.5cm,clip]{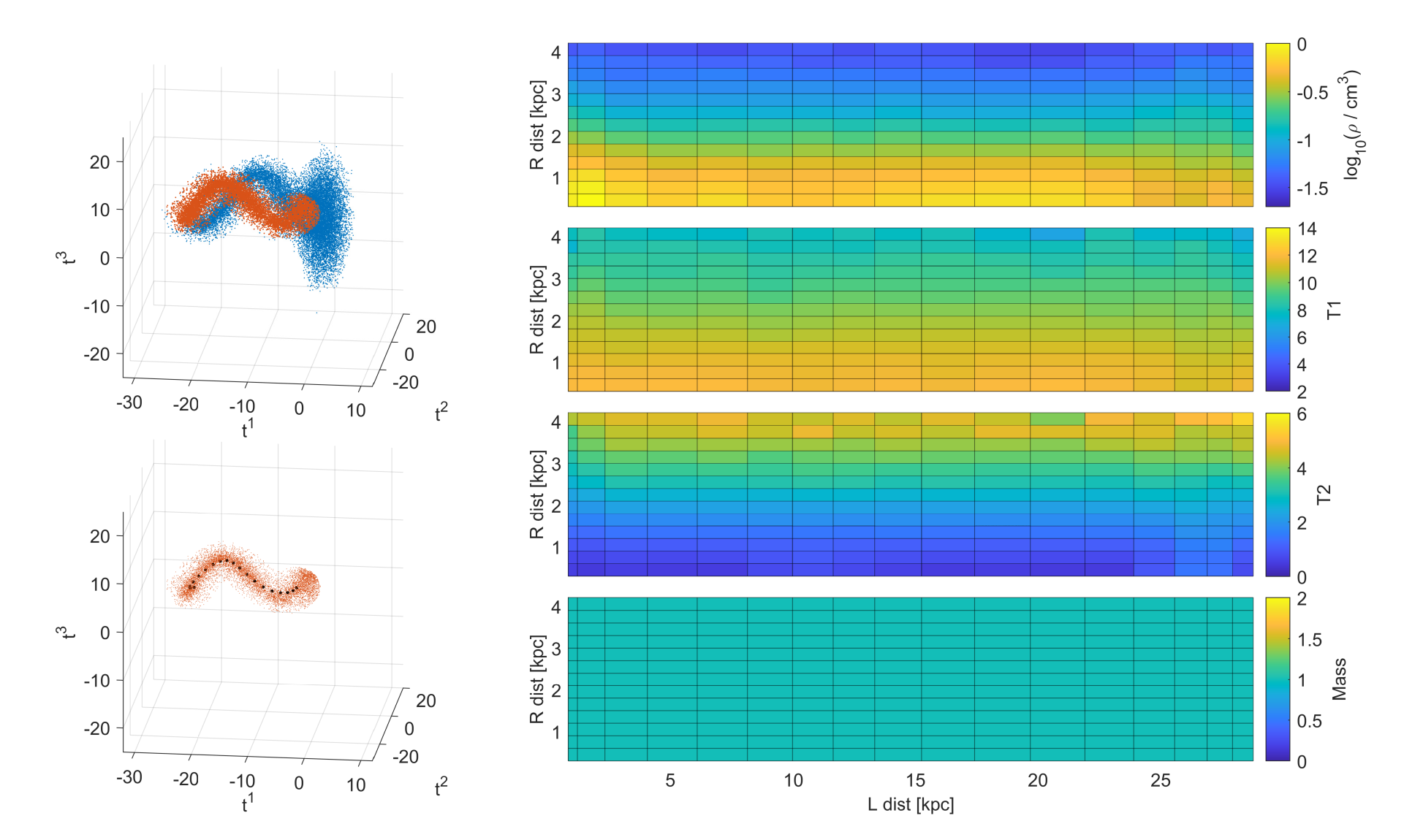}};
\node[] (b_cap) at (a.south west) {\textbf{(b)} Ground Truth manifold $\mathcal{M}_2$, normalized weighted mean. The mean curve of the manifold is the Ground Truth used to construct the data set.};
\node[] (b) at (b_cap.south west) {\includegraphics[width=0.97\textwidth,trim = 1cm 0.5cm 0.5cm 0.5cm,clip]{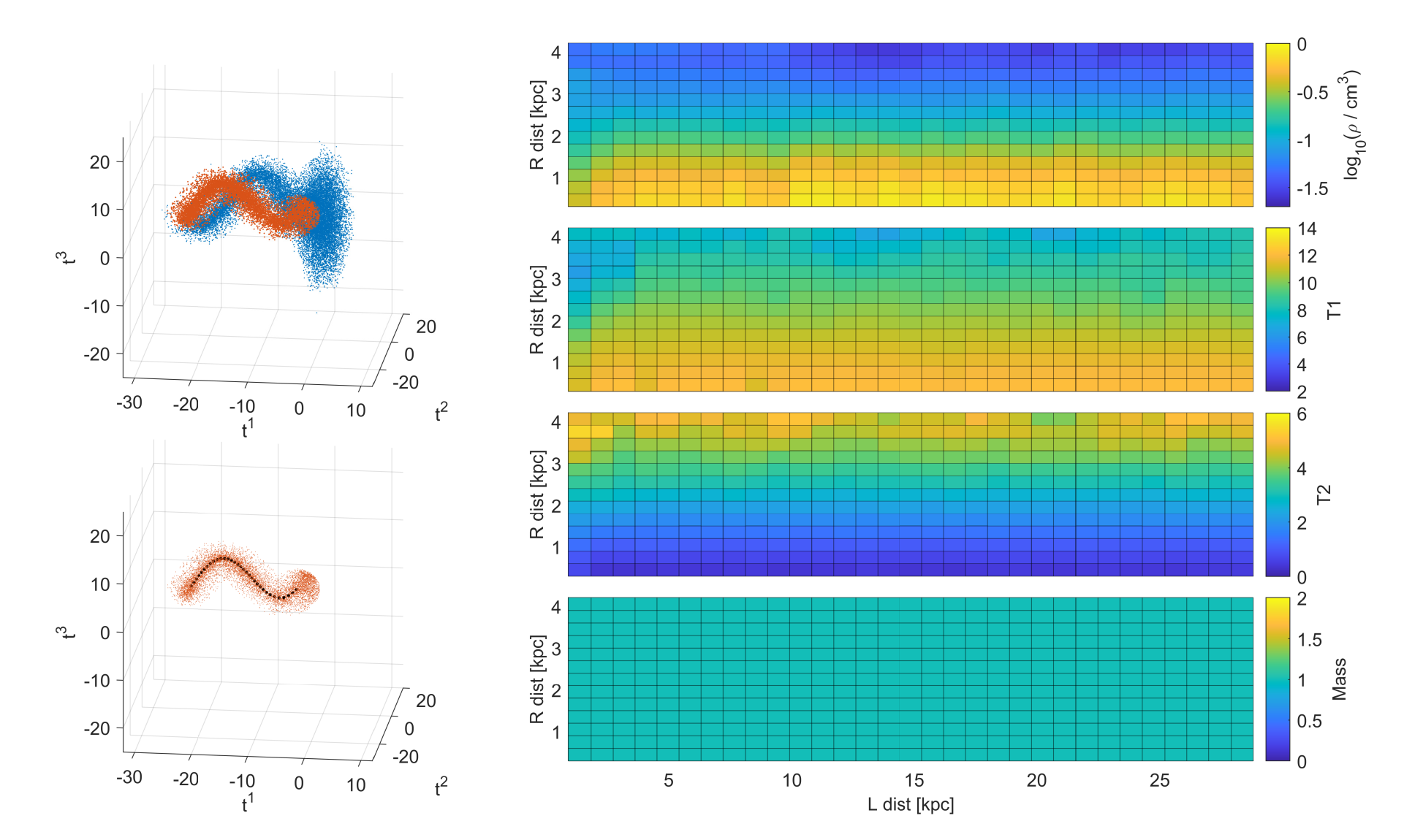}};
\end{tikzpicture}
\caption{
Comparison of the estimated manifold $\mathcal{M}_2$ (a) and its Ground Truth (b) on the synthetic Jellyfish data.
Left: top panels show the noisy data set points (blue) with background noise removed and including the noisy manifold identification (orange) both for SGTM (a) and the Ground Truth (b). Correspondingly, the bottom panels depict the skeleton of $\mathcal{M}_2$ (black) recovered via Crawling and SGTM in (a) and its Ground Truth in (b). Right: contains the estimated (a) and true (b) bi-dimensional profiles (normalized with equation \eqref{eq:WeightedMean}) for 
variables $\overline{\rho}_i, \overline{T}_{1i}, \overline{T}_{2i},$ and $\overline m_i$ (from top to bottom).
} 
\label{fig:MeanPlots_M2}
\end{figure*}

Any point $\vector{p} \in \mathcal{C}$ is then mapped to $\vector{p}' \in \mathcal{C}'$ under the combined operator as:
\begin{equation}\label{eq:cylOperator}
    \vector{p}' = (\vector{p} ~ \bm{S}) ~ \bm{R} + \tivec{t}_\ell.
\end{equation}
Having obtained a point cloud $\mathcal{C}'$ uniformly sampling a thick cylindrical volume with axis tangential to manifold $\mathcal{M}_k$ on point $\tivec{t}_\ell$, we can now compute the weighted mean of any quantity contained in the data set, over the volume sampled by $\mathcal{C}'$. 

Note that the families of indices $\mathcal{I}_1,\dots,\mathcal{I}_{N^r}$, when applied to $\mathcal{C}'$, contain all indices of points in cylindrical shells concentric with respect to point $\tivec{t}_\ell$ and vector $\hat{\vector{\xi}}_{\ell}$, radially partitioning the cylindrical volume sampled by $\mathcal{C}'$. For each linear segment of manifold $\mathcal{M}_k$, parameterised by its corresponding SGTM, we have now obtained a uniformly sampled cylindrical volume aligned along its corresponding local tangent space. Computing $\langle T_m(\vector{p}') \rangle$ for every $\vector{p}' \in \mathcal{C}'$ we can now evaluate the mean value of $T_m$ over the concentric rings defined by the index families $\mathcal{I}_1,\dots,\mathcal{I}_{N^r}$ as:
\begin{equation}\label{eq:MeanVarBiDim}
    \overline{T_{m,i}} = \langle T_m(d^r_{i-1},d^r_i)\rangle = \frac{\sum_{j\in\mathcal{I}_i} \langle T_m(\vector{p}'_j)\rangle}{|\mathcal{I}_i|} \enspace,
\end{equation}
obtaining the mean of $T_m$ over the cylindrical shell between $(d^r_{i-1},d^r_i)$ for every $i=1,\dots,N^r$.

We can iterate the whole process for every center of SGTM, obtaining for each linear segment, the distribution of $T_m$ in concentric cylindrical shells centered on the current center (figure \ref{fig:BDimPlot}, left panel).
By considering both longitudinal (defined recursively by the centers of SGTM) and radial (obtained by the linear operator defined in equation \ref{eq:cylOperator} on point cloud $\mathcal{C}$) profiles, we obtain the plots shown in 
figures \ref{fig:MeanPlots_M1}a 
and \ref{fig:MeanPlots_M2}a. 
In both panels, the vertical axis of each plot contains the radius of the cylindrical shells $d^r$ and the horizontal axis the approximated geodesic distance (computed by summation of the lengths of the individual linear segments) from the head of the manifold. From these plots we can verify that the behaviour of quantities $T_1$ (second panel from top) and $T_2$ (third panel from top) are in agreement with how they were designed when constructing the data set (sec. \ref{sec:Synt_Data}). The decreasing profile for quantity $T_1$ and increasing for $T_2$ along manifold $\mathcal{M}_2$ is detected, as well as the sinusoidal behaviour of quantity $T_1$ along manifold $\mathcal{M}_3$'s thickness and $T_2$'s decreasing profile along its longitudinal elongation. The bottom plot in each panel presents the mass distribution over the radial and longitudinal dimensions of the manifolds. As expected, the mass is constant throughout the sampled volumes and it is everywhere $\overline m_i = 1$.

For each manifold, figures \ref{fig:MeanPlots_M1}b and \ref{fig:MeanPlots_M2}b 
show the true profiles recovered by using the ground truth skeletons described in section \ref{sec:Synt_Data}. As previously described, variable $T_1$ shows a sinusoidal variation along the radial direction of manifold $\mathcal{M}_3$ and decreasing radial profile across manifold $\mathcal{M}_2$, variable $T_2$ is decreasing on the longitudinal direction of manifold $\mathcal{M}_3$ and radially decreasing from the core of manifold $\mathcal{M}_2$. The profiles, when compared with the ones obtained using the skeletons recovered by our methodology, look virtually identical. Small deviations are noticeable for variable $T_1$ halfway through manifold $\mathcal{M}_3$, however even in this case the variation is minimal and does not compromise the overall agreement.
This demonstration is a first quantitative confirmation of the accuracy of our methodology in recovering the underlying structures of the data set.
\begin{figure*}[ht!]
    \centering
    \includegraphics[width = \textwidth]{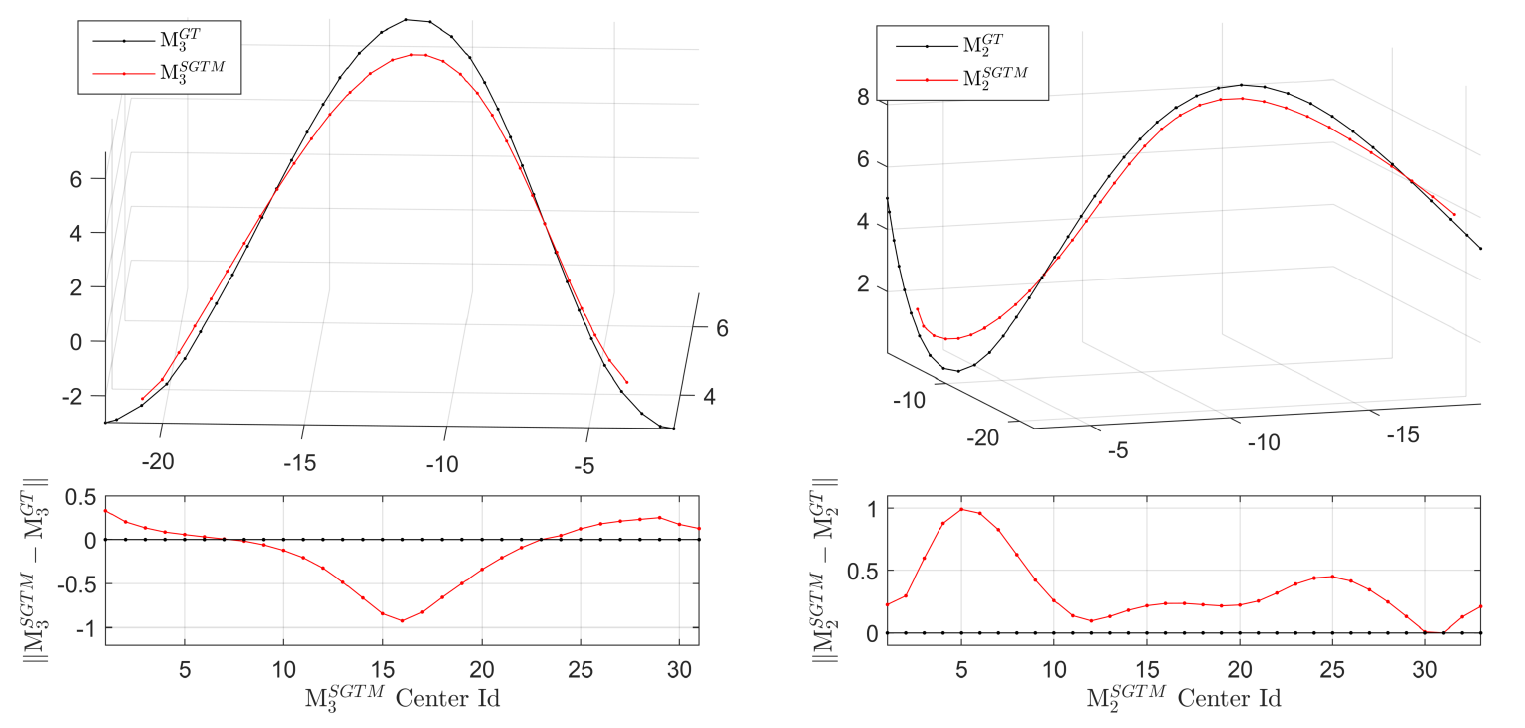}
    \caption{Top row: recovered manifolds (in red) via SGTM and their corresponding Ground Truth (black). Bottom row: Point-by-point orthogonal distance between $\mathcal{M}_k^{GT}$ and $\mathcal{M}_k^{SGTM}$.}
    \label{fig:ManifoldsComparison}
\end{figure*}
Having obtained a parameterization of the two manifolds in the data set ($\mathcal{M}_k^{SGTM}$), it is now possible to compare the recovered structures with their Ground Truth ($\mathcal{M}_k^{GT}$, given in eq. \eqref{eq:Embeddings}).
In order to have a measure that is unbiased with respect to the particular parameterization produced by Crawling and SGTM, we re-sample each manifold by projecting the Ground Truth points onto the recovered curves, along their local orthogonal planes. The module of the projection gives the local orthogonal distance between a point in $\mathcal{M}_k^{GT}$ and its corresponding in $\mathcal{M}_k^{SGTM}$. The two recovered manifolds (red points), together with their Ground Truths (black points) are shown in the top row of figure \ref{fig:ManifoldsComparison}. The point-by-point orthogonal distance is shown in the bottom row.
The maximum distance between the two curves is at the point of maximal curvature of the manifold. This discrepancy has two main contributions. Firstly, the data set obtained with EM3A is slightly misaligned with the Ground Truth. This is probably due to the random sparsity of the transverse and background noise in the data set. The combined effect of these two uncertainty factors may affect the local density estimation, and thus the displacement of nearby points onto local tangent spaces (over- or under-shooting). The second contribution is due to the application of Crawling and SGTM. While the initialization provided by crawling lies on the data set obtained with EM3A and carries the same uncertainty, the SGTM formulation should compensate for possible small deviations. However, the radius $r = 2$ chosen in this example might be too large for capturing the high-curvature regions. Also, the imposed $\sigma$ value (estimated from the RBF setup) used for constraining the model's complexity is possibly too large. It is however generally advisable to avoid over-fitting the data with an overly complex model for the sake of its generalizability to unseen data sampled from the same distribution. Despite this minor discrepancy between the recovered curves, an overall agreement is achieved for both manifolds. 
The discrepancy between the two curves at the peak of the distance is also responsible for the deformation of the bi-dimensional profiles, especially in the central part of the plots (e.g. figure \ref{fig:MeanPlots_M1} panels a and b) 
for variable $T_1$ (sinusoidal profile from the core of manifold $\mathcal{M}_2$). 
Nonetheless, despite the amount of noise corrupting both the structures and the simulated quantities, 1-DREAM still manages to recover a reasonable approximation to the Ground Truth.

\begin{table}[t]
\centering
\def\myCWidth{0.12}
\caption{Recovery accuracy from bi-dimensional profiles}
\label{tab:FracDev}
\begin{tabularx}{\columnwidth}{@{\extracolsep{\fill}}
ll>{\raggedleft}p{\myCWidth\columnwidth}>{\raggedleft\arraybackslash}p{\myCWidth\columnwidth}}
Manifold & $\varsigma^{0.2}(\log^{10}~\rho)$ & $\varsigma^{0.2}(T_1)$ & $\varsigma^{0.2}(T_2)$\\
\toprule
$\mathcal{M}_3$ & $0.9897$ & $0.8615$ & $0.9846$\\
$\mathcal{M}_2$ & $0.9089$ & $1.0000$ & $0.8776$\\
\bottomrule
\end{tabularx}
\end{table}

The centers obtained by projection of the Ground-Truth onto the manifolds recovered via SGTM can also be used to obtain new Ground Truth bi-dimensional profiles. In this case, the two bi-dimensional profiles for each manifold are directly comparable. The mean value of variable $T_m$ in bin $(d_{i-1}^r,d_i^r)$ of the bi-dimensional profile $\overline{T_{m,i}} = \langle T_m(d_{i-1}^r,d_i^r)$ is derived according to equation \eqref{eq:MeanVarBiDim} for the ground-truth, $\overline{T_{m,i}}^{GT}$, and SGTM, $\overline{T_{m,i}}^{SGTM}$, recovered manifolds. For each bin we can now compute the fractional deviation of variable $T_m$ between the two recovered structures:
\begin{equation}\label{eq:FractionalDev}
    \varsigma_{m,i} = \left\lVert\frac{\overline{T_{m,i}}^{SGTM} - \overline{T_{m,i}}^{GT}}{\overline{T_{m,i}}^{GT}}\right\rVert \enspace.
\end{equation}
We estimate the accuracy of the recovered bi-dimensional profiles as the ratio of pixels having fractional deviation lower than $0.2$ and the total number of pixels, such that
\begin{equation}\label{eq:BidimAcc}
    \varsigma^{0.2}(T_m) = \frac{| \{(i,\ell)~ | ~\varsigma_{m,i} < 0.2 \}|}{N^r \times (L^k -1)} \enspace.
\end{equation}
The values of $\varsigma^{0.2}(T_m)$, $T_m$ being $\log_{10}(\rho), T_1$ or $T_2$ for both manifolds $\mathcal{M}_2$ and $\mathcal{M}_3$ are given in table \ref{tab:FracDev}. In agreement with the previous discussion, variable $T_1$ for manifold $\mathcal{M}_3$ and $T_2$ for $\mathcal{M}_2$ show the largest variation with respect to the Ground-Truth, however, only $\sim 14\%$ and $\sim 13\%$ (respectively) of the pixels have larger fractional deviation than $0.2$, while we reach a good agreement between Ground-Truth and SGTM in the other cases. The parameters of the Bi-dimensional profile technique are provided in tab. \ref{tab:Param_BiDim}.
\begin{table}[h]
\centering
\def\myCWidth{0.12}
\caption{Full list of parameters for Bi-dimensional profiles.}
\label{tab:Param_BiDim}
\begin{tabularx}{\columnwidth}{@{\extracolsep{\fill}}
ll>{\raggedleft}p{\myCWidth\columnwidth}>{\raggedleft\arraybackslash}p{\myCWidth\columnwidth}}
\toprule
    $c~^\ast \in \NN$      & Radial number of bins \\
    $r_M~^\ast \in \RR$          &  Maximum radial distance from mean curve\\
\bottomrule
\end{tabularx}
\end{table}
\subsection{Co-Moving orthonormal coordinate frames}\label{subsec:OrthoPlanes}

\begin{figure}[t]
\centering
\includegraphics[width = \columnwidth]{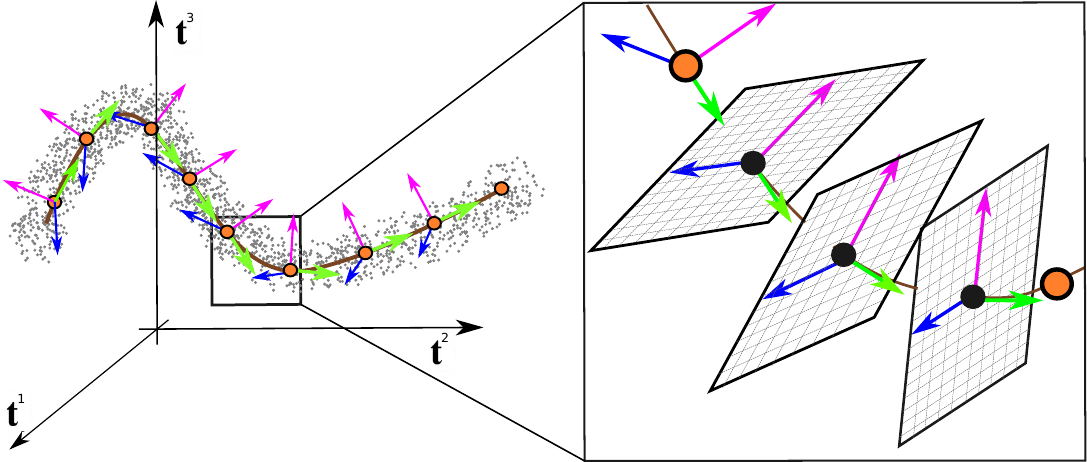}
\caption{Sketch depicting the formation of co-moving orthonormal planes along a portion of manifold $\mathcal{M}_k$, enclosed within two centers of the corresponding SGTM.}
\label{fig:OrthoPlanes}
\end{figure}

For each manifold $\mathcal{M}_k$, we can now obtain a better discretization of its ``spine'', by introducing more points in the latent space and propagating them in the ambient space through the mapping function $\vector{y}(x;\bm{W}): \mathcal{P}^k \longrightarrow \overline{\mathcal{P}}^k$. Consider the linear segment $ \mathcal{I}_\ell \coloneqq [x_\ell, x_{\ell + 1}]$, where $x_\ell, x_{\ell+1} \in \mathcal{P}^k$. Denoting by $d_\ell = \lVert \mathcal{I}_\ell \rVert$ the length of segment $\mathcal{I}_\ell$, let us assume that the number of equidistant points to be inserted in $\mathcal{I}_\ell$ is $N_\ell$. 
We can define the new points via the recursive rule:
\begin{equation}
    x_\ell^m = x_\ell + \frac{m}{N_\ell + 1} d_\ell \quad \mathrm{for} \quad m = 1,\dots,N_\ell \enspace.
\end{equation}
Applying this relation to points $x_\ell, \forall \ell = 1,\dots,L^k-1$,
We obtain the up-sampled latent space $\mathcal{P}^k_{\uparrow} = \left\{x_1,x_1^1,\dots,x_1^{N_\ell},x_2,\dots,x_{L^k-1},x_{L^k-1}^1,\dots,x_{L^k-1}^{N_\ell},x_{L^k}\right\}$, having size $\left| \mathcal{P}^k_\uparrow \right| = N^\ell \left( L^k - 1 \right) + L^k$. Propagation of latent point set $\mathcal{P}^k_\uparrow$ into the ambient space through mapping function $\vector{y}(x_\ell;\bm{W})$, for every $x_\ell \in \mathcal{P}^k_\uparrow$, leads to the up-sampled embedded point-set $\overline{\mathcal{P}}^k_\uparrow$, as depicted in figure \ref{fig:OrthoPlanes}, right panel, black dots. As in section \ref{subsec:BidimProfiles}, the tangent bundle $TM^k$ is updated by applying equation \eqref{eq:TB_Update} to every point in $\overline{\mathcal{P}}^k_\uparrow$. In this section, slightly abusing mathematical notation, we will use index $\ell$ for any latent (and corresponding embedded) point belonging to $\mathcal{P}^k_\ell$ and drop subscript $\uparrow$, for readability purposes.

\begin{figure*}[t!]
\begin{tikzpicture}[node distance = 0cm,nodes = {anchor=north west,inner sep=0cm},font=\footnotesize]
\node[] (a_cap)                   {\textbf{(a)} Manifold $\mathcal{M}_3$, snapshot n.20 of associated movie clip.};
\node[] (a) at (a_cap.south west) {\includegraphics[width=\textwidth,trim = 0.67cm 1.1cm 1.2cm 1cm,clip]{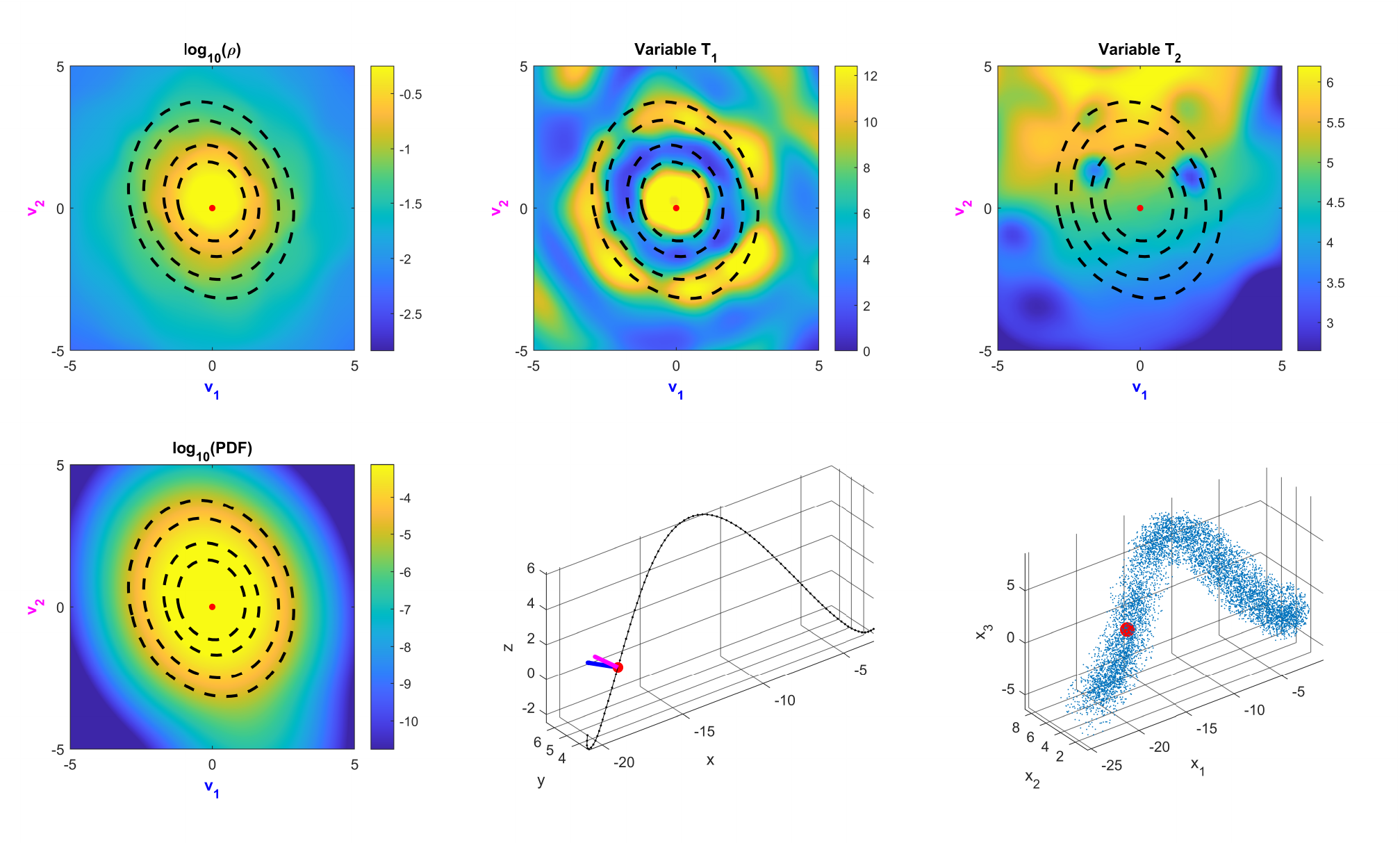}};
\node[] (b_cap) at (a.south west) {\textbf{(b)} Manifold $\mathcal{M}_3$, snapshot n.80 of associated movie clip.};
\node[] (b) at (b_cap.south west) {\includegraphics[width=\textwidth,trim = 0.65cm 1.1cm 0.4cm 1cm,clip]{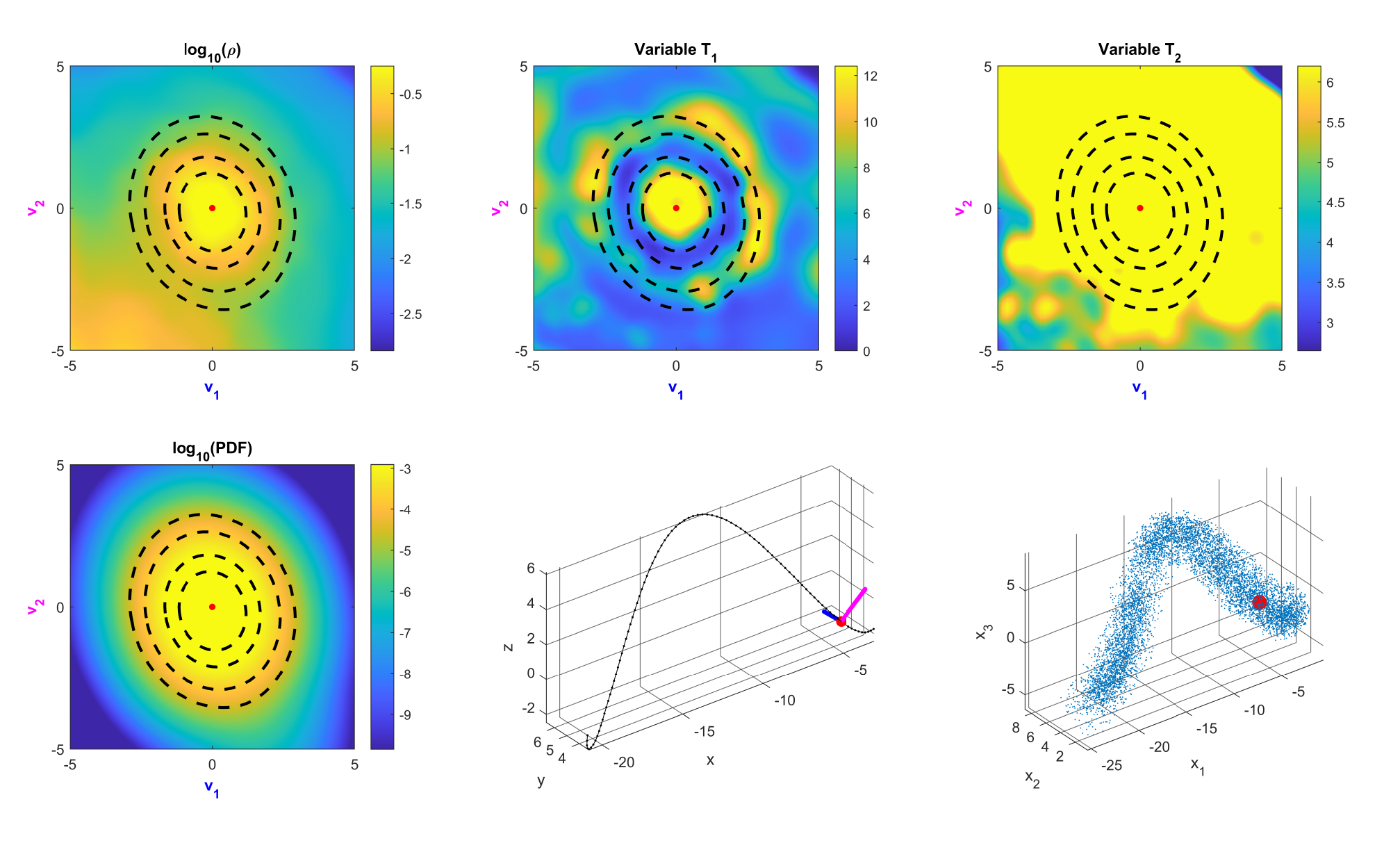}};
\end{tikzpicture}
\caption{Snapshots n.20 (a) and n.80 (b) of the video clip generated by considering co-moving coordinate frames on the elongation of manifold $\mathcal{M}_3$. 
The top row panels show the distribution of variables $\rho,T_1$ and $T_2$ over the coordinate frame $\mathcal{T}^\perp_\ell$ defined on $\vector{t}_\ell$, shown in bottom central and right panels as a red sphere. Bottom left panel presents the Probability Density Function obtained through SGTM on the same plane.}
\label{fig:MoviesM1_1}
\end{figure*}

For every $\hat{\vector{\xi}}_{\ell} \in TM^k$ we can recover a set of two vectors, $\vector{u}_1 = (u^1_1, u^2_1, u^3_1)$ and $\vector{u}_2 = (u^1_2, u^2_2, u^3_2)$, perpendicular to $\hat{\vector{\xi}}_{\ell}$, spanning the perpendicular plane $\mathcal{T}^\perp_\ell$ to manifold $\mathcal{M}_k$ on center $\tivec{t}_{\ell}$. 
This is achieved by solving the system of linear equations given by:
\begin{equation}
\begin{cases}
 \vector{u}_1 \cdot \vector{u}_2 = u^1_1 u^1_2 + u^2_1 u^2_2 + u^3_1 u^3_2 = 0\\     
 \vector{u}_1 \cdot \hat{\vector{\xi}}_{\ell} = u^1_1 \hat{\xi}^1_{\ell} + u^2_1 \hat{\xi}^2_{\ell} + u^3_1 \hat{\xi}^3_{\ell} = 0\\
 \vector{u}_2 \cdot \hat{\vector{\xi}}_{\ell} = u^1_2 \hat{\xi}^1_{\ell} + u^2_2 \hat{\xi}^2_{\ell} + u^3_2 \hat{\xi}^3_{\ell} = 0    
\end{cases}
\end{equation}
Being a degenerate system of linear equations we can recover an infinite number of solutions giving infinite pairs of vectors spanning the perpendicular plane $\mathcal{T}^\perp_\ell$. In order to maintain consistency throughout the manifold's elongation, we choose the solution to be:
\begin{align}
    \vector{u}_1 &= (\hat \xi^2_{\ell}, ~-\hat \xi^1_{\ell}, ~0),\\
    \vector{u}_2 &= (\hat \xi^1_{\ell} \hat \xi^3_{\ell}, ~\hat \xi^2_{\ell} \hat \xi^3_{\ell}, ~-[(\hat \xi^1_{\ell})^2 + (\hat \xi^2_{\ell})^2]),
\end{align}
so that $\mathcal{T}^\perp_\ell = \spn(\hat{\vector{u}}_1, \hat{\vector{u}}_2)$, where $\hat{\vector{u}}_1 = \vector{u}_1/\lVert \vector{u}_1\rVert$ and $\hat{\vector{u}}_2 = \vector{u}_2 / \lVert \vector{u}_2\rVert$. Under this scheme, the two vectors form an orthonormal coordinate frame for the plane locally perpendicular to center $\tivec{t}_{\ell}$. The two vectors are shown in figure \ref{fig:OrthoPlanes} as the magenta and blue arrows, changing direction slightly, between any pair of adjacent centers in $\mathcal{P}^k$. The tangent bundle is here also shown (sampled on points in $\mathcal{P}^k$) as green arrows.
We can now impose a regular $M \times M$ square grid of side $a$ on plane $\mathcal{T}^{\ell}_{\perp}$, taking advantage of the local coordinate frame given by unit vectors $\hat{\vector{u}}_1$ and $\hat{\vector{u}}_2$ (black gridded rotated squares in figure \ref{fig:OrthoPlanes}). We define the set $\mathcal{Y}_\ell = \{\vector{y}_{11},\dots,\vector{y}_{1M}, \vector{y}_{21},\dots,\vector{y}_{2M},\dots,\vector{y}_{M1},\dots,\vector{y}_{MM}\}$, where 
\begin{equation}
\begin{cases} \vector{y}_{ij} = 
\vector{0}_\ell + i~\delta\hat{\vector{u}}_1 + j~\delta\hat{\vector{u}}_2; \\ 
\vector{0}_\ell = \tivec{t}_\ell -2(\hat{\vector{u}}_1 + \hat{\vector{u}}_2)
\end{cases}
\end{equation}
and the increments along vectors $\hat{\vector{u}}_1$ and $\hat{\vector{u}}_2$ are given by $\delta\hat{\vector{u}}_1 = (a/M)\hat{\vector{u}}_1$ and $\delta\hat{\vector{u}}_2 = (a/M)\hat{\vector{u}}_2$ respectively. 
The number of bins $M$ must be chosen as an odd integer in order for $\tivec{t}_\ell$ to be on the origin of the local coordinate frame. In order to represent only local properties of the manifold we perform a selection of relevant particles in the data set based on their position with respect to the new reference frame. We first compute the projection $\vector{t}^{\parallel}_m$ of all particle's original positions $\vector{t}_m$ onto the tangent vector $\hat{\vector{\xi}}_\ell$ (note that $\lVert\hat{\vector{\xi}}_\ell\rVert = 1$) to manifold $\mathcal{M}_k$ at point $\tivec{t}_\ell$: $\vector{t}^{\parallel}_m = \left[(\vector{t}_m - \tivec{t}_\ell) \cdot \hat{\vector{\xi}}_\ell\right] \hat{\vector{\xi}}_\ell$. We assume that the distance $\overline{d}_\ell$ between adjacent points in $\overline{\mathcal{P}}^k_\ell$ is always proportional to the distance $d_\ell$ between corresponding adjacent points in $\mathcal{P}^k_\ell$ and constant\footnote{This might not always be the case, but present small variations due to training SGTM. However, since we interpolate the SGTM model by up-sampling $\mathcal{P}^k_\ell$ in order to increase smoothness of the orthonormal planes, the distance between adjacent centers in $\mathcal{P}^k_\ell$ (and thus its embedding $\overline{\mathcal{P}}^k_\ell$) can always be regularized by up-sampling this set more densely.}. We then select only those particles such that $\lVert \vector{t}^{\parallel}_m \rVert \leq \overline{d}_\ell/2$. 
Additionally, we compute the perpendicular component $\vector{t}_m^\perp$ of position $\vector{t}_m$, by building the projection operator onto $\mathcal{T}^\ell_\perp$ as in section \ref{subsec:Crawling}: $\bm{P} = \bm{V}\bm{V}^\dagger$, where $\bm{V}$ is the matrix having $\vector{u}_1$ and $\vector{u}_2$ as column vectors: $\vector{t}_m^\perp = \bm{P} (\vector{t}_m - \tivec{t}_\ell)$. 
In our analysis we will only consider particles lying within the sphere of radius $b$, centered on $\tivec{t}_\ell$: $\lVert\vector{t}_m^\perp\rVert <= b$, where $b$ can be imposed by the user or automatically selected as $b = a\sqrt{2}/2$: the half-diagonal of the gridded plane $\mathcal{T}_\perp^\ell$. However, this parameter is only defined when a non-SPH weighting (e.g. Gaussian) scheme is applied. In fact, when SPH is in place, there is no need to select a subset of particles surrounding the plane in order to compute the mean value of properties on the plane. This is achieved via the SPH weighting scheme presented in eq. \ref{eq:WeightedMean} through the smoothing length parameter, defined for all particles in the data set.
Identifying with $\mathcal{J}$ the index set of all particles satisfying these two conditions, we can now compute the weighted mean of
 variables $T_1(\mathcal{J},\vector{y}_{ij})$ and $T_2(\mathcal{J},\vector{y}_{ij})$ with respect to points $\vector{y}_{ij} \in \mathcal{Y}_\ell$, 
under the SPH formulation (see section \ref{subsec:WeightScheme}). Top row of figures \ref{fig:MoviesM1_1}a-\ref{fig:MoviesM1_1}b
present the behaviours of variables $T_1, T_2$ and density $\rho$ respectively, for manifold $\mathcal{M}_2$ detected in the synthetic data set, computed at position $\tivec{t}_\ell^k$ ($\tivec{t}_\ell^k$ varying along the manifolds for each group of pictures). 
Having the probabilistic model for manifold $\c{M}_k$ as SGTMs, we also compute the Probability Density Function of the Mixture on plane $\mathcal{T}^\perp_\ell$ as:
\begin{equation}\label{eq:PDF_SGTM}
    p^k(\vector{y}_{ij}) = \sum_{x_\ell \in \mathcal{P}^k} \pi_\ell p(\vector{y}_{ij}|x_\ell,\hat{\bm{\Sigma}}_\ell,\hat{\bm{W}}),
\end{equation}
where $\hat{\bm{\Sigma}}_\ell$ and $\hat{\bm{W}}$ are the parameters found through optimization with the EM algorithm (see section \ref{subsec:SGTM}) and $\pi_\ell$ the component proportion given by $\pi_\ell = 1/|\overline{\mathcal{P}}^k|$. 

The PDFs for manifolds $\mathcal{M}_k$ on plane $\mathcal{T}^\perp_\ell$ at position $\vector{t}_\ell^k$ are shown in the bottom left panel of figures \ref{fig:MoviesM1_1}a-\ref{fig:JF_Movies_M1_2}b,
while central and right panels show the current position on the manifold and with respect to the whole data set respectively.
By applying recursively this procedure to every $\tivec{t}_\ell \in \overline{\mathcal{P}}^k$ we obtain a representation of the variables of interest, on co-moving orthonormal reference frames along each manifold, showing the clear radial behaviour of these variables. 

Figures \ref{fig:MoviesM1_1}a-\ref{fig:MoviesM1_1}b
present the results for selected points on manifold $\mathcal{M}_3$. 
The top central panel in each figure clearly manifests the sinusoidal radial behaviour of variable $T_1$. This implies that the center of the manifold has been correctly identified by SGTM and its centers lie closely to its underlying nature (see equation \ref{eq:Embeddings}). Comparing top right panel in figure \ref{fig:MoviesM1_1}a 
and the one in figure \ref{fig:MoviesM1_1}b, 
the longitudinal dimming of variable $T_2$ can be verified, showing that its true nature has been correctly recovered. The snapshots obtained via the co-moving orthonormal coordinate frames technique are joined sequentially into a movie that shows the evolution of the quantities while moving along the manifold. The presented figures only show selected snapshots of the movie for manifold $\mathcal{M}_3$, however the analysis has been performed for both manifolds in the synthetic data sets and the associated movies can be found at: \url{https://git.lwp.rug.nl/cs.projects/1DREAM}. The snapshots presented in this work are individual frames of the movie clip and they are referred to as ``snaphsot n. \dots'' when addressed in the captions.
\begin{table}[h]
\centering
\def\myCWidth{0.12}
\caption{Full list of parameters for Co-moving Orthonormal coordinate frames.}
\label{tab:Param_Orth}
\begin{tabularx}{\columnwidth}{@{\extracolsep{\fill}}
ll>{\raggedleft}p{\myCWidth\columnwidth}>{\raggedleft\arraybackslash}p{\myCWidth\columnwidth}}
\toprule
    $N_\ell \in \NN~(N_\ell = 5)$      &  Latent interval upsample size \\
    $M \in \NN ~(M = 25)$          &  Pixels on plane\\
    $a~^\ast \in \RR ~(a > 0)$  &  Length of plane \\
    $b \in \RR ~(b = a\sqrt{2}/2)$  &  Distance from plane\\
\bottomrule
\end{tabularx}
\end{table}

\subsection{Weighting schemes}\label{subsec:WeightScheme}

Depending on the data at hand, it is possible to implement multiple different weighting schemes. Here, we focus on two main approaches: an SPH formulation and a classical Gaussian smoothing, with variable length-scale. 
If the data set is obtained via an SPH simulation, the weighted mean of any simulated quantity has to be computed following the formulation of the code used. In the following, we will consider the weighting scheme implemented in GADGET2 and a general Gaussian smoothing technique. We provide a detailed description of the two formulations for a given center of SGTM and the associated point-set on whose points the mean is intended to be computed (either the uniformly distributed cylindrical shells or the sampled perpendicular plane centered on $\tivec{t}_\ell$). We will identify the query point-set by $\mathcal{C}' = \{\vector{p}_1',\vector{p}_2',\dots,\vector{p}_N'\}$.

\subsubsection{SPH-like weights}

Consider a point $\vector{p}_i' \in \mathcal{C}'$. We need to compute the weighted mean, under the SPH formalism, of variable $T_m$ summing through all the particles $\vector{t}_j \in \mathcal{Q}$. The spline approximation of the Gaussian kernel (usually referred to as \emph{smoothing kernel}) on a finite support is:
\begin{equation}\label{eq:KernelSPH}
    W(q_j,h_j) = \frac{1}{\pi h_j^3} 
    \begin{cases}
    \frac{1}{4} (2 - q_j)^3 - (1 - q_j)^3 & 0 \leq q_j < 1 \\
    \frac{1}{4} (1 - q_j)^3  & 1 \leq  q_j < 2 \\
    0   & q_j \geq 2
    \end{cases}
\end{equation}
where $q_j = \lVert \vector{t}_j - \vector{p}_i' \rVert/h_j$ ($h_j$ being the smoothing length defined in sec. \ref{sec:Quantities}). 
Using the kernels, the exact weighted mean of quantity $T_m$ at point $\vector{p}'$ is:
\begin{equation}\label{eq:WeightedMean}
    \overline{T}_{m,i} \coloneqq \langle T_m(\vector{p}_i') \rangle = \frac{\sum\limits_{\vector{t}_j \in \mathcal{Q}} \frac{m_j}{\rho_j} T_m(\vector{t}_j) W(q_j,h_j)}{\sum\limits_{n =1}^{|\mathcal{Q}|} \frac{m_n}{\rho_n}W(q_n,h_n)}
\end{equation}
\begin{figure*}[t!]
\begin{tikzpicture}[node distance = 0cm,nodes = {anchor=north west,inner sep=0cm},font=\footnotesize]
\node[] (a_cap)                   {\textbf{(a)} Manifold $\mathcal{M}_3$, weighted mean without normalization.};
\node[] (a) at (a_cap.south west) {\includegraphics[width=\textwidth,trim = 0.6cm 1.2cm 0.5cm 1cm,clip]{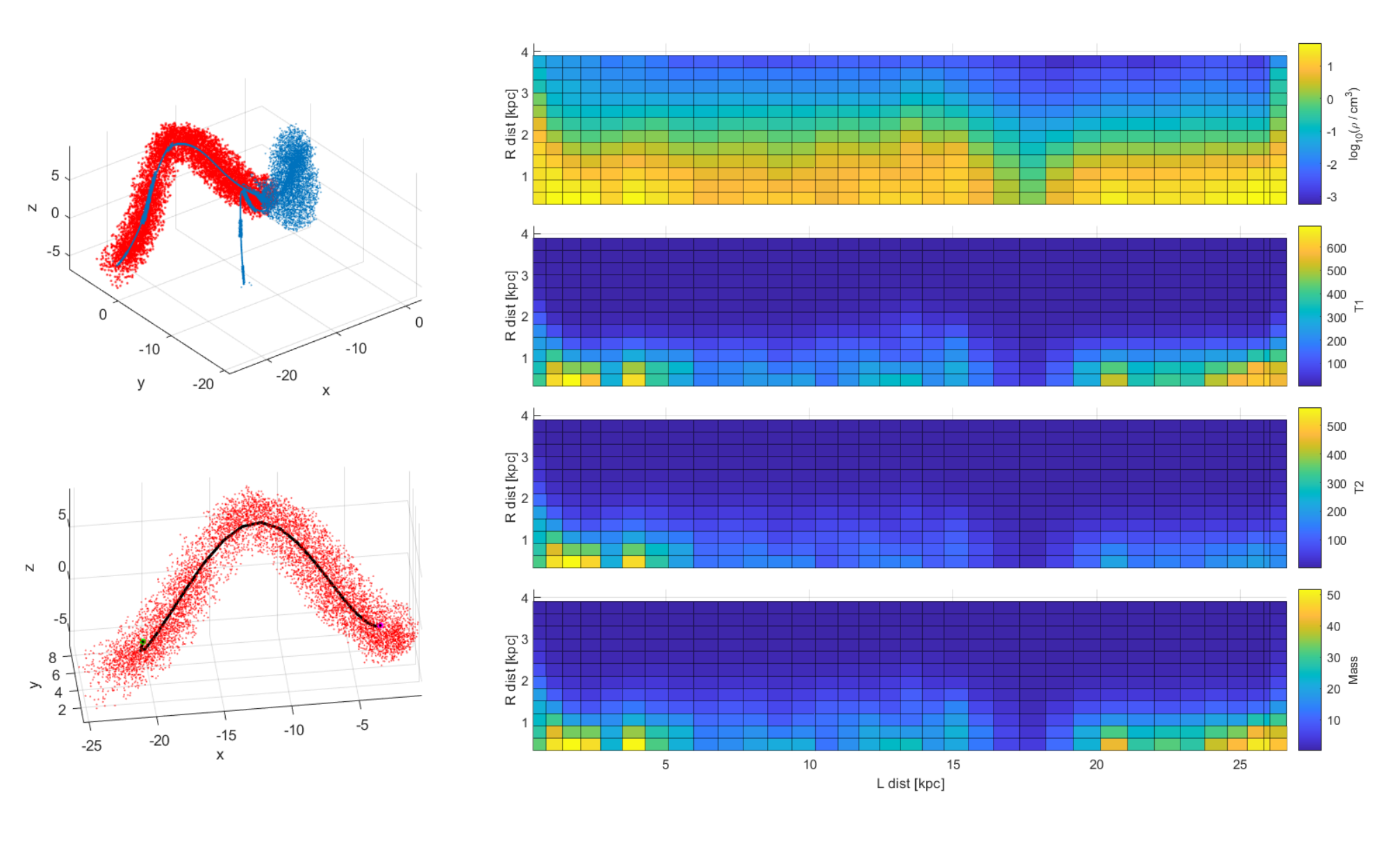}};
\node[] (b_cap) at (a.south west) {\textbf{(b)} Manifold $\mathcal{M}_2$, weighted mean without normalization.};
\node[] (b) at (b_cap.south west) {\includegraphics[width=\textwidth,trim = 0.6cm 1.2cm 0.5cm 1cm,clip]{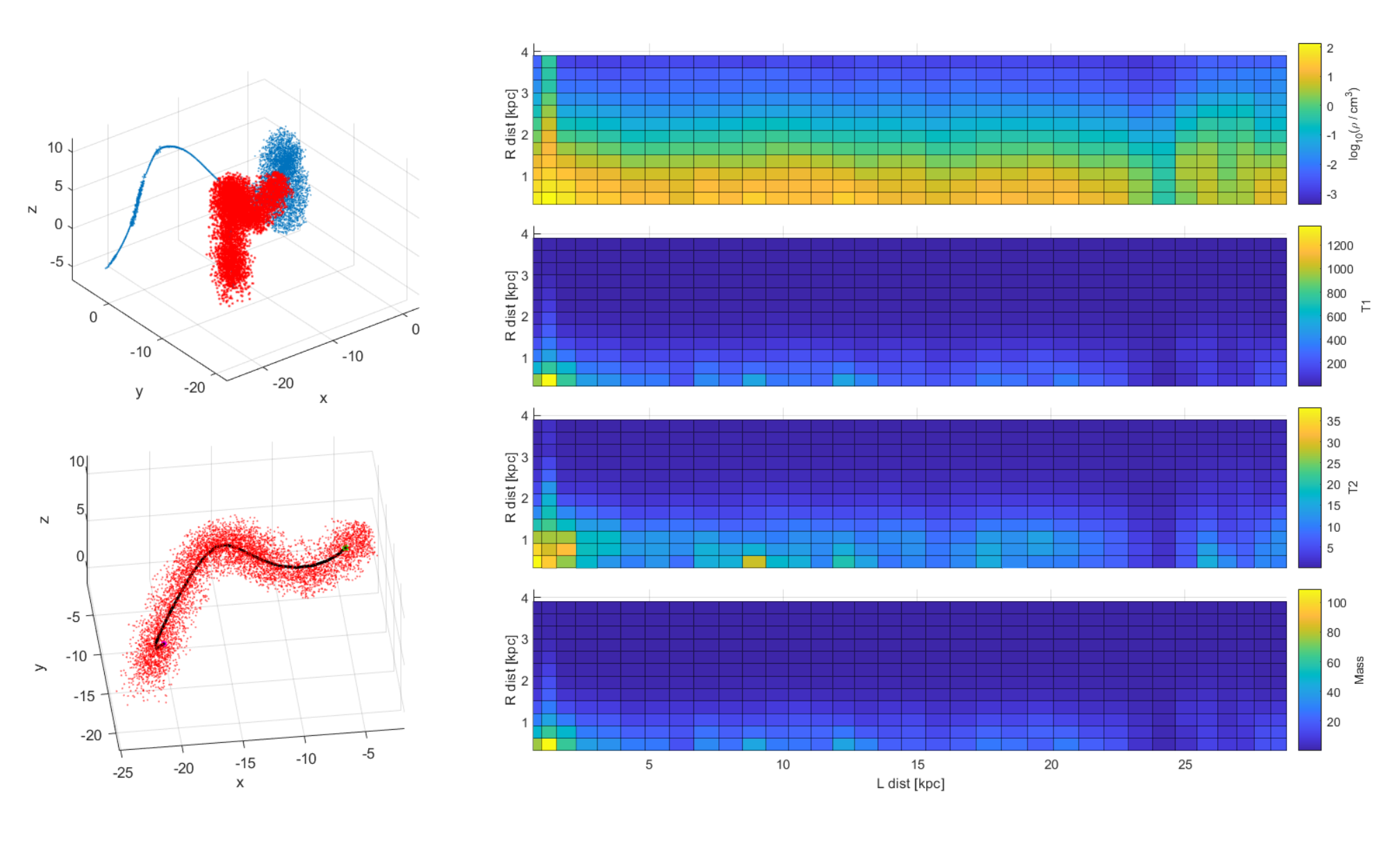}};
\end{tikzpicture}
\caption{Same as panels figures \ref{fig:MeanPlots_M1}a and \ref{fig:MeanPlots_M2}a but without normalization in the SPH weighting scheme.} 
\label{fig:MeanPlotsNoNorm}
\end{figure*}
The term in the denominator is generally considered to be approximating unity when the particles in a data set are distributed uniformly; however, this is not often the case in practice. 
Each particle $\vector{t}_j$ of an SPH data set samples a spherical volume of radius $h_j$. Since all particles are evolved following the equations of motion defined by the Lagrangian formulation of fluids, their distribution at a given evolutionary stage is far from uniform. Thus, the \emph{approximation to unity} assumption is generally incorrect at an advanced evolutionary stage. The role of the normalization term $\sum_{n =1}^{|\mathcal{Q}|} \frac{m_n}{\rho_n}W(q_n,h_n)$ in the denominator of equation \eqref{eq:WeightedMean} is to eliminate the interpolation's dependence from the particles' distribution. As such, it can not be disregarded when computing the weighted mean of quantity $T_m$ on any point in $\mathcal{C}'$.\\
In order to check the influence of the normalization term, we show in figure \ref{fig:MeanPlotsNoNorm} panel a and b 
the corresponding behaviours of $\overline{\rho}_i, \overline{T}_{1,i}, \overline{T}_{2,i}$ and $\overline{m}_i$ when the normalization term is omitted in equation \eqref{eq:WeightedMean}. While the ranges of the first three variables exceed their true respective values, it is striking the difference of the mass distribution with respect to the normalized versions (figures \ref{fig:MeanPlots_M1}a and \ref{fig:MeanPlots_M2}a). 
The radial sinusoidal profile of variable $T_1$ on manifold $\mathcal{M}_3$ is also completely lost (right side, 
second panel from the top), as the sparser distribution in the further regions from the core of the manifold does not carry enough weight to compensate for the inner regions.

\subsubsection{Gaussian smoothing}

If the data set has not been generated via an SPH simulation, the weighted mean of any variable on point-set $\mathcal{C}'$ is obtained by imposing a Gaussian isotropic kernel on each point $\vector{p}_i' \in \mathcal{C}'$, with scale length $\kappa$, obtaining for each point $\vector{t}_j \in \mathcal{Q}$, weight:
\begin{equation}\label{eq:GaussWeight}
p(\vector{t}_j|\vector{p}_i',\kappa) = \frac{1}{2\pi \kappa^2} \exp\left(-\frac{\left\lVert\vector{t}_j - \vector{p}_i' \right\rVert^2}{2\kappa^2} \right).
\end{equation}
The weighted mean of variable $T_m$ on every point $\vector{p}_i' \in \mathcal{C}'$ is then obtained by summing through all particles in data set $\mathcal{Q}$ as:
\begin{equation}\label{eq:GaussMean}
\overline{V}_{m,1} \coloneqq \langle T_m(\vector{p}_j')\rangle = \frac{\sum\limits_{j =1}^{|\mathcal{Q}|} p(\vector{t}_j|\vector{p}_i',\kappa)T_m(\vector{t}_j)}{\sum\limits_{n =1}^{|\mathcal{Q}|} p(\vector{t}_n|\vector{p}_i',\kappa)}.\\
\end{equation}

\section{Experiments on different data sets} \label{sec:experiments}

This section is devoted to the application of the proposed methodology to three different data sets. In order to avoid confusion, each subsection indicates the data set by the same notation used in the previous sections ($\mathcal{Q}$ and $\tilde{\mathcal{Q}}$ for the noisy and diffused data sets respectively). The first data set (section \ref{subsec:JellyFish}) is a simulated dwarf galaxy interacting with its host galaxy cluster. In particular, we examine a single simulated snapshot of the dwarf's evolution. The choice of the snapshot is motivated by the presence of multiple gaseous filamentary structures, located mainly at the back of the simulated box and forming a gaseous tail. In this case we are only considering the gas particles' distribution, disregarding Dark Matter and Stellar particles. Here, we analyse the gas temperature ($T$), gas density ($\rho$), neutral gas fraction (which is the ratio of neutral, or atomic, gas mass to total gas mass) and metallicity (the iron abundance [Fe/H]) of the extracted streams. We note that for [Fe/H], which is defined as a logarithmic ratio of concentrations, we first recover the linear ratio then we use eq.~\eqref{eq:WeightedMean}, and finally we go back to logarithmic scale. 
The second data set (section \ref{subsec:CosmicWeb}) is obtained via a Dark Matter simulation of a sample volume of the Universe's Large Scale Structure (LSS). We focus here on the kinematic properties of the filaments of dark matter extracted from the Cosmic Web. 
The third and last data set is the observed stellar spatial distribution of a sky-region enclosing the globular cluster $\omega-$Centauri. 
We aim at recovering the two stellar streams detected by \cite{2019NatAs...3..667I} and describing their radial and longitudinal density profiles. 

\subsection{Simulated jellyfish: dwarf galaxy in Fornax cluster}\label{subsec:JellyFish}

\textbf{Methodologies:} LAAT $\rightarrow$ EM3A $\rightarrow$ Dimensionality Index $\rightarrow$ Crawling $\rightarrow$ SGTM $\rightarrow$ Bi-dimensional profiles $\rightarrow$ Co-moving Orthonormal coordinate frames.\\

The simulations initially consider a dwarf galaxy evolving in isolation for 8 billion years. 
Here, isolation means that the galaxy was assembling its mass through mergers but was not absorbed by a more massive structure, such as a galaxy cluster, where its internal properties could be affected by external processes such as gravitational interactions with other galaxies and ram-pressure stripping. A full catalogue and detailed study of these galaxies can be found in \cite{2015ApJ...815...85V,2017A&A...607A..13V}.

Taking the end product of this initial evolutionary stage, \cite{Mastropietro2021} study the evolution of these galaxies when injected on different orbits in the gaseous halo of a Fornax-like galaxy cluster. During this stage, filamentary structures form in different orbital epochs.
In our analysis we will consider a single temporal snapshot of a dwarf galaxy evolving on a generic orbit.\\

\subsubsection{Extracted manifolds}\label{subsec:ExManifolds}
We recover $15$ streams of gas with varying lengths. In the following, for the sake of clarity, we discuss only the most elongated manifold recovered through the methodology. The manifolds are visualized via the two methodologies described in sec. \ref{subsec:BidimProfiles} and \ref{subsec:OrthoPlanes}, respectively. The values of the free parameters adopted for this analysis are shown in tab. \ref{tab:JF_Params}
\begin{table}[h]
\centering
\def\myCWidth{0.12}
\caption{Adopted values for Experiments on Jellyfish galaxy}
\label{tab:JF_Params}
\begin{tabularx}{\columnwidth}{X | X | X}
\toprule
    LAAT     &  $r = 1$ & $F^j_{Th} = 5$ \\
    EM3A     &  $R_{min} = 0.5$ & $R_{max} = 1.5$ \\
    Crawling & $r = 1$  & \\
    SGTM     & $r = 1$ & $S = L/2$ \\
    Bi-dim profile & $r_M = 3$ & $c = 19$\\
    Moving frames & $a = 2$ & \\
\bottomrule
\end{tabularx}
\end{table}

\subsubsection*{Bi-dimensional profiles on jellyfish}
\begin{figure*}[th]
\centering
\includegraphics[width = \textwidth,trim = 0.4cm 0.5cm 0.4cm 0.7cm,clip]{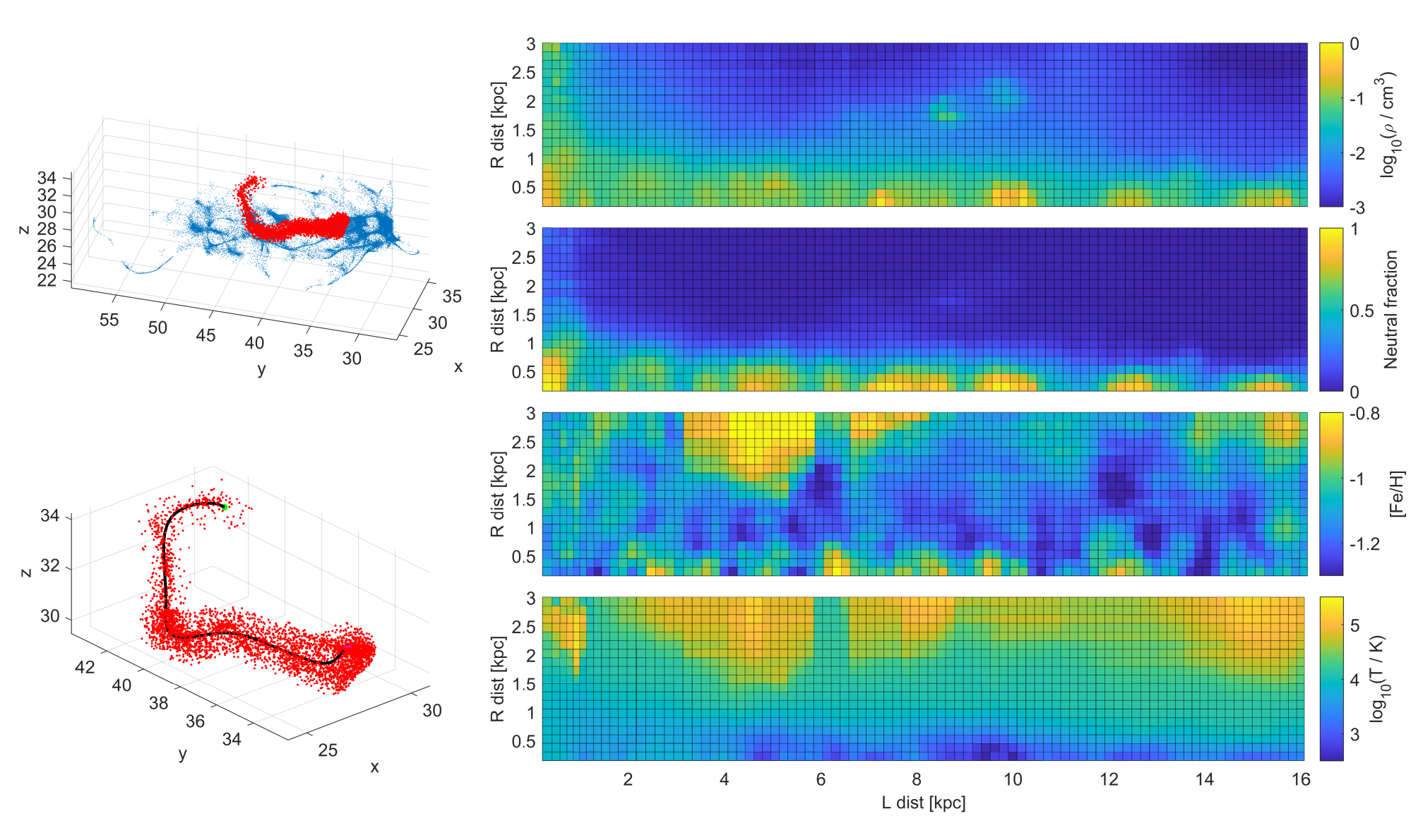}
\caption{Bi-dimensional profiles of one elongated manifold recovered via the methodology discussed in this work.}
\label{fig:BiDimJF}
\end{figure*}
Using the visualization tool described in sec. \ref{subsec:BidimProfiles} we present, in figure \ref{fig:BiDimJF}, one of the manifolds recovered via our methodology, departing from the head of the jellyfish and extending throughout its tail.
Overall, while the elongation of individual manifolds varies, the inspected properties behave similarly for all detected structures. In particular, the neutral fraction (second bi-plot from the top), along with density (top bi-plot), is generally higher in the inner regions of the manifolds, across roughly their whole elongation. At the same time, temperature (bottom bi-plot, in logarithmic scale) is consistently lower in the same regions. A similar behaviour can not be found for the metallicity (third bi-plot from the top). It can be argued that this quantity is generally higher in the inner regions as well, however the non-uniformity of its distribution discourages an accurate inference on the causes of this anomaly. It is possible that, as the manifolds are from different regions of the galaxy, their interaction with the gas of the galaxy cluster's halo in previous epochs heavily influenced their evolution. 
Nonetheless, since the gas's metallicity does not directly affect the chances of its collapse, we can safely argue that the streams in jellyfish galaxies are effective loci of star formation. 
Furthermore, the core of these streams are more likely to contain newly born stars, and thus be observed via optical observations.

\subsubsection*{Co-moving orthonormal coordinate frames on jellyfish}

\begin{figure*}[!t]
\begin{tikzpicture}[node distance = 0cm,nodes = {anchor=north west,inner sep=0cm}]
\node[] (a_cap)                   {\textbf{(a)} Jellyfish manifold (Manifold $\mathcal{M}_4$ out of $15$), snapshot n.34 of associated movie clip.};
\node[] (a) at (a_cap.south west) {\includegraphics[width=\textwidth,trim = 0.4cm 1cm 0.2cm 1cm,clip]{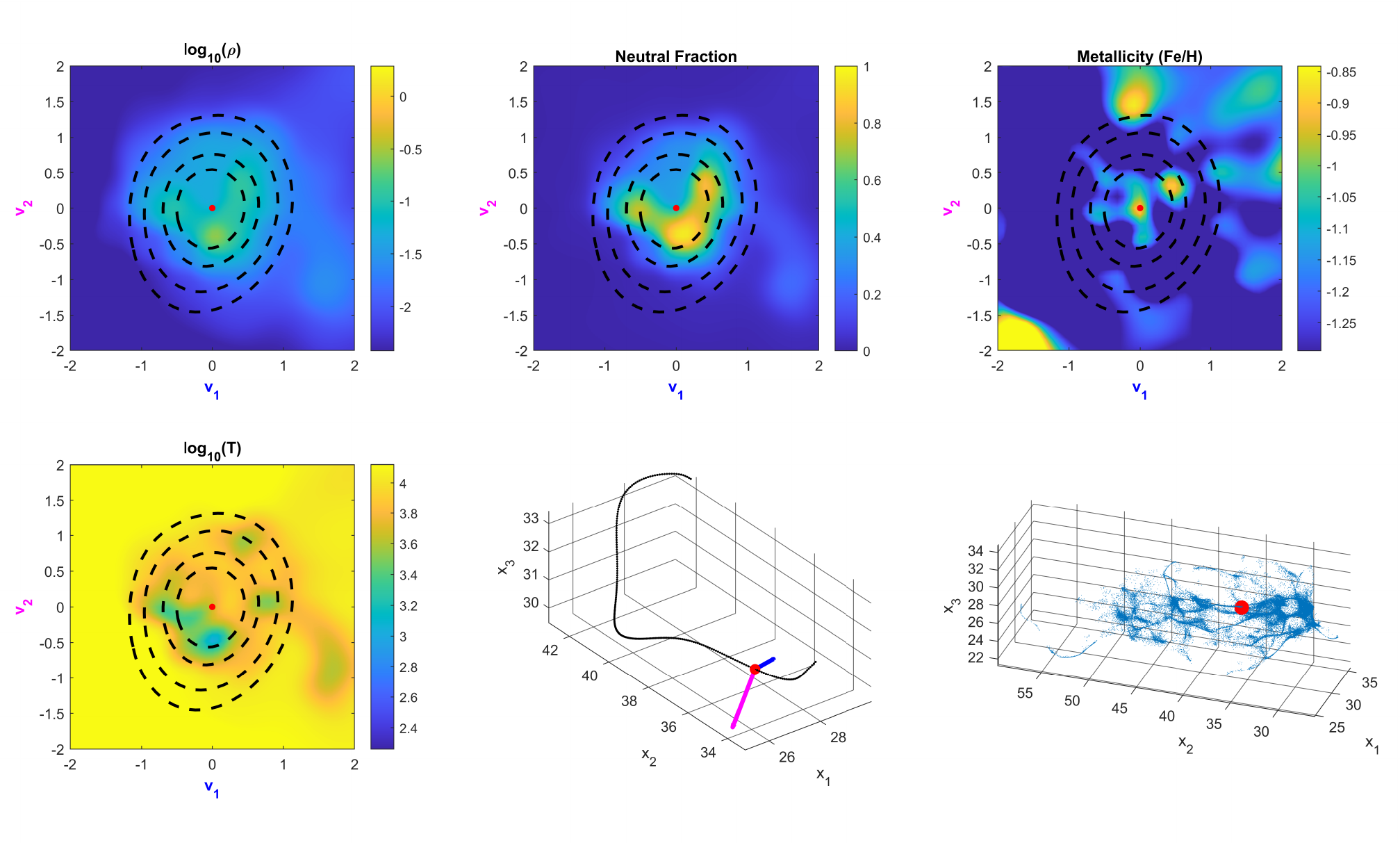}};
\node[] (b_cap) at (a.south west) {\textbf{(b)} Jellyfish manifold (Manifold $\mathcal{M}_4$ out of $15$), snapshot n.90 of associated movie clip.};
\node[] (b) at (b_cap.south west) {\includegraphics[width=\textwidth,trim = 0.4cm 1cm 0.2cm 1cm,clip]{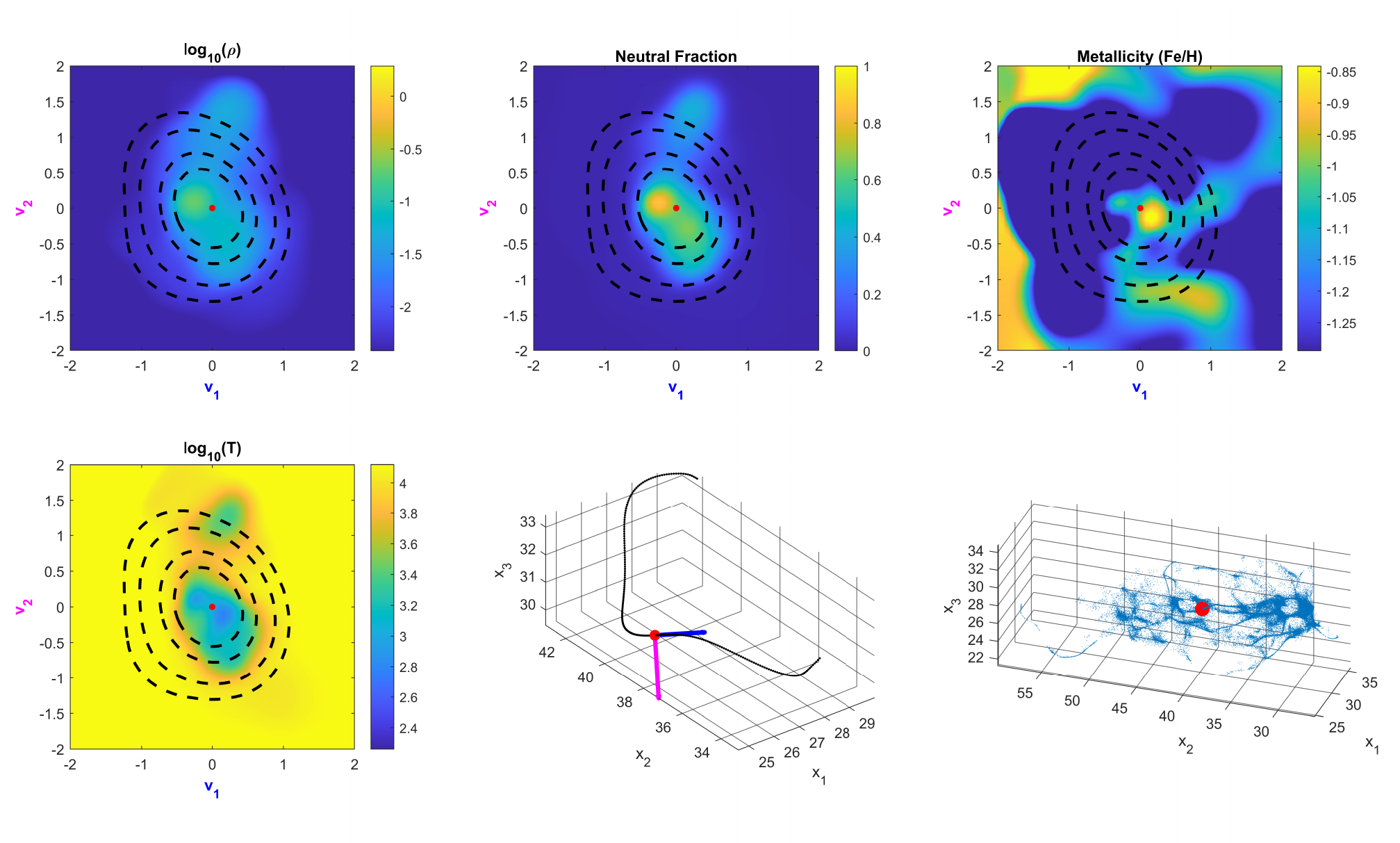}};
\end{tikzpicture}
\caption{Equally spaced snapshots of the extracted manifold from jellyfish data set. Each panel in each sub-figure shows the distribution of a variable across the current orthonormal plane. The two bottom panels show the position of the current center (red sphere) on the detected manifold (black curve) and on the global diffused data set (blue dots, right panel).}
\label{fig:JF_Movies_M1_1}
\end{figure*}

\begin{figure*}[th!]
\begin{tikzpicture}[node distance = 0cm,nodes = {anchor=north west,inner sep=0cm}]
\node[] (a_cap)                   {\textbf{(a)} Jellyfish manifold (Manifold $\mathcal{M}_4$ out of $15$), snapshot n.136 of associated movie clip.};
\node[] (a) at (a_cap.south west) {\includegraphics[width=\textwidth,trim = 0.4cm 1cm 0.2cm 1cm,clip]{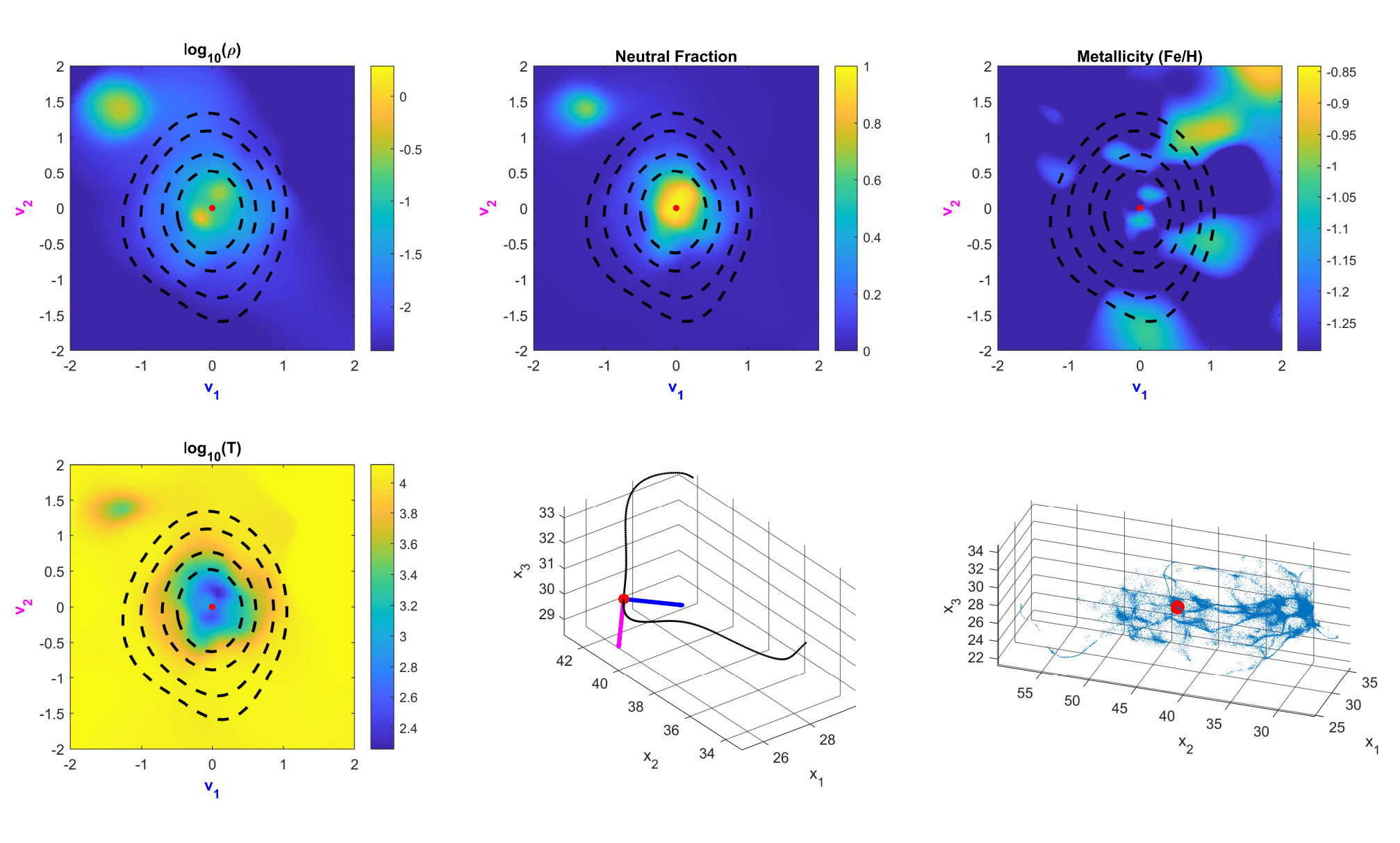}};
\node[] (b_cap) at (a.south west) {\textbf{(b)} Jellyfish manifold (Manifold $\mathcal{M}_4$ out of $15$), snapshot n.199 of associated movie clip.};
\node[] (b) at (b_cap.south west) {\includegraphics[width=\textwidth,trim = 0.4cm 1cm 0.2cm 1cm,clip]{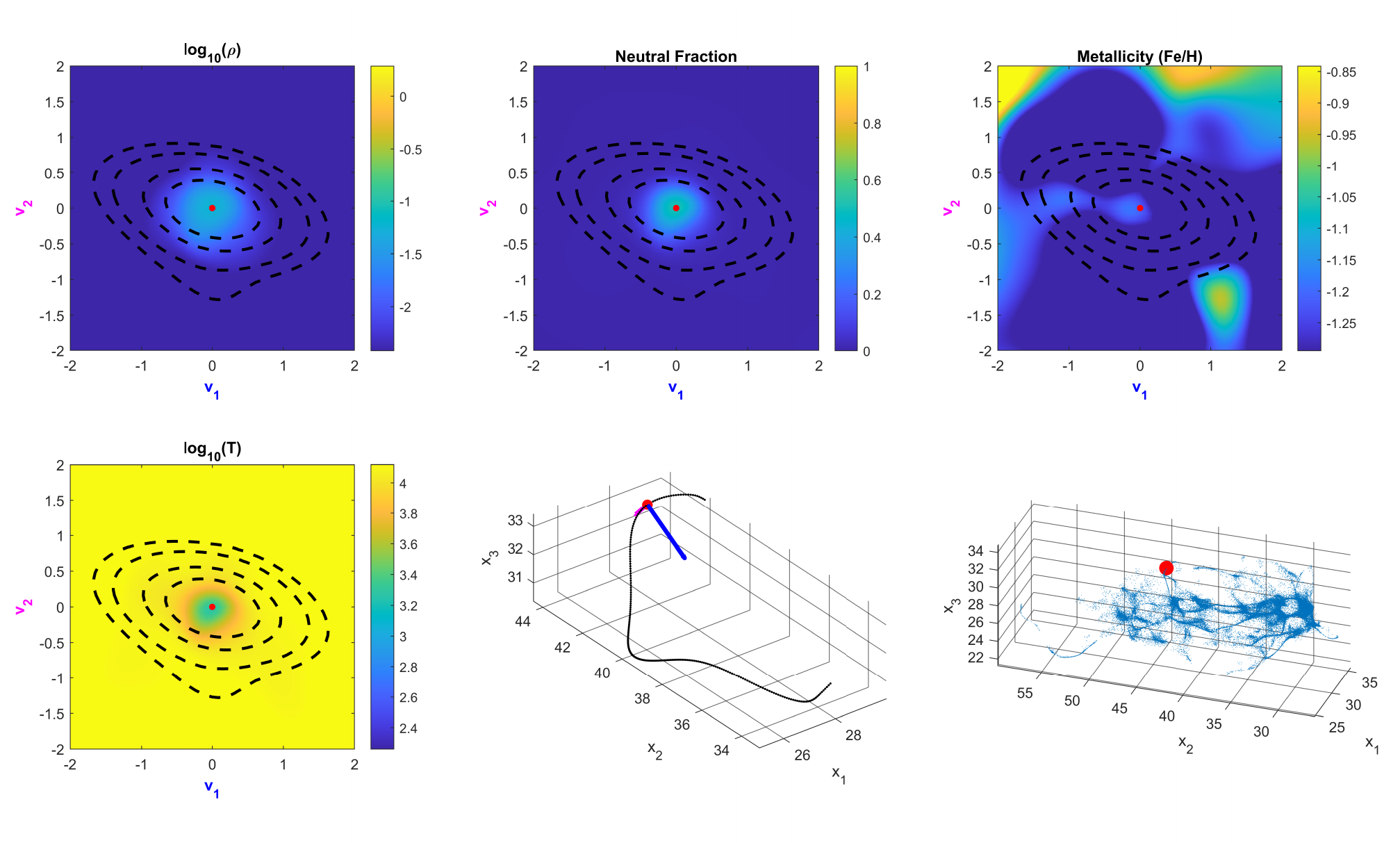}};
\end{tikzpicture}
\caption{Same as in figure \ref{fig:JF_Movies_M1_1} for additional centers on the manifold.}
\label{fig:JF_Movies_M1_2}
\end{figure*}

The same quantities ($\rho$, neutral gas fraction, [Fe/H], and $T$) have been studied via the Co-moving orthonormal coordinate frames technique and here presented for the same manifold previously identified. The results are shown in figures \ref{fig:JF_Movies_M1_1} and \ref{fig:JF_Movies_M1_2} for four subsequent centers on the examined manifold. The three panels on top (left to right) and the one on the bottom left, show the distribution of these quantities on the perpendicular planes to the tangent bundle of the manifold.
Over-plotted to the quantities distribution, in every panel we show the PDF of the probabilistic model obtained via SGTM, as (red) iso-curves. The red dot represents the center of the plane, corresponding to the embedded skeleton of the manifold. The bottom central and right panel show the location of the current plane with respect to the manifold and the whole data set respectively. For consistency, we will show in this work snapshots from the same manifold shown in figure \ref{fig:BiDimJF}.\\
Throughout the manifold's elongation we verify, via the snapshots presented in figures \ref{fig:JF_Movies_M1_1} and \ref{fig:JF_Movies_M1_2}, the centering of the recovered skeleton with respect to the regions having the highest density. This region is also always associated to a higher neutral fraction and lower temperature. In the case of the studied manifold in agreement with the bi-dimensional profile (see figure \ref{fig:BiDimJF}), little can be said about the behaviour of the metallicity, although there is a tendency of creating cores with local peaks surrounded by lower metallicity regions.

\subsection{Kinematic study on cosmic web's filaments}\label{subsec:CosmicWeb}

On the many mega-parsec cosmological scales of the Universe, the spatial distribution of galaxies as well as clusters of galaxies is not uniform. In fact, looking at the results of the Sloan Digital Sky Survey (SDSS) \citep{Fukugita1998, GunnEtal1998}, one can see that there is an intricate, interconnected pattern that emerges at such a scale. This pattern forms a network now famously known as the cosmic web \citep{Bond1996}. As described in \citet{Peebles1980}, the cosmic web emerges as the outcome of the anisotropic nature of gravitational collapse. The latter is the driving force behind structure formation including the emergence of the cosmic web's different morphological components namely: clusters, filaments, and walls. The connection between these components can be summarized as follows: clusters are regions of intersection of filaments, and filaments are regions of intersection of walls \citep{DoroshkevichEtal1980, ShapiroEtal1983, PaulsMelott1995, SathyaprakashEtal1996, CautunEtal2013}.

The growing interest in studying the cosmic web lies not only in its involvement in the cosmology domain, but also in its important influences on the evolution and properties of galaxies. For instance, in works such as \citet{AragonCalvoEtal2007},  \citet{HahnEtal2007b, HahnEtal2007a}, and \citet{PazEtal2008} it has been shown that the spin-orientation and shape of dark matter halos are distinctly influenced by the cosmic web environment they occupy (whether it is filamentary or sheetlike in nature). It has also been demonstrated that galaxies tend to have an alignment with the filaments that they inhabit \citep{JonesEtal2010, TempelEtal2013, GaneshaiahVeenaEtal2018, WelkerEtal2020}. Moreover, the influence of the cosmic web environment extends to other properties of galaxies such as their colours, gas content, and star formation rates (SFRs). Different studies have pointed out general trends of these properties whereby galaxies closer to the cosmic web structures exhibit a lower specific SFR (are redder in colour) and tend to be older, more metal rich and $\alpha$-enhanced when compared to galaxies that have a larger distance to the structures \citep{RojasEtal2004, BeyguEtal2016, ChenEtal2017, KarljikEtal2018, WinkelEtal2021}.    

Given what has been presented, we illustrate the applicability of our methods by applying our pipeline on data sets pertaining to the cosmic web. The nature of the data is further described in the following section. We also reserve the detailing of the robustness of the extracted and modelled morphologies to a second paper, where we narrow  our focus to the cosmic web and the ability of our tool to trace out its different structures.       

\subsubsection{Generation of the cosmic web dataset}

We use a dark matter-only $N$-body cosmological simulation that was run using the GADGET-3 code. The initial conditions were generated at redshift $z=200$ using the Multi Scale Initial Condition software (MUSIC; \citet{HahnAbel2011}). The CAMB package (Code Anisotropies in the Microwave Background; \citet{LewisEtal2000}) is used to calculate the linear power spectrum. We study a single cosmological volume with dimensions $120\times120\times120$~Mpc/h. The dark matter (DM) particles have a fixed mass of $1.072\times10^9 M_{\odot}/h$, and a cosmology of $\Omega_m = 0.3$, $\Omega_{\Lambda} = 0.7$, $\Omega_b = 0.047$ and $h_0 = 0.684$ was assumed for the initial conditions and for the simulation itself. In this project, we consider the redshift zero output file, which consists of the masses, velocities and positions of all the dark matter particles in the present.

\subsubsection{Extracted manifolds}\label{subsubsec:CW_Manifolds}
\textbf{Methodologies:} LAAT $\rightarrow$ EM3A $\rightarrow$ Dimensionality Index $\rightarrow$ Crawling $\rightarrow$ SGTM $\rightarrow$ Co-moving Orthonormal coordinate frames.\\

\begin{table}[h]
\centering
\def\myCWidth{0.12}
\caption{Adopted values for Experiments on Cosmic Web}
\label{tab:CW_Params}
\begin{tabularx}{\columnwidth}{X | X | X}
\toprule
    LAAT     &  $r = 1.5$ & $F^j_{Th} = 5$ \\
    EM3A     &  $R_{min} = 1$ & $R_{max} = 2$ \\
    Crawling & $r = 1.5$  & \\
    SGTM     & $r = 1.5$ & $S = L/2$ \\
    Moving frames & $a = 1$ & \\
\bottomrule
\end{tabularx}
\end{table}
Adopting the methodology presented previously, we isolate two manifolds in the simulated volume, connected by a node of the Cosmic Web. 
By building orthonormal co-moving frames over the skeletons of these manifolds, we study their DM distribution and kinematic properties. The values of the free parameters adopted for this analysis are shown in tab. \ref{tab:CW_Params}.
For each particle in the simulation, at each location over the manifold's skeleton, we compute its tangential and orthogonal velocity with respect to the local reference frame. 
We look at the DM particles distribution as a discrete sample of the actual dark matter, where each particle is representative of the kinematic state of a fixed size neighborhood.

Given a specific point $\tivec{t}_\ell \in \overline{\mathcal{P}_{\uparrow}}^k$ belonging to manifold $\mathcal{M}_k$'s up-sampled skeleton, as a product of our methodology we recover the corresponding tangent vector $\hat{\vector{\xi}}_\ell$ and perpendicular plane $\mathcal{T}_\ell^{\perp} = \spn\{\hat{\vector{u}}_1,\hat{\vector{u}}_2\}$. As discussed in section \ref{subsec:OrthoPlanes}, we obtain the index set $\mathcal{J}$ of all particles falling within the box of size $\overline{d}_\ell \times 2b$ centered on $\tivec{t}_\ell$. Each $\vector{t}_m$ such that $m \in \mathcal{J}$, has a velocity $\vector{\nu}_m \in \RR^3$ describing its motion. The velocity vector is projected onto $\mathcal{T}_\ell^{\perp}$ and along the tangent vector $\hat{\vector{\xi}}_\ell$ to manifold $\mathcal{M}_k$ on $\tivec{t}_\ell$:
\begin{equation}
    \vector{\nu}_m^{\perp} = \bm{P} \vector{\nu}_m; \qquad \vector{\nu}_m^{\parallel} = (\vector{\nu}_m \cdot \hat{\vector{\xi}}_\ell) \hat{\vector{\xi}}_\ell,
\end{equation}
for every $m \in \mathcal{J}$. We can now compute the weighted mean of the two projected velocities over points on $\mathcal{T}_\ell^\perp$:
\begin{align}
    \overline{\vector{\nu}}(\vector{y}_{ij})^\perp &= \frac{\sum_{m \in \mathcal{J}}p(\vector{t}_m^\perp|\vector{y}_{ij},\delta) \vector{\nu}_m^\perp}{\sum_{q \in \mathcal{J}}p(\vector{t}_q^\perp|,\vector{y}_{ij},\delta)}\\
    \overline{\vector{\nu}}(\vector{y}_{ij})^\parallel &= \frac{\sum_{m \in \mathcal{J}}p(\vector{t}_m^\perp|\vector{y}_{ij},\delta) \vector{\nu}_m^\parallel}{\sum_{q \in \mathcal{J}}p(\vector{t}_q^\perp|\vector{y}_{ij},\delta)},
\end{align}
where $p(\vector{t}_m^\perp|\vector{y}_{ij},\delta)$ is the Gaussian kernel defined in equation \eqref{eq:GaussWeight}.
Note that parameter $\delta$ can be fixed by the user to match a desired smoothing, however in our experiments we fix it to $\delta = \frac{\sqrt{a/M}}{4}$: the half-diagonal of the square formed by four adjacent points sampling $\mathcal{T}^\perp_\ell$ in $\mathcal{Y}_\ell$. Having obtained a velocity field on $\mathcal{T}^\perp_\ell$, we are now able to compute its rotor and divergence with respect to the local coordinate frame. We define the $\nabla$ operator on the local reference frame given by $\spn\{\hat{\vector{u}}_1,\hat{\vector{u}}_2,\hat{\vector{\xi}}_\ell\}$ as $\nabla = (\partial/\partial u_1, \partial/\partial u_2, \partial/\partial \xi_\ell)$, so that the rotor and divergence of vector field $\overline{\vector{\nu}}$ can be written as
\begin{align}
    \nabla \times \overline{\vector{\nu}}^\perp &= 
    \left( \frac{\partial \overline{\vector{\nu}}^{\perp,3}}{\partial u_1} - \frac{\partial \overline{\vector{\nu}}^{\perp,1}}{\partial \xi_\ell}\right)\hat{\vector{u}}_1 \nonumber\\
    &+ \left( \frac{\partial \overline{\vector{\nu}}^{\perp,1}}{\partial \xi_\ell} - \frac{\partial \overline{\vector{\nu}}^{\perp,3}}{\partial u_1}\right)\hat{\vector{u}}_2 \nonumber\\
    & + \left( \frac{\partial \overline{\vector{\nu}}^{\perp,2}}{\partial u_1} - \frac{\partial \overline{\vector{\nu}}^{\perp,1}}{\partial u_2}\right)\hat{\vector{\xi}}_\ell \nonumber\\
    &=  \left( \frac{\partial \overline{\vector{\nu}}^{\perp,2}}{\partial u_1} - \frac{\partial \overline{\vector{\nu}}^{\perp,1}}{\partial u_2}\right)\hat{\vector{\xi}}_\ell;\\ 
    \nabla \cdot \overline{\vector{\nu}}^\perp &= \frac{\partial \overline{\vector{\nu}}^{\perp,1}}{\partial u_1} + \frac{\partial \overline{\vector{\nu}}^{\perp,2}}{\partial u_2} + \frac{\partial \overline{\vector{\nu}}^{\perp,3}}{\partial \xi_\ell} = \frac{\partial \overline{\vector{\nu}}^{\perp,1}}{\partial u_1} + \frac{\partial \overline{\vector{\nu}}^{\perp,2}}{\partial u_2}.
\end{align}
The final forms of the rotor and divergence are obtained by noting that only the third term in the expansion is non-null in the first case, while it is the only null element in the second (being the vector field over the plane).\\
The results for one manifold selected from the simulated volume are shown in figures \ref{fig:CW_M1_Movies_1}-\ref{fig:CW_M1_Movies_2}. 

In each sub-figure, information is presented by the following scheme:

Skeleton (solid, black line) recovered by up-sampling the embedded graph obtained via SGTM and isosurface of the corresponding model's PDF (red, opaque surface). Extracted simulated particles obtained via soft assignment on the whole data set, given the model (grey dots). Current center $\tivec{t}_\ell$ (red sphere) of the perpendicular plane $\mathcal{T}_\ell^\perp$ and vectors $\hat{\vector{u}}_1$ (blue arrow) $\hat{\vector{u}}_2$ (magenta arrow) spanning it (top left panel).
Heatmap of the weighted mean projected tangential velocity $\lVert \overline{\vector{\nu}}^{\parallel}\rVert$, at rest frame: the velocity of the central point $\overline{\vector{\nu}}^\parallel(\tivec{t}_\ell)$ has been removed in order to represent local velocities with respect to the local frame. Over-plotted are the iso-contours of the model's PDF (dashed black lines) and a mask hiding sparsely populated regions of plane $\mathcal{T}_\ell^\perp$ (top middle panel).

Quiver plot (vector plot) of the weighted mean velocity field (black arrows) over the plane with over-plotted iso-contours of the model's PDF. Again the central velocity $\overline{\vector{\nu}}^\perp(\tivec{t}_\ell)$ has been subtracted to the velocity field (top right panel).

Number of particles within the sphere of radius $\kappa$ for each point $\vector{y}_{ij} \in \mathcal{Y}_\ell$, sampling regularly plane $\mathcal{T}_\ell^\perp$, with over-plotted mask (bottom left panel).

Tangential component of the rotor (also called ``curl'') of the weighted mean velocity field computed over the perpendicular plane (bottom middle panel).

Divergence of the weighted mean velocity field computed over the perpendicular plane (bottom right panel).

A few main results can be drawn from the visualization of the kinematic properties of the two manifolds using our technique:
\begin{figure*}[th!]
\begin{tikzpicture}[node distance = 0cm,nodes = {anchor=north west,inner sep=0cm}]
\node[] (a_cap)                   {\textbf{(a)} Manifold n.1, snapshot n.9};
\node[] (a) at (a_cap.south west) {\includegraphics[width=\textwidth,trim = 3.8cm 1cm 3cm 0.8cm,clip]{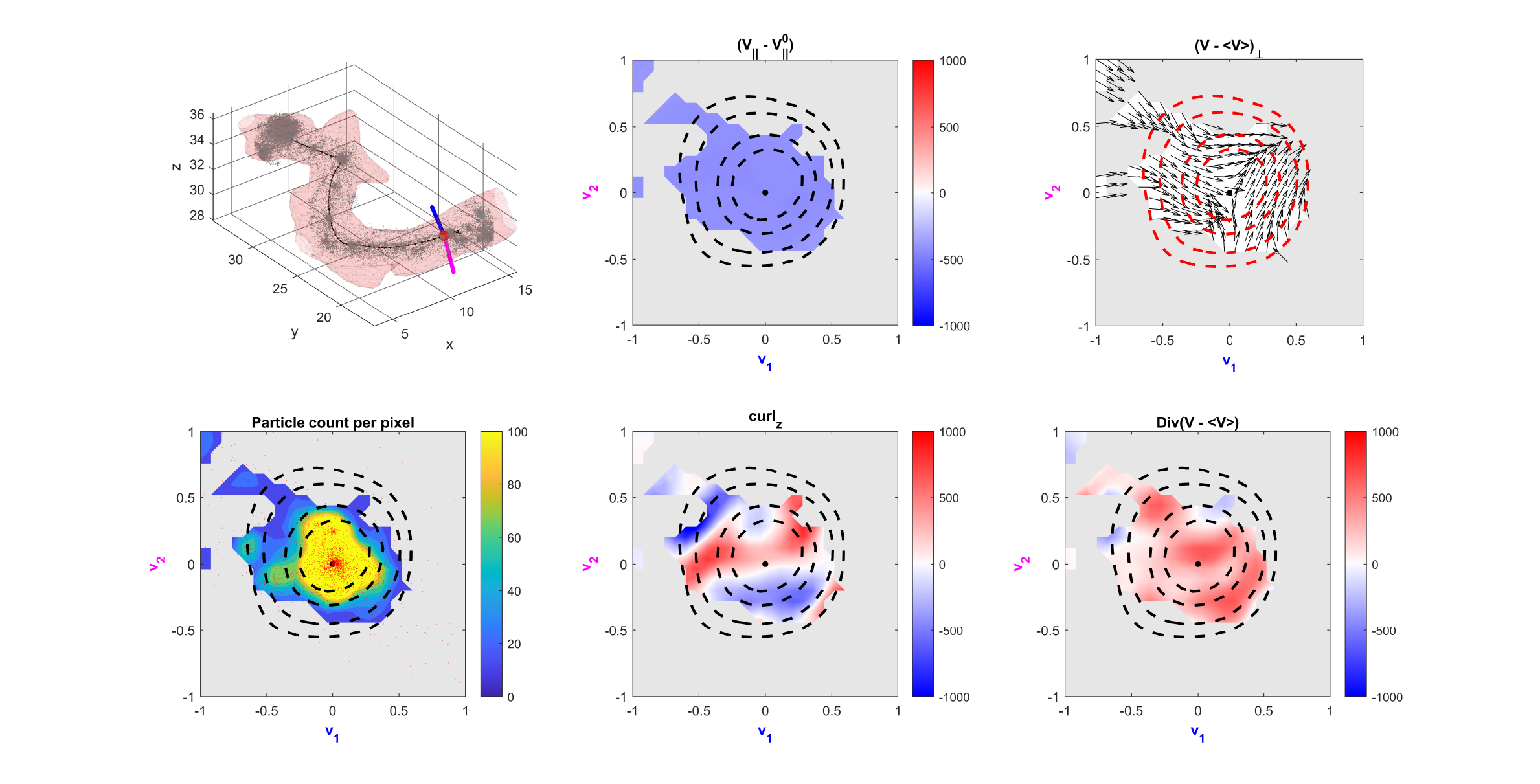}};
\node[] (b_cap) at (a.south west) {\textbf{(b)} Manifold n.1, snapshot n.20};
\node[] (b) at (b_cap.south west) {\includegraphics[width=\textwidth,trim = 3.8cm 1cm 3cm 0.8cm,clip]{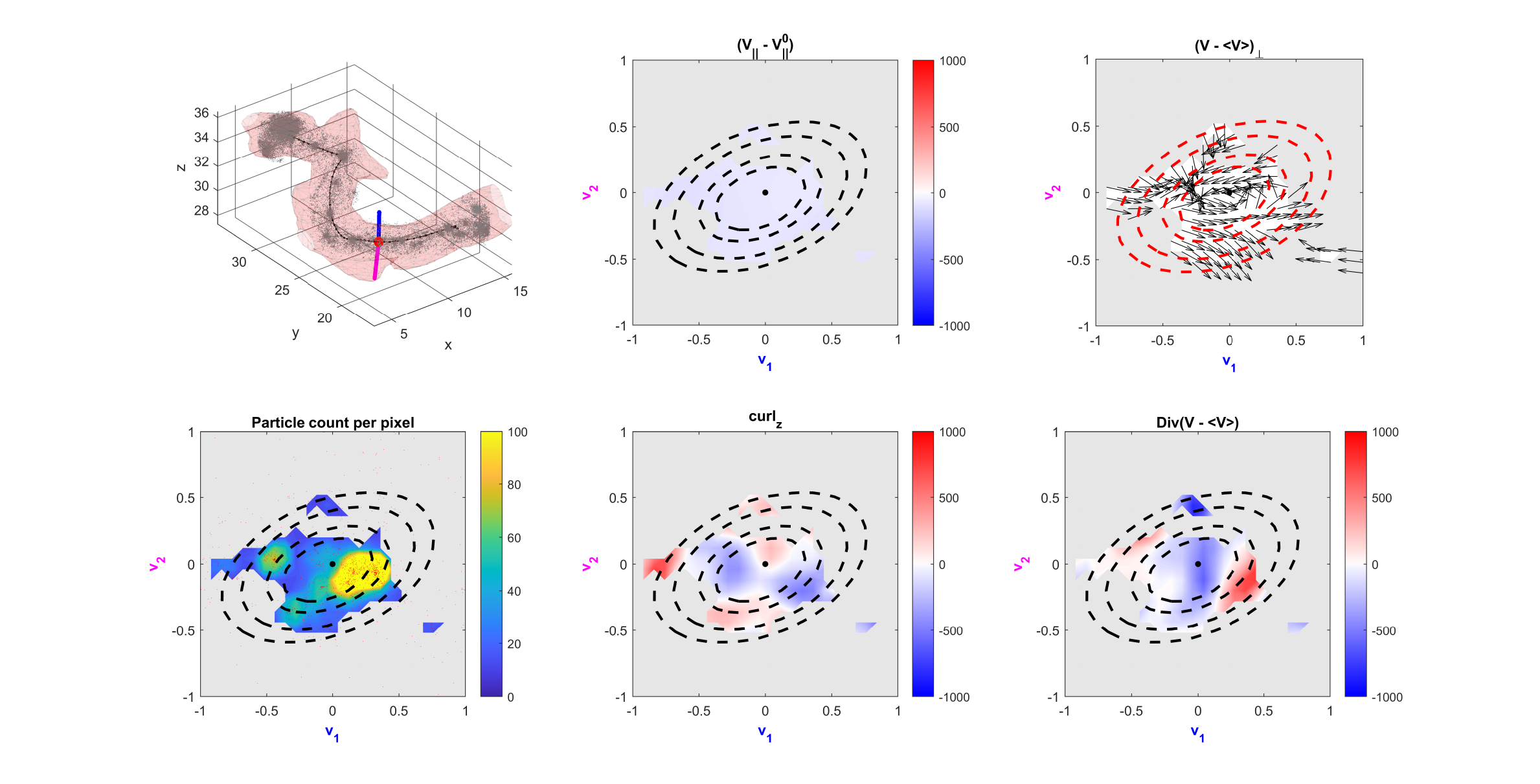}};
\end{tikzpicture}
\caption{Different snapshots of orthonormal coordinate frame movie for manifold n.1 of the Cosmic Web.}
\label{fig:CW_M1_Movies_1}
\end{figure*} 
\begin{figure*}[th!]
\begin{tikzpicture}[node distance = 0cm,nodes = {anchor=north west,inner sep=0cm},square/.style={regular polygon,regular polygon sides=4}]
\node[] (a_cap)                   {\textbf{(a)} Manifold n.1, snapshot n.39};
\node[] (a) at (a_cap.south west) {\includegraphics[width=\textwidth,trim = 3.8cm 1cm 3cm 0.8cm,clip]{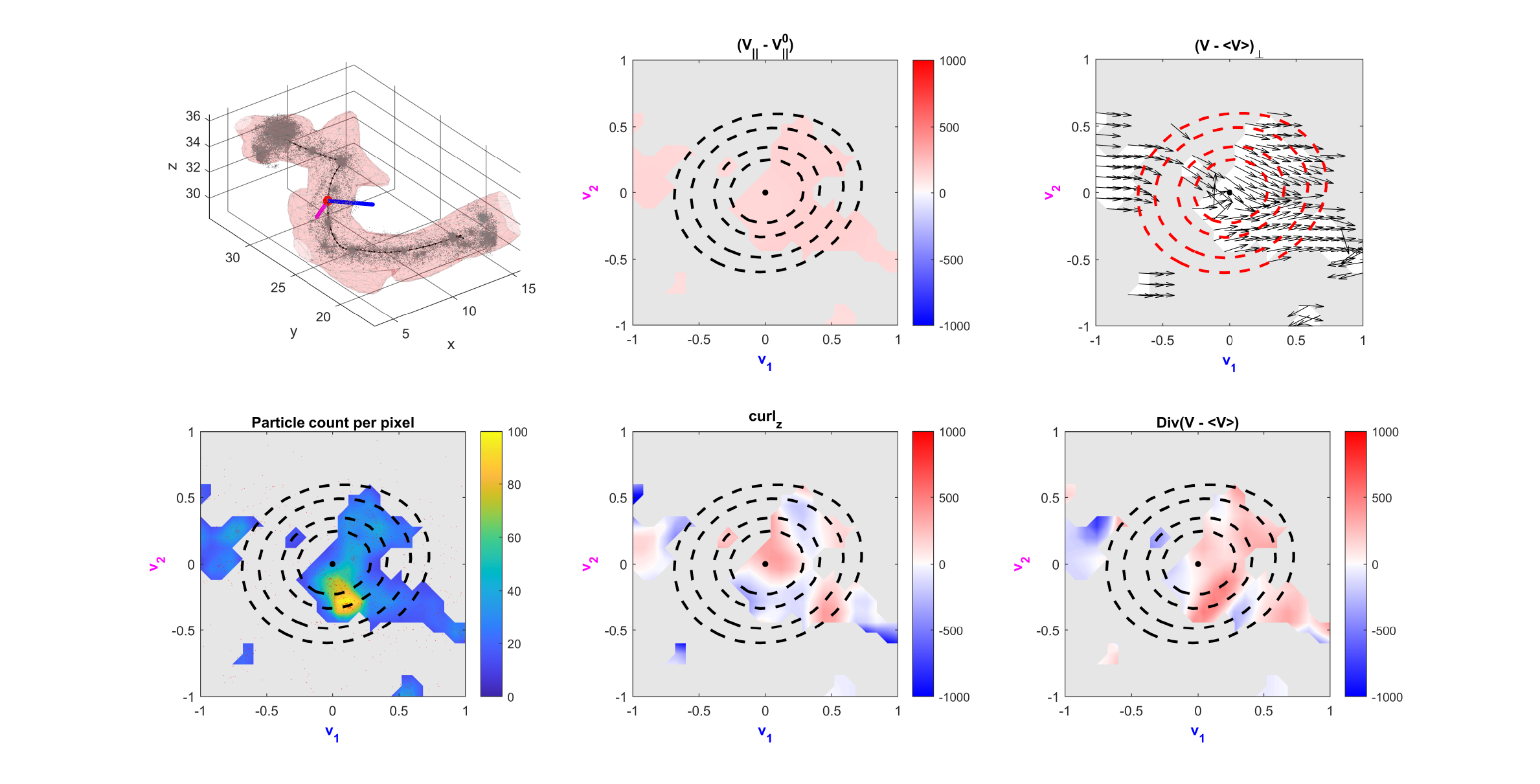}};
\node[] (b_cap) at (a.south west) {\textbf{(b)} Manifold n.1, snapshot n.50};
\node[] (b) at (b_cap.south west) {\includegraphics[width=\textwidth,trim = 3.8cm 1cm 3cm 0.8cm,clip]{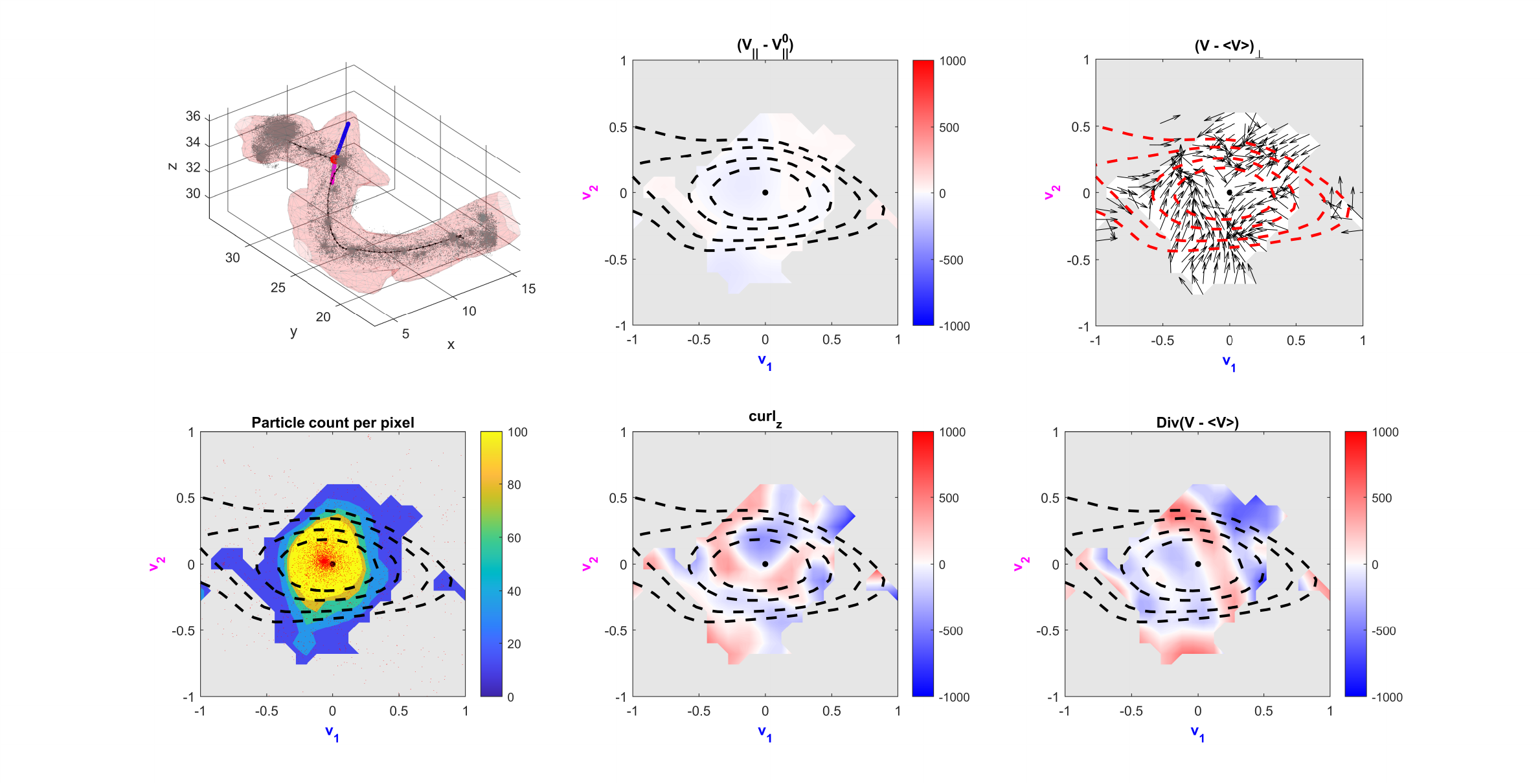}};
\node[line width=0.5mm,square,draw=magenta,minimum size=0.45cm] (curl) at ([shift={(-0.42,2.52)}]b.south) {};
\node[draw=none,anchor=north west,text=magenta] at ([shift={(-0.65,2.6)}]b.south) {$\mathbf{1}$};
\node[line width=0.5mm,square,draw=violet,minimum size=0.45cm] (vecm) at ([shift={(5.52,-1.9)}]b.north) {};
\node[draw=none,anchor=north west,text=violet] at ([shift={(5.3,-1.82)}]b.north) {$\mathbf{1}$};

\node[line width=0.5mm,square,draw=magenta,minimum size=0.45cm] (curl2) at ([shift={(-0.2,2.6)}]b.south) {};
\node[draw=none,anchor=north west,text=magenta] at ([shift={(0.2,2.8)}]b.south) {$\mathbf{2}$};
\node[line width=0.5mm,square,draw=violet,minimum size=0.45cm] (vecm2) at ([shift={(5.75,-1.82)}]b.north) {};
\node[draw=none,anchor=north west,text=violet] at ([shift={(6.1,-1.6)}]b.north) {$\mathbf{2}$};

\node[line width=0.5mm,square,draw=magenta,minimum size=0.45cm] (curl3) at ([shift={(0.1,3.25)}]b.south) {};
\node[draw=none,anchor=north west,text=magenta] at ([shift={(0.5,3.5)}]b.south) {$\mathbf{3}$};
\node[line width=0.5mm,square,draw=violet,minimum size=0.45cm] (vecm3) at ([shift={(6.05,-1.15)}]b.north) {};
\node[draw=none,anchor=north west,text=violet] at ([shift={(6.4,-0.9)}]b.north) {$\mathbf{3}$};
\end{tikzpicture}
\caption{Same as in figure \ref{fig:CW_M1_Movies_1} for further centers on the manifold. The bottom central panel (curl) and top right panel (velocity field) of fig. \ref{fig:CW_M1_Movies_2} also present numbered square regions in shades of purple. The three numbered squares identify regions presenting opposite sign in curl and their respective velocity field. Zoomed in pictures of the velocity field in the three regions are presented in fig. \ref{fig:Zooms}.}
\label{fig:CW_M1_Movies_2}
\end{figure*}
\begin{figure*}[t!]
\begin{tikzpicture}[node distance = 0cm,nodes = {anchor=north west,inner sep=0cm},square/.style={regular polygon,regular polygon sides=4}]
\node[] (a_cap) at (0,0)                   {\textbf{(a)} Zoom-in region $1$};
\node[] (a) at (a_cap.south west) {\includegraphics[width=0.4\textwidth]{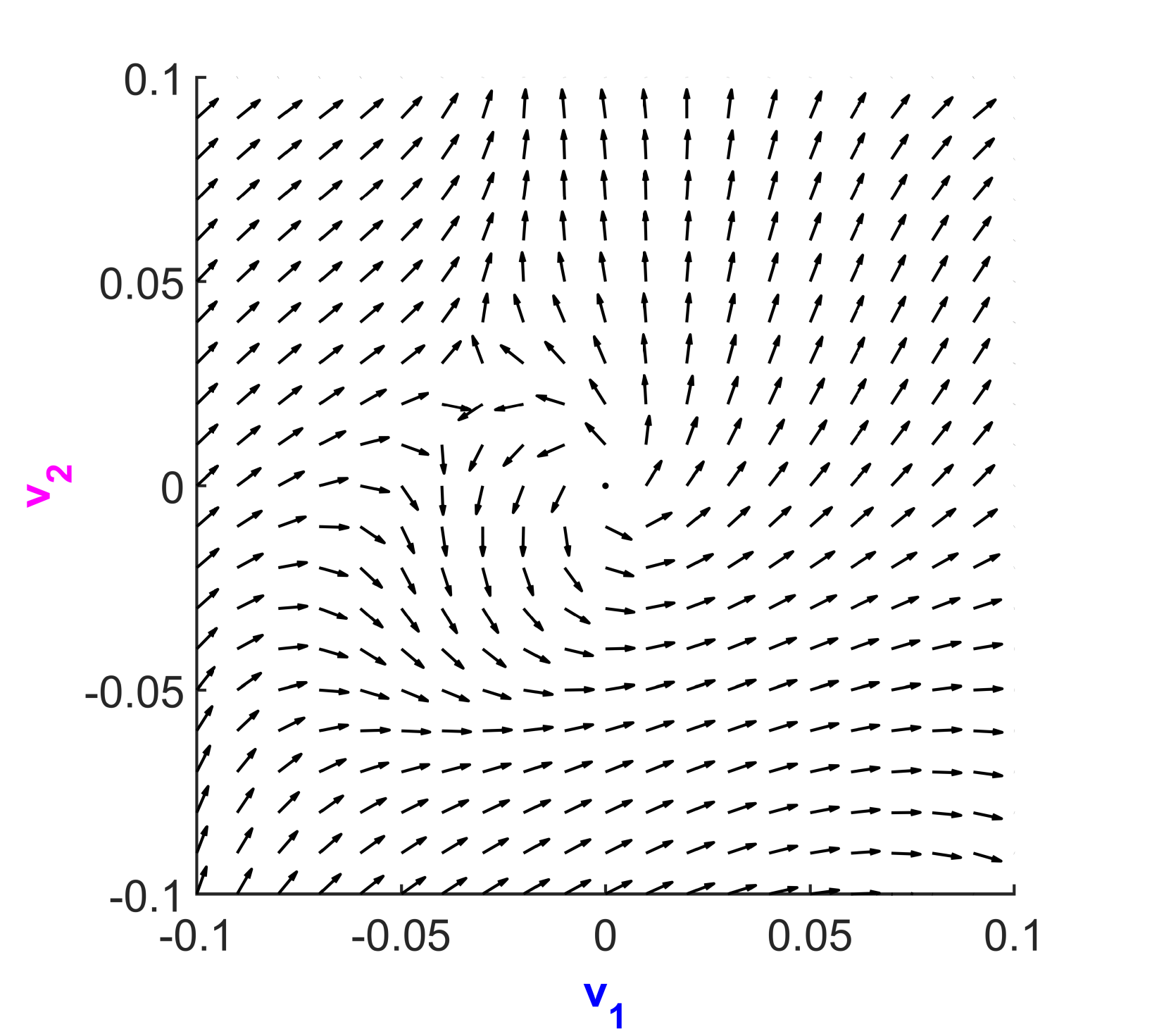}};
\node[] (b_cap) at (5.5cm,0) {\textbf{(b)} Zoom-in region $2$};
\node[] (b) at (b_cap.south west) {\includegraphics[width=0.4\textwidth]{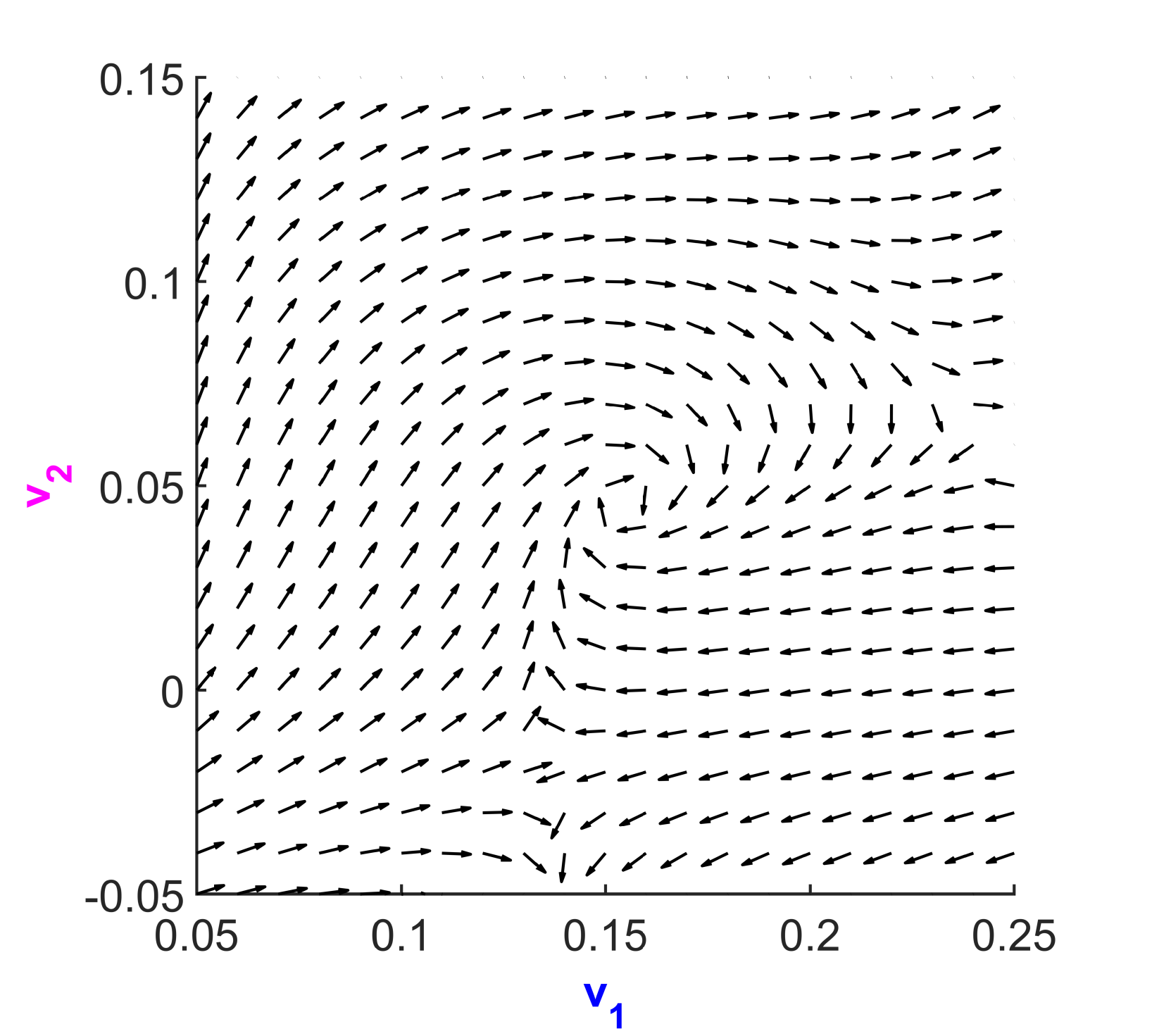}};
\node[] (c_cap) at (11cm,0) {\textbf{(c)} Zoom-in region $3$};
\node[] (c) at (c_cap.south west) {\includegraphics[width=0.4\textwidth]{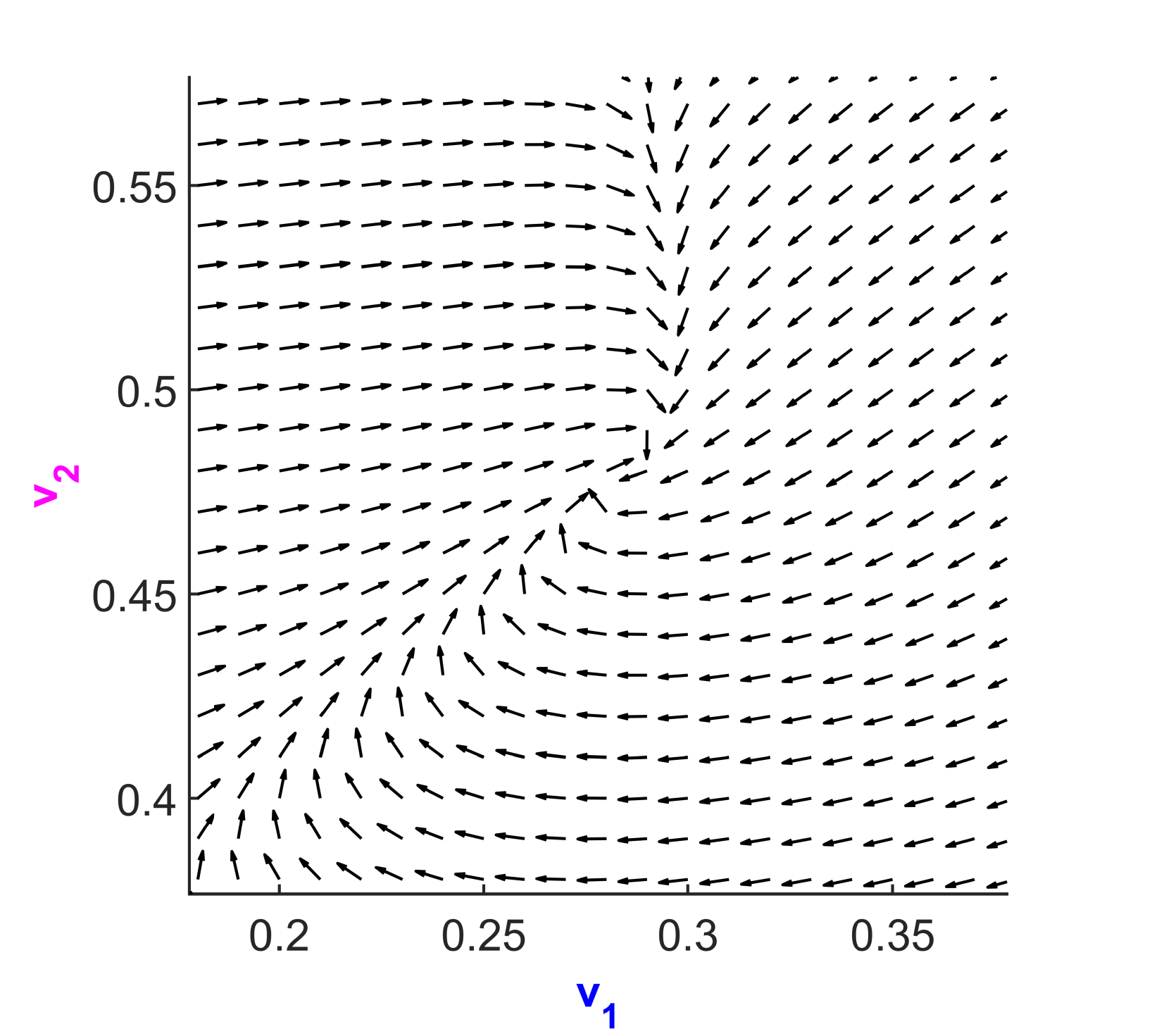}};
\end{tikzpicture}
\caption{Zoomed-in regions of high vorticity identified in fig. \ref{fig:CW_M1_Movies_2}b. Note how local vortices can be identified within each panel. The corresponing values for the curl can be identified in the bottom central panel in figure \ref{fig:CW_M1_Movies_2}b. The zoom-in region $1$ curl (blue) is opposite to zoom-in $2$ and $3$ (red) and this is reflected in the orientation of the vortices: counter clockwise for $1$ and clockwise for $2$ and $3$.}
\label{fig:Zooms}
\end{figure*}
\begin{itemize}
\item The skeletons of the manifolds are generally aligned with their corresponding densest regions, when these are unique. In cases where mass has a multi-modal distribution, the center of the plane is usually placed in order to include all modes. This is visible in the bottom left panel of each figure. The regions containing the largest amount of particles (and thus the highest density) are usually centered on the plane when presenting one peak, or slightly shifted when more than one peak can be found.
\item The tangential velocity of particles on the manifold (top central panel) has the tendency to change sign at a certain distance from the two extremities. The extremities of each manifold are the nodes of the cosmic web (i.e the clusters). In particular, starting from one node and moving towards the other, the tangential velocity has a negative sign at first (overall blue color), meaning that particles are pulled towards the starting node. As we move towards the second node, the tangential velocity's module tends to decrease until it reaches zero. After this ``saddle'' point, the tangential velocity is aligned with the crawling direction, meaning that the particles tend to be pushed towards this second node (the parallel velocity map turns red). Figure  \ref{fig:CW_M1_Movies_2}a clearly shows the flip in sign of the tangential velocity.  
Figure \ref{fig:CW_M1_Movies_2}b, shows a slight deviation from this behaviour. This is due to the presence of a mass concentration within the filament (also identifiable in the bottom left panel of the figure) locally pulling particles towards itself.
\item The parallel curl (perpendicular to the plane) of the velocity field shows clear islands of opposite signs. This result is in agreement with what found in \cite{10.1093/mnras/stu2289}, although the analysis is only performed on one filament here. In their work, the number of regions with opposite curl shows a large variability throughout the elongation of the stream.
It is however also pointed out that four regions are most commonly found. In our analysis we generally agree with the findings of \cite{10.1093/mnras/stu2289} in terms of the variability in the number of these regions, but further analysis is needed to confirm the most common occurrence. However, in some iterations, we are in agreement with their prediction (e.g. figure \ref{fig:CW_M1_Movies_1}b). Future studies will focus on applying the proposed methodology to a larger set of filaments in order to obtain a robust estimate about the variability in the number of these regions across a larger population. 
\item Particles in the outskirts of the manifolds are attracted towards their cores. This result transpires from a visual inspection of the top right panels of all figures. The weighted mean perpendicular velocity field is generally oriented towards the center of the plane (the manifold's core). 
\item In figure \ref{fig:CW_M1_Movies_2}b we identify three regions of high vorticity from the bottom central panel. The three square regions are numbered and shown correspondingly on the top right panel. The same regions are presented in fig. \ref{fig:Zooms}a-c respectively. The velocity map in the three regions confirms the behaviour presented in the colored regions in bottom central panel of fig. \ref{fig:CW_M1_Movies_2}b. In particular, the ``blue-curl'' region $\mathbf{1}$ presents a vortex oriented counter clockwise, while the vortices in the ``red-curl'' regions $\mathbf{2}$ and $\mathbf{3}$ are oriented clockwise. 
\end{itemize}

To summarize, our tool was applied to a simulated cosmic web volume, and it was able to recover from the complex point clouds streams of complicated morphology. The proposed analysis of the structures is in agreement with previous findings regarding the dynamics of the filaments. This example of application of our toolbox is further proof of the validity of the proposed methodologies. Furthermore, it is still possible to extend the analysis to other non-dynamical properties of the cosmic filaments (e.g. as shown in section \ref{subsec:JellyFish} for the jellyfish galaxy), as well as their morphology, with no further effort in the development of new tools. As this will be subject of future studies, we believe that the toolbox here presented and demonstrated may prove extremely useful in the understanding of these structures.

\subsection{\texorpdfstring{$\omega$}{omega}-Centauri's stellar stream from GAIA-DR2}\label{subsec:GAIA}

The study of the stellar galactic halo of the Milky Way is of great importance to astronomers interested in the archaeological aspects of the Galaxy, particularly because of the halo’s key role in characterizing the galaxy’s formation history. A crucial component to the formation of galaxies in a hierarchical formation scenario is the growth by tidal disruption or mergers with external astronomical objects. This led to the deposit of merger debris in the Milky Way’s halo in the form of stellar streams or stellar overdensities \citep{Helmi2020}. In order to characterize the interaction history of the Milky Way, astronomers have tracked the stars found within stellar streams in the halo allowing them to trace the stars' origins back to the early phases of the Galaxy’s formation \citep{Helmi2020}. 

Moreover, dynamically cold stellar streams provide an opportunity to probe the acceleration field of the Galaxy both locally and globally \citep{JohnstonEtal1999, IbataEtal2002, JohnstonEtal2002, Carlberg2012}. This gives great insight onto the nature of the gravitational force and the distribution of dark matter both of which are encoded in the Milky Way's acceleration field \citep{IbataEtal2021}.    

Deep wide-field photometric surveys including the SDSS \citep{YorkEtal2000}, PanSTARRS \citep{ChambersEtal2016}, and DES \citep{AbbottEtal2018} have increased our knowledge of the Galaxy’s stellar halo by revealing many of the narrow streams and overdensities belonging to it \citep{Helmi2020}. However, much greater clarity was obtained with the Gaia mission following the second Gaia data release (DR2; \citet{GaiaCollaboration2018}) and the third early data release (EDR3; \citet{GaiaCollaboration2020}).

Given the multidimensional data provided in Gaia EDR3, the Milky Way's stellar streams constitute another natural application of our pipeline. For a test subject, we have chosen $\omega$-Centauri as it is the largest cluster known and has been extensively proven to be tidally disrupted. In particular we base our work on the information provided in \citet{IbataEtal2019_OmegaCen} for spotting the  tidal arms of $\omega$-Centauri. Further detailing of this procedure will be provided in section \ref{subsubsec:OC_Filtering}. Through this third demonstration, we show that our modeling is applicable not only to simulation outputs, but also to the ever-increasing amounts of observational data. 

\subsubsection{Isolation of \texorpdfstring{$\omega$}{omega}-Centauri's stream}\label{subsubsec:OC_Filtering}

\textbf{Methodologies:} EM3A $\rightarrow$ Dimensionality Index $\rightarrow$ Crawling $\rightarrow$ SGTM $\rightarrow$ Co-moving Orthonormal coordinate frames (mod.). \\

In this section, we outline the different steps followed to spot the tidal-arms of Omega-Centauri ($\omega-$Cen). This will act as the preprocessing stage before applying our algorithms to extract and model the targeted streams. Through the $N$-body simulations conducted in \citet{IbataEtal2019_OmegaCen}, the ``Fimbulthul structure'' \citep{IbataEtal2019} in Gaia DR2 was identified as part of the tidal arm of the cluster. The properties of the system that were guided by the results of their $N$-body simulations then served as a selection filter applied to the stars in an area around the cluster. In order to obtain a distribution of the stars that show the two streams, we follow their selection criteria and refer the reader to \citet{IbataEtal2019_OmegaCen} for a detailed motivation of the selection basis.

From the Gaia archive, we choose a rectangular region spanning $l=[-70^\circ, -30^\circ]$ and $b=[5^\circ,50^\circ]$ where $l$ and $b$ are the galactic longitude and latitude respectively. In this region, we select the stars that have a parallax uncertainty less than $1$ $mas$ and those with parallax measurements consistent within $1\sigma$ with distances between 4 and 6~kpc. 

We then apply a filter on the kinematic behaviour where we select the stars that have proper motions along the declination direction $\mu_{\delta}$ similar to that of the cluster, and proper motions along the right ascension direction $\mu_{\alpha}$ that show a decreasing linear gradient as a function of $b$. The rate of decrease is taken as $0.125$ mas/yr for every degree in $b$. For these requirements, $\mu_{\alpha} = -3.1925 \pm 0.0022$ mas/yr and $\mu_{\delta} = -6.7445 \pm 0.0019$ mas/yr are chosen as the reference values for the cluster, and the stars within $1$~mas/yr of the two kinematic criteria are selected.

Furthermore, to correct for interstellar extinction, we use the dust maps provided in \citet{SchlegelEtal1998} and recalibrated by \citet{SchlaflyFinkbeiner2011} to modify the brightness and colors of the remaining stars. The extinction-corrected magnitudes are obtained assuming foreground-only interstellar extinction with $R_V = 3.1$. For choosing the stars belonging to the Color-Magnitude Diagram (CMD) of the cluster, we draw the polygons shown in Figure 4a in \citet{IbataEtal2019_OmegaCen} and select the stars belonging to the regions within those polygons. This selection rule allows for the filtering out of a large number of background stars while minimizing the bias against selecting stars belonging to $\omega$-Centauri \citep{IbataEtal2019_OmegaCen}. 

\subsubsection{Modelling of stream via SGTM}

\begin{table}[h]
\centering
\def\myCWidth{0.12}
\caption{Adopted values for Experiments on $\omega-$Centauri}
\label{tab:OC_Params}
\begin{tabularx}{\columnwidth}{X | X | X}
\toprule
    EM3A     &  $R_{min} = 1$ & $R_{max} = 2$ \\
    Crawling & $r = 1.5$  & \\
    SGTM     & $r = 1.5$ & $S = L/2$ \\
    Moving frames & $a = 2.5$ & \\
\bottomrule
\end{tabularx}
\end{table}

\begin{figure}[t!]
\begin{tikzpicture}[node distance = 0cm,nodes = {anchor=north west,inner sep=0cm}] 
\node[] (a_cap) {\textbf{(a)} GAIA DR2 stars (adapted from \cite{2001Natur.412...49I})
};
\node[] (a) at (a_cap.south west) {\includegraphics[width=\columnwidth,trim = 1.25cm 0.25cm 1.1cm 1.0cm,clip]{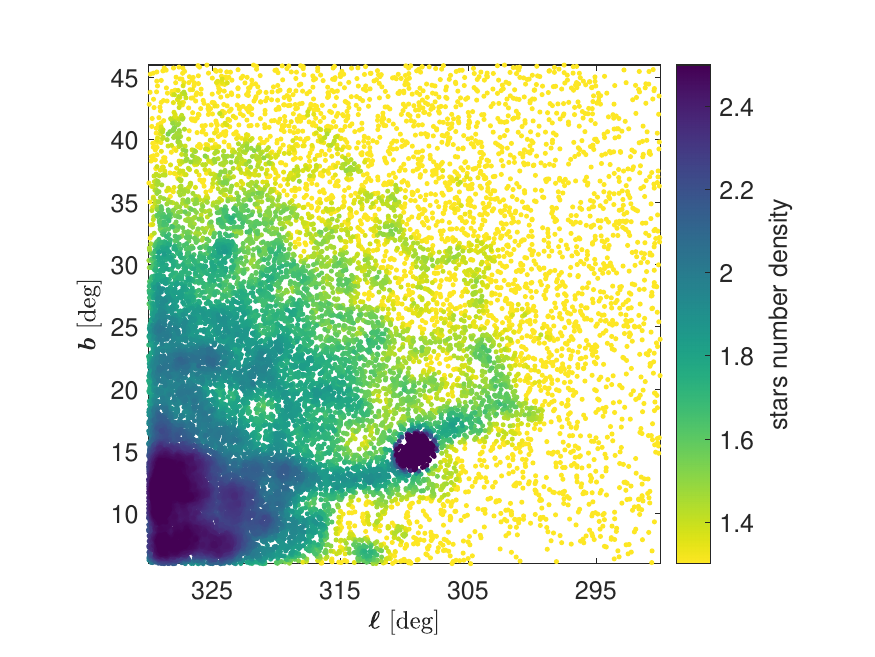}};
\node[] (b_cap) at (a.south west) {\textbf{(b)} Probabilistic model of Omega-Centauri stream
};
\node[] (b) at (b_cap.south west) {\includegraphics[width=\columnwidth,trim = 1.5cm 0.5cm 1.5cm 1.1cm,clip]{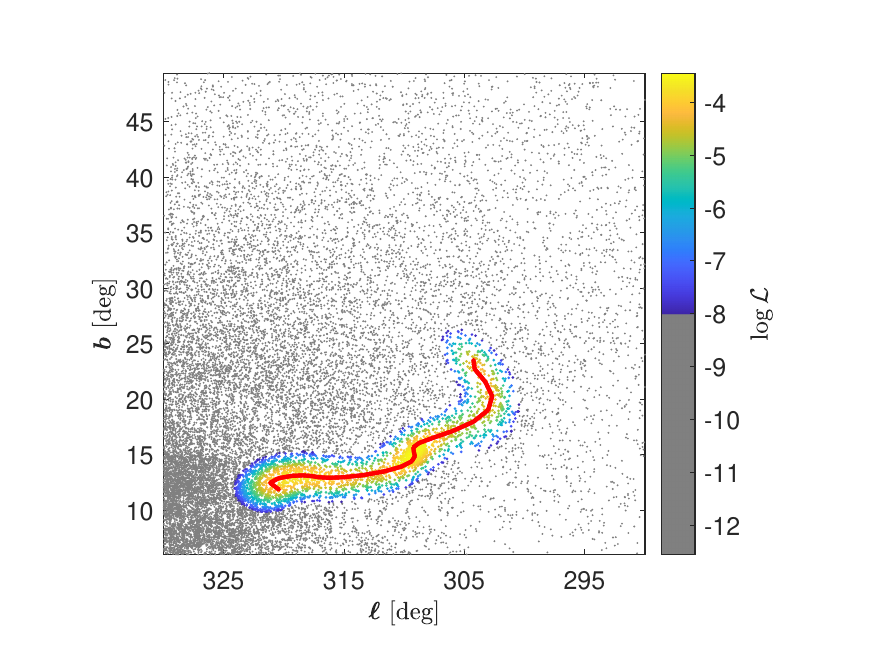}};
\end{tikzpicture}
\caption{Panel (a) depicts the GAIA DR2 stars after selection by criteria explained in section \ref{subsubsec:OC_Filtering}, coloured by local stellar log-density (adapted from \cite{2001Natur.412...49I}). 
Panel (b) shows the probabilistic model of Omega-Centauri's stream as obtained from crawling and SGTM. 
The red line marks the trained skeleton 
and the stars are coloured by the 
log likelihood with a threshold to gray to visualize isocurves. 
}
\label{fig:OmegaCen}
\end{figure}

All the selected particles are collected in data set $\mathcal{Q}$. We first apply EM3A \ref{subsec:diffusion}.
In fact, when removing particles within a circular region centered on the core of $\omega_{Cen}$, the point distribution is irregular in its vicinity, causing a sharp discontinuity between the dense (fig. \ref{fig:OmegaCen}a, lower left) and sparse (fig. \ref{fig:OmegaCen}b, upper right) filaments. Furthermore, the over-density of the galaxy on the bottom left region of the panel, has a higher chance of pheromone being deposited there than on the filaments. This results in LAAT not being able to equally distribute pheromone along the regions of interest. However, the application of EM3A to the data set obtained by kinematical and color-magnitude filtering proved successfull, being EM3A less influenced by global densities and more focused on local anysotropies.
Since the data set has large variations of the local densities, particular care has to be taken when applying the methodology. In particular, adopting a large number of iterations with a large radius for neighborhood search may result in the production of spurious structures, unrelated to the filament we require to extract. On the other hand, a small radius may disrupt the main structure of interest prematurely, fragmenting it in multiple clumpy regions. 
It is thus advisable to monitor the advancement of the methodology at each iteration and to test different values for its hyper-parameters.
Furthermore, since $\omega_{Cen}$'s main body has the highest star's number density, we remove from data set $\mathcal{Q}$ all the stars lying within the sphere of radius $r = 0.8~\mathrm{deg}$, centered at $[309.10202, +14.96833]$ in the galactic coordinate reference system.
Following the application of the proposed methodologies, we recover the skeleton and SGTM model of $\omega_{Cen}$'s stream. The resulting SGTM is shown in figure \ref{fig:OmegaCen}b, together with the stars selected by the criteria in section \ref{subsubsec:OC_Filtering} (figure \ref{fig:OmegaCen}a), as obtained by \cite{2001Natur.412...49I}. The parameters used for the methodologies are shown in tab. \ref{tab:OC_Params}. The red line is the skeleton after training SGTM, the noise model is represented by the isocurves at different isovalues of the model's likelihood over the data space. The stream's model resembles very closely the one highlighted in figure \ref{fig:OmegaCen}a. Grey dots in figure \ref{fig:OmegaCen}b are the same points presented in figure \ref{fig:OmegaCen}a, 
however we omit the colouring in order to enhance visibility of the noise model's iso-contours.
The stream is analyzed via the methodology presented in section \ref{subsec:OrthoPlanes}, but adapted for the 1D case. 

\subsubsection*{Co-moving orthonormal reference frame}
\begin{figure*}[!t]
\begin{tikzpicture}[node distance = 0cm,nodes = {anchor=north,inner sep=0cm}] 
\node[] (a)              {\includegraphics[width=\textwidth,trim={0.5cm 0 1.8cm 0},clip]{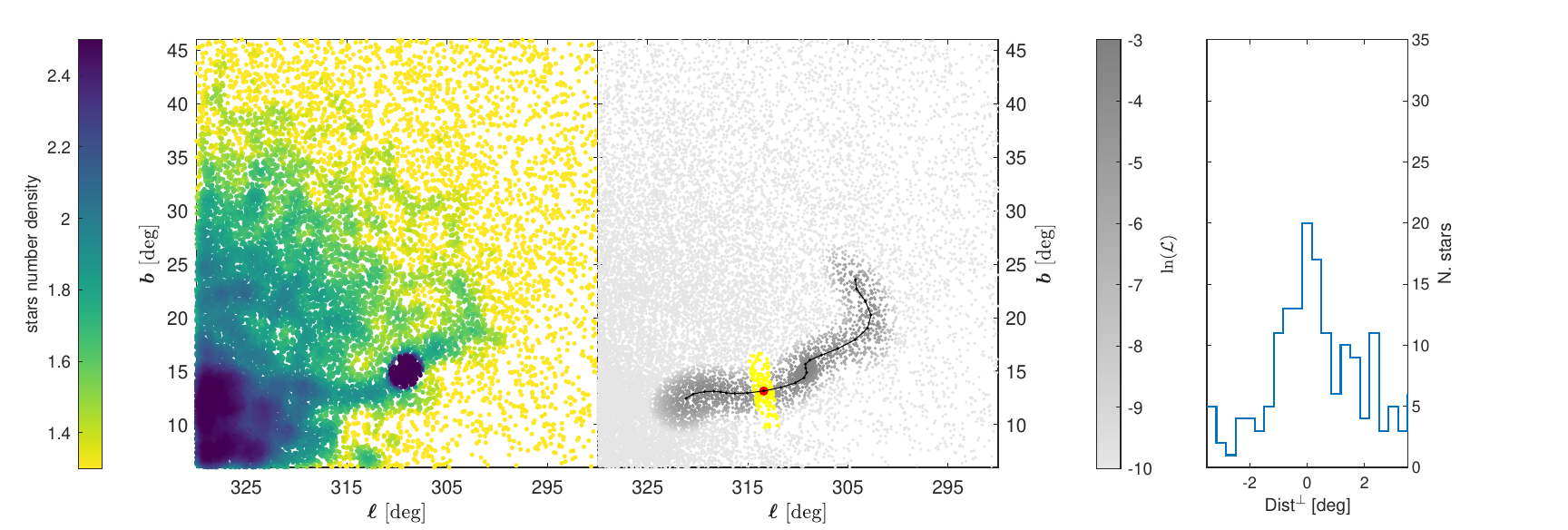}};
\node[] (b) at (a.south) {\includegraphics[width=\textwidth,trim={0.5cm 0 1.8cm 0},clip]{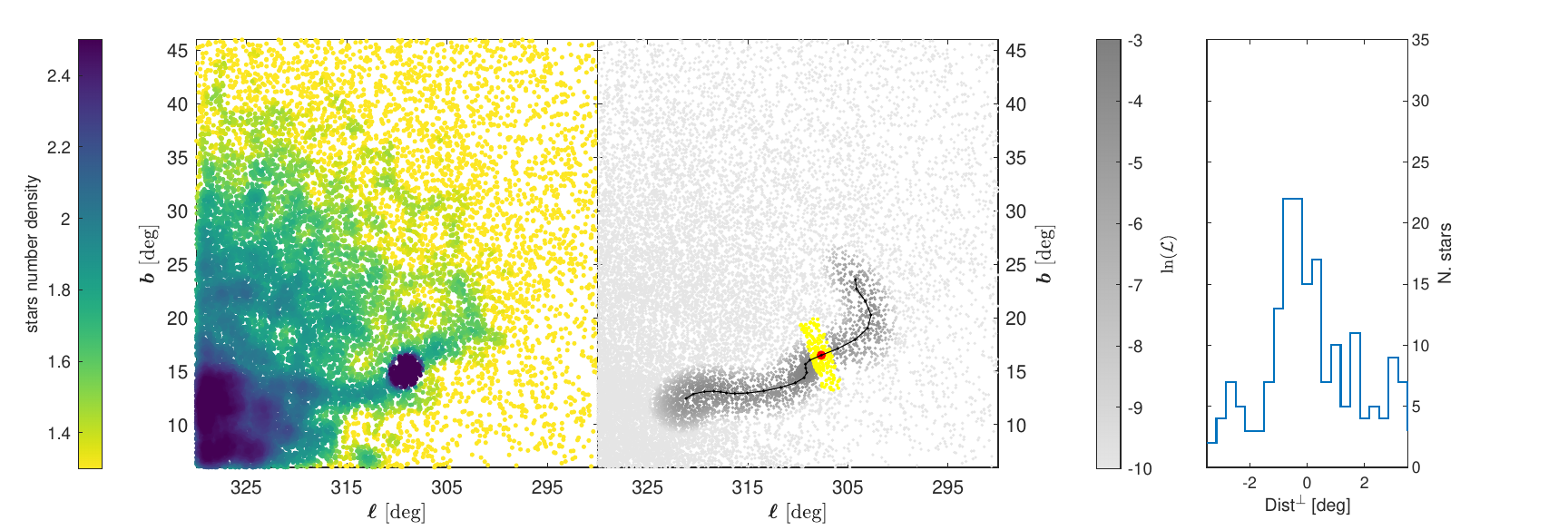}};
\node[] (c) at (b.south) {\includegraphics[width=\textwidth,trim={0.5cm 0 1.8cm 0},clip]{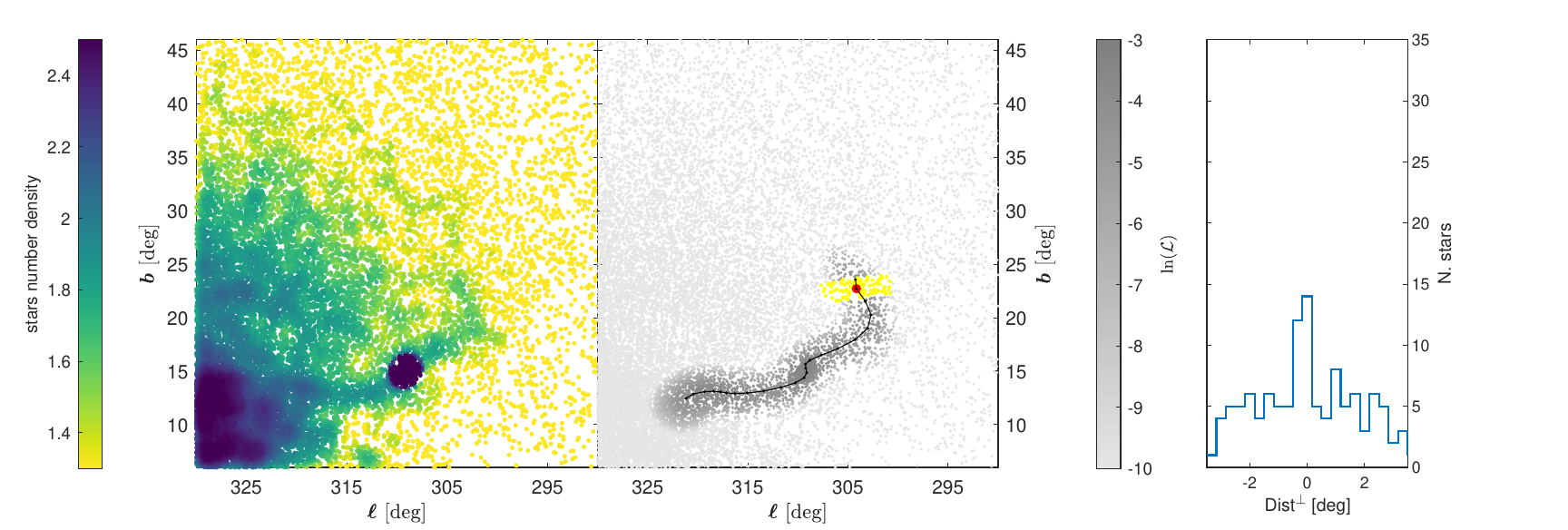}};
\node[anchor=north west] (a_cap) at (a.north west) {\textbf{(a)} Reference frame 9 of $\omega_{Cen}$ stream};
\node[anchor=north west] (b_cap) at (a.south west) {\textbf{(b)} Reference frame 17 of $\omega_{Cen}$ stream};
\node[anchor=north west] (c_cap) at (b.south west) {\textbf{(c)} Reference frame 23 of $\omega_{Cen}$ stream};
\end{tikzpicture}
\caption{Visualizations of $\omega_{Cen}$'s stream recovered from GAIA DR2. 
Each row contains the original density map (left), the model likelihood with the visualization frame position marked by a red dot (middle). The yellow window expands perpendicular to the stream and can be used to inspect properties at different positions by analyzing sources within, such as distribution perpendicular to the stream (right histogram). 
(a) and (b) shows regions where the stream peaks clearly, while the reference frame in panel (c) exemplifies a fainter stream part exhibiting a flatter histogram.}
\label{fig:OC_Movies}
\end{figure*}
We now consider the co-moving orthonormal reference frame technique, for this one-dimensional case. Figures \ref{fig:OC_Movies}a-c show the results on $\omega_{Cen}$'s streams. The yellow strip on the top-right panel of each figure, shows the selection of stars used for computing local density of the stream. The selection is made on a geometrical basis, by projecting all stars onto the tangential and perpendicular spaces of the manifold at each node in the trained SGTM. 
For each point in $\tivec{t}_\ell \in \overline{\mathcal{P}}$ (we drop index $k$, since we only consider one manifold), we know the tangent vector to the manifold at $\tivec{t}_\ell$: $\hat{\vector{\xi}}_\ell = [\xi_\ell^1,~ \xi_\ell^2]$. Its orthogonal complement is given by either $\vector{\xi}_\ell^{\perp +} = [-\xi_\ell^2,~ \xi_\ell^1]$ or $\vector{\xi}_\ell^{\perp -} = [\xi_\ell^2,~ -\xi_\ell^1]$, giving the vectors pointing towards increasing or decreasing $b$ respectively. 
We chose here to adopt $\vector{\xi}_\ell^{\perp} = \vector{\xi}_\ell^{\perp+}$ as the orthogonal complement to tangent vector $\hat{\vector{\xi}}_\ell$. 

At each $\tivec{t}_\ell$, we project all stars in $\mathcal{Q}$ onto the tangent vector $\vector{\xi}_\ell$. To define the width of the selection strip, we impose a maximum module for the orthogonal projection onto $\vector{\xi}_\ell$. In this case the maximum tangential distance is fixed at the distance between two adjacent centers in the SGTM. We then project all stars onto $\vector{\xi}_\ell$'s orthogonal complement $\vector{\xi}^{\perp}$ and impose a maximum perpendicular distance of $8^\circ$. The stars satisfying these two restrictions are plotted in the central panel of each plot (yellow points) in figure \ref{fig:OC_Movies} and it can be verified that they form indeed a rectangular shape with shorter edge aligned with the tangential space to the manifold and the longer edge aligned with its orthogonal complement. We then partition the orthogonal axis in bins and form a histogram of the number of particles per bin, as shown in the right panels of fig. \ref{fig:OC_Movies}, at varying $\tivec{t}_\ell \in \overline{\mathcal{P}}$. The histogram shows how the stellar number density is generally higher nearby the middle of the perpendicular axis, where the current center of SGTM is located. 

\begin{figure*}[ht!]
    \centering
    \includegraphics[width = \textwidth]{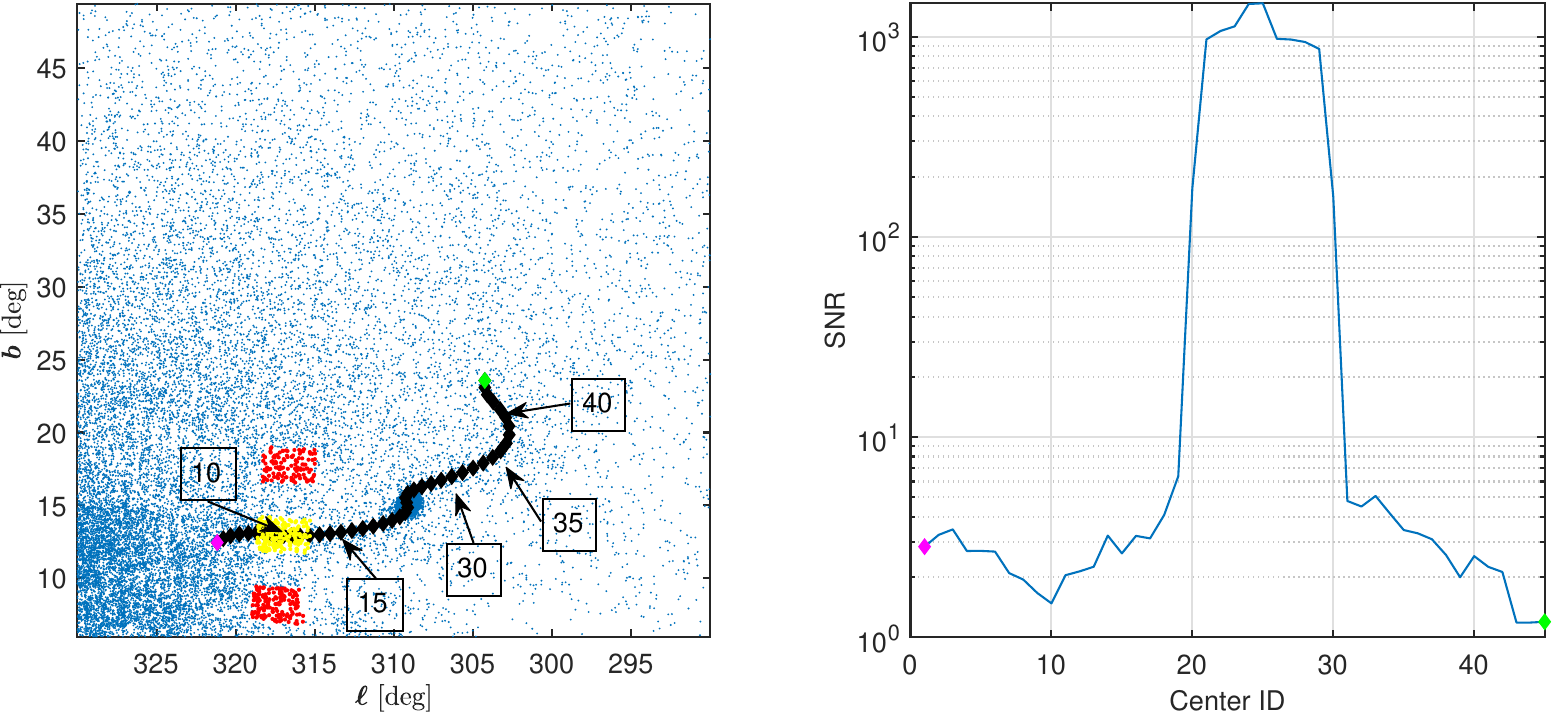}
\caption{Left panel: sky background selection (red circle), overlaid to the box containing the filament from $\omega_{Cen}$. Black diamonds show the position of the centers for SGTM (Center ID), after training. Right panel: SNR for each corresponding center of SGTM. We removed the central centers identifying the core of $\omega_{Cen}$, since our interest is on the quality of detection of the stream.}
\label{fig:SNR}
\end{figure*}

In order to estimate the detection quality of the stream we evaluate the Signal-to-noise ratio (SNR) of the central counts with respect to a co-moving background. For each point on the stream, we define three apertures of fixed area. The first one lies on the streams and is centered on the current position along the manifold (figure \ref{fig:SNR} left plot, yellow dots). This is used to estimate $c^{Stream}$ the count of stars belonging to the stream. However, the resulting count is biased by the sky contribution. In order to remove this bias we define two other apertures (figure \ref{fig:SNR} left plot, red dots) and estimate their respective counts as $c^{Sky}_1$ and $c^{Sky}_2$. The average sky count is obtained by $\langle c^{Sky} \rangle = 0.5\times(c^{Sky}_1 + c^{Sky}_2)$. The counts per pixel are obtained by normalizing the counts by the area of the Sky aperture $A$ (fixed for all centers).
We now compute the Signal of the stream by subtracting the average sky count from the stream count and obtain the SNR for the stream detection at point $\tivec{t}_\ell \in \overline{\mathcal{P}}$:
\begin{equation}\label{eq:SNR}
    SNR(\tivec{t}_\ell) = \frac{S(\tivec{t}_\ell)}{N(\tivec{t}_\ell)} = \frac{c^{Stream} - \langle c^{Sky} \rangle}{\sqrt{\langle c^{Sky} \rangle}} \frac{\sqrt{A}}{A}.
\end{equation}
Particularly important for the estimation of the SNR is the area of both sky and on-object apertures. In this case, we fix the width of the rectangles to be double the distance between neighboring centers on the stream. However, for future analysis, the multiplicative factor can be changed and imposed by the user in order to optimize the result. We believe a factor $2$ is sufficient in our case to obtain a reasonable SNR throughout the whole stream.
The SNR for each point (Center ID in the figure) on the stream is shown in figure \ref{fig:SNR}, right plot. In order to allow visibility of both the SNR of the streams and the core of $\omega-$Centauri, we present the SNR in logarithmic scale. Representative Centers IDs are also shown on the left plot for an easier visual correspondence with the x-axis of the right plot. The SNR shows a sharp peak at around $\mathrm{Center~ID} = 25$ in correspondance to the core of $\omega-$Centauri. 
The left-most edge of the plot also shows a slight increase in SNR, due to the proximity to the hosting galaxy ($\mathrm{Center~ID} = 0$, purple diamond). After the peak in the core, the SNR gradually decreases up to the right-most edge of the manifold (green diamond), where the filament is barely detectable.


\section{Conclusions}\label{sec:Conclusions}

We present a coherent, semi-automatic toolbox for the analysis of noisy data sets with underlying complicated one-dimensional structures convoluted with transverse noise. The toolbox is built on five methodologies, each one addressing different challenges in this kind of data sets. The first methodology (LAAT) aims at recovering high-density regions distributed along elongated filaments within sparse background noise. It is based on ant-colony optimization techniques and via the assignment of a scalar field over the data set it enables selection of relevant features depending on a user-specified threshold. The second methodology (EM3A) enhances over-density along filaments, by pushing points towards the perpendicular complement of the manifolds. It is again inspired by ant-colony optimization and uses principles of game theory for parameter tuning. The third methodology (Dimensionality Index) is devoted to defining a dimensionality index to all particles in the data set. Through the dimensionality index, it is possible to define partitions of the data set containing only points belonging to structures of defined dimensionality. A smoothed version of the index is proposed as a way to take into account the global structure to which particles locally belong. Having separated one-dimensional points from the rest of the data set, a manifold crawling algorithm is proposed that traces the skeletons of the hidden structures. In order to describe the transverse noise to the manifolds, a constrained Gaussian Mixture Model is devised in the form of Stream Generative Topographic Mapping. 

We also presented two visualization techniques that take full advantage of the manifold formulation. The Bi-dimensional profile gives a global view over the manifold, showing concisely the behaviour of desired quantities along the mean curve and across the radial direction of the filaments. The orthonormal coordinate frame technique gives a detailed depiction of the same quantities over local frames perpendicular to the manifold's tangent direction at each desired location. 

The aim of this work is to demonstrate how the various methodologies can be combined in different ways for a range of astronomical applications, both on simulated and observed data sets. Particular care is dedicated to describing in detail the various aspects of each methodology, with the user experience in mind. 

Initially, the toolbox has been applied to a mock data set, where two main filamentary structures have been defined having non-linear shapes. The point cloud sampling the manifolds includes additional transverse and background noise. Each particle in the data set has a specific value for two simulated physical quantities. The quantities are designed to follow specific profiles along and across the two filaments. The application of the toolbox to this case is used as evaluation of its accuracy in recovering the hidden structures as well as the variation of the two simulated quantities along them. Knowing the ground-truth of the two filaments, a quantitative comparison with the structures recovered via the toolbox was possible, univocally proving the quality of its performance. We also showed how removing the normalization by the ``approximation to unity'' from SPH interpolations, is prone to producing artifacts in the final visualization of simulated quantities, misguiding inference from simulations.

The toolbox has also been applied to
two simulated data sets: a dwarf galaxy interacting with its host galaxy cluster and a filament from the cosmic web. In the first case, particular attention has been devoted to the onset of Star Formation in the arms generated by mixing of the dwarf's interstellar gas and the gas from the cluster. After the recovery of a dominant filament in the data set, its density, neutral fraction, metallicity and temperature have been analyzed with both our visualization techniques, finding favorable Star Formation conditions in the inner parts of the manifolds, along its whole elongation. In the second application, we studied the dynamics of a filament extracted from the simulated Cosmic Web. Focusing only on Dark Matter, we show that islands of opposite curl may appear in the core of the filament, confirming previous findings. Given the orthonormal coordinate plane visualization technique, we are able to estimate the number of these islands while scrolling along the manifold. Furthermore, to prove the wide range of possible applications of our toolbox, we studied the stellar filament of the $\omega$-Centauri globular cluster, recovered from the GAIA DR2 data set. The application to this particular data set proved successful, enabling a detailed study of the stellar number density along the manifold allowing for the computation of the local Signal-to-Noise Ratio, for a further constraint on its detectability. 

For future continuation of this work, we showcase the applicability and fitness of the proposed toolbox for extensively studying the properties of the cosmic web structures. It is then possible to explore the physical properties of the Dark Matter halos, gas, or individual galaxies in relation to these structures. As also mentioned in the introduction, the ability of these methodologies to handle very large point clouds opens the possibility of investigating unexplored regions of the Milky Way that are dense in stars in pursuit of  signs of merger debris. Aside from the examples mentioned so far, the proposed toolbox can be applied to a variety of cases and will help in the detection and analysis of filamentary structures with very different natures, hidden within noisy environments. 
It is essential to note that this also includes non-physical manifolds, (e.g. in high-dimensional parameter spaces). This work showcases the potential of 1-DREAM on manifolds in physical spaces, either simulated or observed. However, with no additional modifications or assumptions, it is possible to extend it to a variety of different spaces, such as the multidimensional parameter space of astronomical observations, or the color-magnitude diagram of stellar systems and many more. Some of these studies will be the subject of future works, but we believe that 1-DREAM will become a useful tool in many different fields.

\section{Acknowledgments}
This project has received financial support from the European Union's Horizon 2020 research and innovation program under the Marie Sklodowska-Curie grant agreement No. 721463 to the SUNDIAL ITN Network. Additional funding was also provided by the Alan Turing Institute, within the Fellowship 96102.


\bibliographystyle{elsarticle-harv.bst} 
\typeout{}
\bibliography{Bibliography}





\end{document}